\documentclass[a4paper, USenglish, cleveref, autoref, thm-restate]{lipics-v2021}
\hideLIPIcs
\usepackage{amsmath,amssymb,amsfonts, amsthm}
\usepackage{dsfont}
\usepackage{subcaption} 
\usepackage{multicol}
\usepackage{amssymb} 
\usepackage{cite}
\usepackage{tikz} 
\usetikzlibrary{calc, arrows.meta}
\usetikzlibrary{decorations.pathreplacing} 
\usetikzlibrary{arrows,decorations.markings}
\usetikzlibrary{backgrounds}

\nolinenumbers

\tikzset{
    mid arrow/.style={
        line width=0.75pt, 
        postaction={
            decorate,
            decoration={
                markings,
                mark=at position 0.5 with {\arrow{>}}
            }
        }
    }
}

\usepackage{hyperref}
\usepackage{csquotes}
\usepackage{float} 

\newtheorem*{objective}{Objective}

\newcommand{\CC}{\mathcal{C}}
\newcommand{\FF}{\mathcal{F}}

\newcommand{\Cbo}{\mathcal{C}_{b}^{\scriptscriptstyle <}}
\newcommand{\Cao}{\mathcal{C}_{a}^{\scriptscriptstyle <}}
\newcommand{\Cabo}{\mathcal{C}_{ab}^{\scriptscriptstyle <}}
\newcommand{\Cabco}{\mathcal{C}_{abc}^{\scriptscriptstyle <}}
\newcommand{\Cabcdo}{\mathcal{C}_{abcd}^{\scriptscriptstyle <}}

\input{rounding-rectangle-trick}

\definecolor{g6}{RGB}{220,220,220}
\definecolor{g5}{RGB}{200,200,200}
\definecolor{g4}{RGB}{180,180,180}
\definecolor{g3}{RGB}{150,150,150}
\definecolor{g2}{RGB}{130,130,130}
\definecolor{g1}{RGB}{100,100,100}
\definecolor{gris_clair}{RGB}{220,220,220}
\definecolor{gris}{RGB}{180,180,180}
\definecolor{gris_fonce}{RGB}{100,100,100}
\definecolor{gg}{rgb}{0.35,0.35,0.35}
\definecolor{ggg}{rgb}{0.6,0.6,0.6}
\definecolor{gggg}{rgb}{0.7,0.7,0.7}

\title{Constrained intersection representations and forbidden patterns}
\author{Laurent Feuilloley}{Université Claude Bernard Lyon 1, CNRS, LIRIS}{laurent.feuilloley@cnrs.fr}{ 0000-0002-3994-0898}{}

\author{Michel Habib}{Université Paris Cité, CNRS, IRIF}{habib@irif.fr}{ 0000-0002-8564-2314}{}

\author{Amalia Sorondo}{Université Paris Cité, CNRS, IRIF}{sorondo@irif.fr}{0009-0003-2077-242X}{}

\authorrunning{L. Feuilloley, M. Habib, A. Sorondo} 


\acknowledgements{We thank István Tomon for pointing us to references~\cite{PachT19}, \cite{PachT20}, \cite{MiddendorfP93} and \cite{Sinden66}.
We also thank the reviewers of a previous version for useful comments on definitions, references, and writing.  
}

\funding{This work was supported by the French ANR project ANR-24-CE48-4377 (GODASse).}

\begin{document}

\maketitle{}

\begin{abstract}
    We study the interplay between two types of graph classes: the ones defined by intersections of geometric objects and the ones defined by the absence of a substructure. More precisely, we study intersection graphs that are grounded, that is, the objects lie above a horizontal line that they touch, and ordered graphs with forbidden patterns, that is, ordered graphs that avoid a set of ordered induced subgraphs. 
    These two approaches are fundamentally different, but it is known that several interesting graph classes have both types of characterizations, and we aim to understand whether this is a general phenomenon. Our contributions are the following.
    
    First, we identify a family of patterns on four vertices and a natural hierarchy of intersection graphs of strings whose classes are nicely interleaved. This raises hopes to prove connections between the two perspectives. 
    
    Second, we dampen these hopes by proving that a class of (ordered) intersection graphs that we study cannot be characterized by a finite set of forbidden patterns.
    
    Third, we show a dual impossibility result: there exists a pattern-defined class of our hierarchy that admits no intersection representation, if we require this representation to be closed under natural operations such as translation and scaling. However, it turns out that if we allow slightly more exotic intersection representations, then we can find one for this pattern-defined class. 
\end{abstract}

\vfill{}

\paragraph*{AI disclosure}
An LLM has been used to suggest improvements to the writing and the pictures, and to assist with literature review. 
The authors take full responsibility for the paper. 

\newpage{}

\setcounter{page}{1}

\section{Introduction}
\label{sec:intro}

\subsection{Context, problem and motivation}

There are various ways of defining and characterizing graph classes. 
One can bound some structural parameter, describe how to build the graphs from a few basic graphs and a few operations, forbid some structure, translate geometric configurations into graphs, and so on.
In this paper, we focus on the connection between forbidden structures and geometric intersections, and unveil a rich and promising landscape. 

\begin{figure}[!ht]
\centering
\begin{subfigure}{0.4 \textwidth}
	\centering
	\includegraphics[scale=0.95]%
	{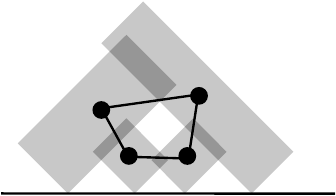}
	\vspace{0.2cm}
	\caption{\label{fig:grounded-rectangles} Grounded rectangles.}\end{subfigure}
\begin{subfigure}{0.4 \textwidth}
	\centering
	\vspace{1cm}
	\input{pictures-and-tikz/rectangle.tex}
	\vspace{1cm}
	\caption{\label{fig:rectangle-patterns} Associated pattern $P_b$.}
\end{subfigure}
\caption{\label{fig:rectangle}Grounded rectangle graphs and their forbidden pattern on four vertices.}
\end{figure}

\paragraph*{A motivating example}
Let us start from a concrete example that will serve to introduce our setting and motivate our study. 
Consider Figure~\ref{fig:rectangle}. 
On the left is a \emph{grounded rectangle configuration}: there is a horizontal line, called the \emph{grounding line}, and a set of rectangles whose sides are tilted by 45° with respect to the horizontal, and which touch the grounding line at their bottom corner. 
From this configuration we obtain a graph by taking one vertex per rectangle, with an edge between two vertices whenever the corresponding rectangles intersect (see the overlay graph in Figure~\ref{fig:grounded-rectangles}).
For this graph, there is a natural ordering on the vertices which is the ordering in which their rectangles touch the grounding line; we call this the \emph{grounding ordering}. 
In this paper, we will mostly consider graphs with a specified vertex ordering. 

\begin{definition}
An \emph{ordered graph} is a graph $G$ together with a total ordering $<$ of its vertex set, and will be denoted by $(G, <)$.
\end{definition}

It was proved several times independently~\cite{CorreaFPS15,Hixon13,CatanzaroCFHHHS15,SotoC15} (with four different terminologies and motivations, which we discuss in Section~\ref{sec:related-work}) that an ordered graph corresponds to a grounded rectangle configuration if and only if it avoids a specific four-vertex configuration. More precisely, there are no four vertices $a<b<c<d$ such that $(a,c)$ and $(b,d)$ are edges and $(b,c)$ is not an edge. Note that the four vertices do not need to be contiguous. Figure~\ref{fig:rectangle-patterns} depicts this forbidden structure, known in the literature as a \emph{pattern}~\cite{FeuilloleyH21}. This pattern will be called $P_b$ in this paper (for reasons that will be spelled out later). When a pattern is not present in an ordered graph (as in the characterization above) we say that the ordered graph \emph{avoids it}.

For future reference, let us restate this theorem with the terminology we use in this paper, where we write \emph{constrained} instead of \emph{grounded} when referring to ordered graph classes, to specify that the grounding ordering of the representation must coincide with the ordering of the graph.

\begin{theorem}[\cite{CorreaFPS15,Hixon13,CatanzaroCFHHHS15,SotoC15}]\label{thm:grounded-rectangle}
An ordered graph $(G,<)$ has a constrained rectangle representation if and only if it avoids the pattern $P_b$. 
\end{theorem}

\paragraph*{A general phenomenon?}

Theorem~\ref{thm:grounded-rectangle} is striking for two reasons. First, it establishes a clean equivalence between two a priori unrelated classes of ordered graphs. 
But more surprisingly, it reduces a 2-dimensional object (the rectangles in the plane) to a 1-dimensional object: a total ordering on the vertices. 
Of course, this happens because the 2-dimensional objects are grounded on a 1-dimensional object (the grounding line), yet it means that the complex interactions that could happen above the grounding line are captured by looking only at the order and adjacency of quadruples of vertices.  

An important clarification is needed here: Theorem~\ref{thm:grounded-rectangle} is stated for ordered graphs, but in the literature one finds the unordered version: for an unordered graph $G$, $G$ has \emph{an} ordering with a grounded rectangle representation if and only if \emph{there exists} an ordering that does not contain the pattern $P_b$. 
Unordered graphs are more standard than ordered graphs, but going from one to the other via the existential quantification on the ordering is artificial: the proof naturally considers the grounding ordering and the ordering avoiding the pattern to be the same.
In our paper, we will mainly consider ordered graphs, because they are the natural underlying object. But in the last section, we adapt some of our theorems to the unordered case, for easier comparison with the literature and to facilitate future work. 

Getting back to our discussion, the appeal of Theorem~\ref{thm:grounded-rectangle} is strengthened by the fact that there are other geometric classes that can be characterized by similar ``forbidden patterns''.\footnote{The notion of pattern will be formalized below. 
Patterns are simply compact representations for sets of forbidden induced ordered subgraphs.} 
Interval graphs, which are defined by intersections of intervals on a line, can also be captured by a forbidden pattern~\cite{Olariu91,RamalingamR88}, and the so-called grounded L-shape graphs, which fall within the scope of this study, can be characterized by a pair of patterns~\cite{JelinekT19}.  
If we allow two grounding lines instead of one, we can also capture permutation graphs and cocomparability graphs \cite{GolumbicT04}.

At that point, we ask: Is it a coincidence that for a few classes two different points of view align exactly, or is there a structural reason why this happens? 
In particular, the known equalities are for rather simple geometric objects and rather simple patterns; what happens for more complex classes? 

\begin{objective}
Understand the relationships between the classes that are characterized by grounded intersection configurations and those defined by patterns.
\end{objective}

\paragraph*{Structural and algorithmic motivations}

Beyond the curiosity of understanding the relation between two graph definition paradigms, we claim that this research direction can have an impact on intersection graphs and pattern-defined classes independently, and also has potential for algorithmic applications. 

For the field of intersection graphs, first, having characterizations by forbidden structures is very handy: they transform objects that in general are encoded via reals into finite, discrete objects. 
A very basic application is for recognition: while intersection graphs in general can be hard to recognize because they may require large encodings (which can be captured formally by hardness for the existential theory of the reals), the ones that are characterized by a finite forbidden structure naturally land in NP (which is contained in latter). 
Even better and more concretely, the core of the algorithms to recognize interval graphs and permutation graphs (two prominent intersection graph classes) consists in building an ordering avoiding the corresponding patterns~\cite{CorneilOS09, McConnellS99}. 
Pattern characterizations are also used to prove structural results on intersection graphs, for example \cite{Wood06} obtains bounds on the bandwidth this way. 

For interval and permutation graphs the ordering also permits linear-time greedy algorithms for classic NP-complete problems such as maximum independent set, coloring~\cite{Golumbic2004}
and hamiltonicity for interval graphs and cocomparability graphs \cite{CorneilDH13}. Also to be noted here: for some algorithms on geometry-defined graphs, the geometry is not even needed: they just use combinatorial properties. 
For example, there exists a polynomial-time approximation scheme (PTAS) for maximum independent set in unit disk graphs that does not require the knowledge of the disk centers~\cite{NiebergHK04}.  

In the opposite direction, a geometric representation can also be very useful. 
First, if a graph is known to be some kind of intersection graph of grounded strings, then it inherits the results on outer-string graphs, for example the subexponential-time algorithm for maximum independent set~\cite{FoxP11}. Closer to our paper, the algorithm of~\cite{CorreaFPS15} for computing a maximum independent set on constrained rectangle graphs uses the geometry characterization and not the ordering: it is a dynamic program whose induction relation corresponds to rectangles being included in some area of the plane. More generally, forbidden pattern characterizations can be quite mysterious: it was noted in \cite{FeuilloleyH21} that patterns on three vertices are well-understood, but basically nothing can be said about patterns on four vertices or more.

Finally, independently of geometry, patterns on ordered graphs are themselves a very active topic in algorithms, with recent results for matching and coloring using such structures~\cite{BaratFT25, HajebiLS24, PiecykR26, AxenovichRU18}. 
At a more abstract level, the \emph{twin-width} of ordered graphs was characterized by forbidden structures in~\cite{BonnetGMSTT24}, and twin-width is one of the key modern algorithmic parameters, for example FO model checking
is fixed-parameter tractable on classes of bounded twin-width (given a contraction sequence)~\cite{BonnetKTW22}. 

\paragraph*{A formal definition of pattern}

Before we discuss our contributions, we give a formal definition of patterns, since the definition above was informal, and it is a crucial notion of the paper.

A \emph{pattern} $P$ is an ordered graph on vertices $p_1<\dots<p_k$ in which
every pair of vertices is labelled as an \emph{edge}, a \emph{non-edge}, or
\emph{undecided}. An ordered graph $(G,<)$ \emph{contains} $P$ if there exist
vertices $v_1<\dots<v_k$ of $G$ such that, for all $i<j$, $(v_i,v_j)\in E(G)$
whenever $(p_i,p_j)$ is an edge of $P$, and $(v_i,v_j)\notin E(G)$ whenever
$(p_i,p_j)$ is a non-edge of $P$; undecided pairs are unconstrained, and the
vertices $v_1,\dots,v_k$ need not be consecutive in $<$.
Otherwise, $(G,<)$ \emph{avoids} $P$.
For a family of patterns $\FF$, we denote by $\CC^{\scriptscriptstyle <}_{\FF}$
the class of ordered graphs that avoid every pattern of $\FF$, and by
$\CC_{\FF}$ the class of (unordered) graphs $G$ that admit an ordering $<$ such
that $(G,<)\in\CC^{\scriptscriptstyle <}_{\FF}$. We say that $\FF$
\emph{characterizes} a class $X$ of ordered graphs if
$X=\CC^{\scriptscriptstyle <}_{\FF}$, and a class $Y$ of unordered graphs if
$Y=\CC_{\FF}$. When $\FF=\{P\}$, we write $\CC^{\scriptscriptstyle <}_{P}$ and
$\CC_{P}$.

\subsection{Contributions}
Let us now discuss our contributions. 
As mentioned before, except for the last set of results, all the graphs considered are \emph{ordered graphs}. 

\paragraph*{Identifying the relevant classes}

\begin{figure}[!h]
\begin{center}
\input{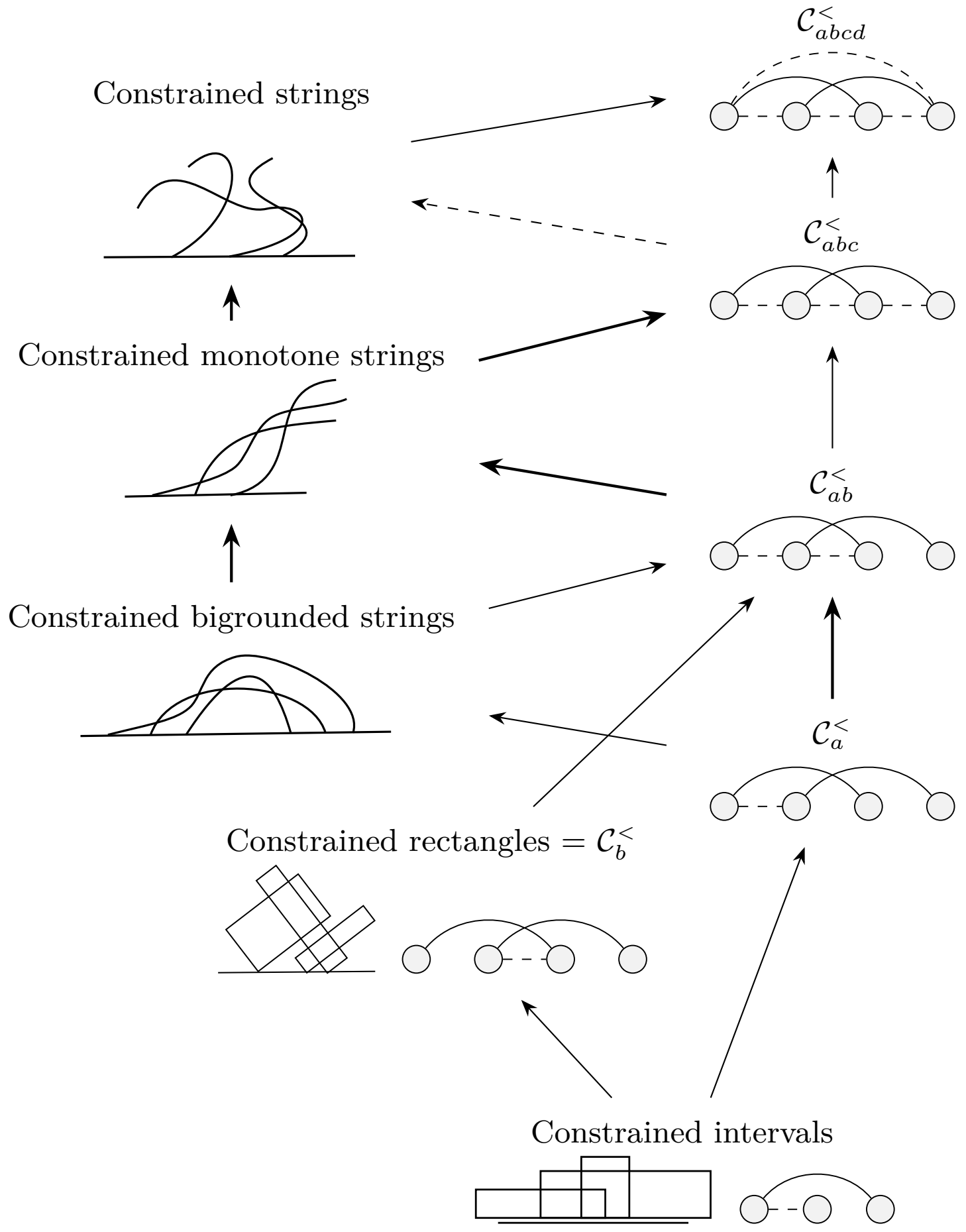}
\end{center}
\caption{
\label{fig:diagram-hierarchy} Diagram representing the interleaving of pattern-defined and geometry-defined classes we study. The arcs represent inclusions in the natural order, thick arcs represent strict inclusions. The dashed edge is a conjectured inclusion. We place here the constrained interval graphs and the constrained rectangle graphs for easy reference, but the discussion focuses on the  classes above.
}
\end{figure}

Our first contribution is to \emph{identify a hierarchy of geometric configurations and a hierarchy of patterns}, and show that the classes defined by \emph{these two hierarchies are interleaved}. 
We refer to Figure~\ref{fig:diagram-hierarchy} for a visual aid in this discussion. 

Let us define the objects we use. These definitions are stated in more detail in Appendix~\ref{sec:definitions}.

\begin{figure}[!h]
\centering
\scalebox{1}{
\input{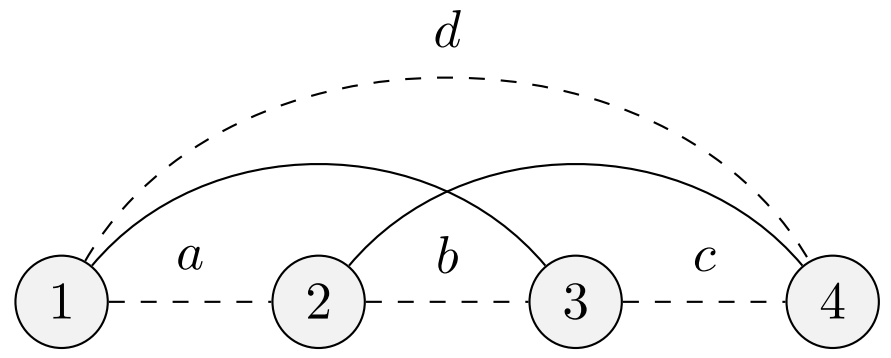}}
\caption{\label{fig:abcd-notations}
Illustration of the patterns we consider.}
\end{figure}

Let us start with the classes defined by patterns. The patterns we consider all have the same general form, depicted in Figure~\ref{fig:abcd-notations}. 
They have four vertices $v_1<v_2<v_3<v_4$, and the edges $(v_1,v_3)$ and $(v_2,v_4)$. What differentiates these patterns is which remaining pairs of vertices are non-edges and which are unconstrained. We use the following shortcuts $a=(v_1,v_2)$, $b=(v_2,v_3)$, $c=(v_3,v_4)$ and $d=(v_1,v_4)$. Then, each pattern is denoted by $P_{S}$ where $S$ is the concatenation of the labels of the non-edges considered. The class $\mathcal{C}_{S}^{\scriptscriptstyle <}$ is the class of ordered graphs that avoid $P_S$. For example, $\Cabo$ is the class of ordered graphs $(G=(V,E),<)$ where there are no four vertices $u<v<w<x$ with $(u,w)$, $(v,x)\in E$, and $(u,v), (v,w)\notin E$. These patterns appear naturally when one considers grounded geometric intersections, in particular when starting from the pattern $P_b$ characterizing the constrained rectangle graphs.  

Now for the geometrically-defined graph classes. We consider intersections of shapes that are \emph{grounded on a line}, that is, they all touch a horizontal line in at least one point, and are only present above this line. These configurations naturally define ordered graphs, where every vertex represents a shape, every edge represents an intersection, and the ordering is the ordering of the leftmost intersection with the horizontal line. (These leftmost intersections, the grounding points, are unique.)  
We use names containing ``\emph{constrained}'' to indicate that we consider ordered grounded configurations. 
Some classes we consider are known under different names in the unordered case; more background will be given in the related work section (Section~\ref{sec:related-work}).

Our most general class is the class of \emph{constrained string graphs}, which are the graphs corresponding to curves that touch the grounding line only on their grounding point, without further constraints. (The unordered class is known as outer-string graphs.) 
We also consider \emph{constrained monotone string graphs}, where in addition the curves must be increasing and monotone in both axes.
Finally, we consider \emph{constrained bigrounded string graphs} which correspond to $x$-monotone curves (i.e. functions) that touch the grounding line exactly on their two  endpoints. (The unordered case is known as \emph{interval filament graphs}.) See Figure~\ref{fig:diagram-hierarchy} for illustrations of the geometric configurations.  

Both the pattern-defined classes and the geometry-defined classes form hierarchies. For patterns: $\Cao \subseteq \Cabo \subseteq \Cabco \subseteq \Cabcdo$ follows from the definitions of pattern-defined classes, see \cite{FeuilloleyH21} (one can also replace $\Cao$ by $\Cbo$). 

For geometric classes, it is clear that constrained string graphs contain both constrained monotone string graphs and constrained bigrounded string graphs. 
The fact that constrained monotone string graphs contain constrained bigrounded string graphs is less direct, but can be derived by geometric arguments (see Corollary~\ref{cor:bigrounded-in-monotone} in Section~\ref{sec:no-characterization-by-patterns}). 

\paragraph*{Proving the interleaving of the hierarchies}

We now state our first set of results, which establish the interleaving of the classes. 
We start with the inclusions of the form: a geometry-defined class included in a pattern-defined class. 

\begin{restatable}{theorem}{thmGeometryInPatterns}
The following three inclusions hold:
\begin{enumerate}
    \item Constrained bigrounded string graphs are included in $\Cabo$.
    \item Constrained monotone string graphs are included in $\Cabco$.
    \item Constrained string graphs are included in $\Cabcdo$.
\end{enumerate}
\label{theo:geometry-in-patterns}
\end{restatable}

These statements follow from Jordan-curve arguments: one considers an arbitrary configuration of the given type, and notes that the pattern cannot appear, because of the geometric constraints. 
The last inclusion was already known, see \emph{e.g.} \cite{PachT19}; we reprove it for completeness.
See Subsection~\ref{subsec:geometry-in-pattern} for the full proofs.

We now move on to the other type of inclusion, where a pattern-defined class is proved to be included in a geometric class. 

\begin{restatable}{theorem}{thmCaIntervalFilament}
\label{thm:Ca-Interval-Filament}
The class $\Cao$ is included in constrained bigrounded string graphs.
\end{restatable}

\begin{restatable}{theorem}{ThmCabInGstairs}
\label{thm:Cab-in-Gstairs} 
The class $\Cabo$ is included in constrained monotone string graphs. 
\end{restatable} 

The way we prove them is by incrementally constructing a geometric representation of an arbitrary pattern-avoiding graph. These proofs are more challenging, because one needs to describe precisely the geometric objects, to make sure that the end result corresponds exactly to the original graph, and more importantly, that the only obstruction to the construction is the pattern that is avoided.

The full proofs appear, respectively, in Subsection~\ref{subsec:Ca-Interval-Filament} and~\ref{subsec:Cab-in-Gstairs}. 

An interesting question is whether $\Cabco$ is included in constrained string graphs. The literature tells us that the co-holes equipped with the natural ordering are the only known obstructions to constrained string graphs~\cite{Sinden66, antic2026two}. If there are no other obstructions, then the inclusion would hold, because forbidding $P_{abc}$ rules out all these ordered co-holes. We state this as a conjecture. 

\begin{conjecture}
The class $\Cabco$ is included in constrained string graphs. 
\end{conjecture}

\paragraph*{No pattern characterization for constrained monotone string graphs}

Now that we have identified pattern-defined and geometry-defined classes that are tightly connected, we would now like to prove an equality between a pair of such classes.   
A promising candidate here is the pair made of constrained monotone string graphs and $\Cabco$. 
Indeed, we have proved that constrained monotone string graphs are included in $\Cabco$, and at first sight, the only obstacle to an incremental construction (as for Theorems~\ref{thm:Ca-Interval-Filament} and~\ref{thm:Cab-in-Gstairs}) should be the pattern $P_{abc}$.

It turns out constrained monotone string graphs are actually \emph{strictly included} in $\Cabco$. The following theorem can be deduced from a paper by Pach and Tomon~\cite{PachT20}.

\begin{theorem}[Adapted from~\cite{PachT20}]
\label{thm:pach-tomon-graph}
There exists an ordered graph that avoids $P_{abc}$ but cannot be represented as a constrained monotone string graph.
\end{theorem}

\begin{figure}[!h]
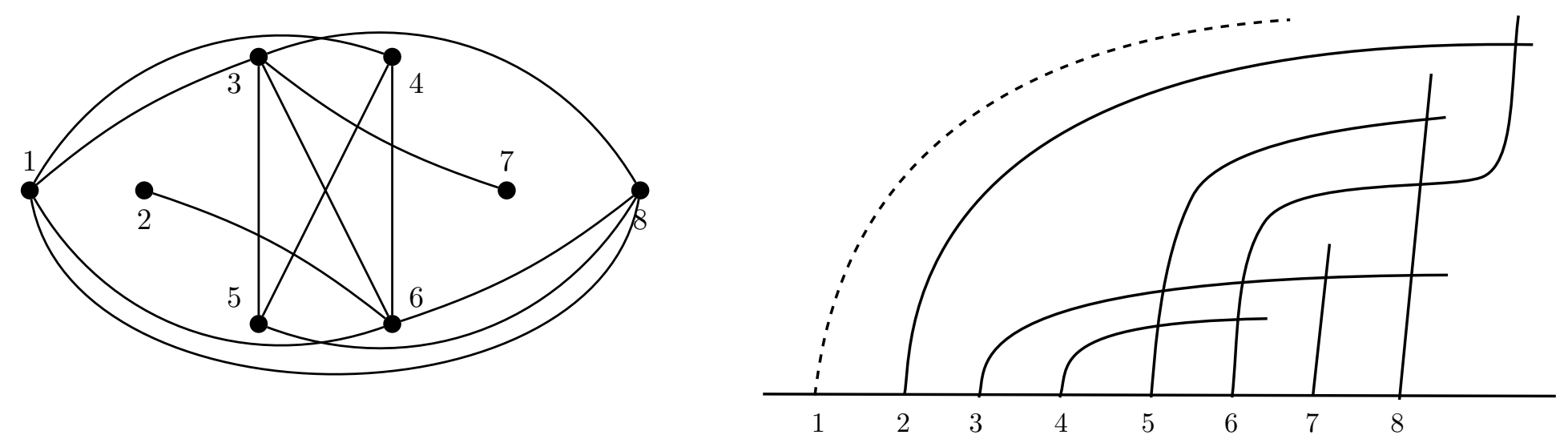

\centering
\begin{subfigure}{0.45 \textwidth}
	\centering
	\scalebox{0.7}{
    \input{pictures-and-tikz/pach-tomon-graph}}
	\caption{\label{fig:pach-tomon-graph-intro}The Pach-Tomon graph}
	\end{subfigure}
\begin{subfigure}{0.5 \textwidth}
	\centering
    \scalebox{0.65}{
    \input{pictures-and-tikz/pach-tomon-try}}	\caption{\label{fig:pach-tomon-try} Tentative representation.}
\end{subfigure}
\caption{On the left, the Pach-Tomon graph, with its vertex ordering. One can check that the pattern $P_{abc}$ does not appear in this ordering. 
On the right, the result of a tentative incremental construction, starting from the rightmost vertex: one can place the strings of vertices 8 down to 2 easily, but it is impossible to draw the string of vertex 1: it has to go above the string of 2, but cannot reach the string of vertex 4, as the graph requires.}
\end{figure}

Theorem~\ref{thm:pach-tomon-graph} can be proved by studying what we call the \emph{Pach-Tomon graph},  defined in Figure~\ref{fig:pach-tomon-graph-intro} and adapted from a construction in~\cite{PachT20}. 
\footnote{Readers interested in this specific statement might be confused by the difference of vocabulary and focus between this paper and  \cite{PachT20}. In Section~\ref{sec:no-characterization-by-patterns}, we establish very precisely the connection between the two. For this introduction, we stay at the level of intuition with Figures~\ref{fig:pach-tomon-graph-intro} and~\ref{fig:pach-tomon-try}.}

At that point, the natural strategy is: (1) to take the Pach-Tomon graph, transform it into a pattern by replacing the edges that are not in the graph by forbidden edges, (2) to add it to $P_{abc}$ in the list of forbidden patterns, and (3) to hope that this new set of forbidden patterns characterizes constrained monotone string graphs.

We prove that not only this would not work, but \emph{there is no characterization of constrained monotone string graphs using a finite number of patterns}!  

\begin{restatable}{theorem}{thmNoCharacterizationStairs}
\label{thm:no-characterization-monotone}
There is no finite set of patterns that characterizes constrained monotone string graphs. 
\end{restatable}

We note that the word \emph{finite} is important: the geometry-defined classes we consider are hereditary (that is, closed by vertex removal), and such classes can always be defined by an \emph{infinite} set of forbidden induced subgraphs or patterns. 
Just as for unordered graphs, one can define these forbidden induced subgraphs by simply collecting all the inclusion-minimal obstructions. 

Of course, Theorem~\ref{thm:no-characterization-monotone} is disappointing, since it rules out our most promising candidate for a new equality between pattern-defined and geometry-defined classes.
But it is also an exciting result, because we have very few results about classes that cannot be characterized by a finite number of patterns. We will get back to this at the end of the introduction. 

The proof proceeds in several steps. 
First, we adapt a theorem from~\cite{PachT20} to our setting, which characterizes constrained monotone string graphs with \emph{two} orderings. 
(The adaptation is not very complicated, but discussing it here would require introducing the terminology of~\cite{PachT20}. It is detailed in Appendix~\ref{sec:no-characterization-by-patterns} and consists of re-grounding on a horizontal line, replacing $x$-monotone curves by monotone ones, and taking
complements.)

Before we state our theorem, let us introduce the pattern used for characterizing the class of cocomparability graphs and therefore known as the \emph{cocomparability pattern}. Given an ordered graph $(G,<)$, with $G=(V,E)$, a triplet of ordered vertices $x<y<z$ will create the cocomparability pattern if $(x,z) \in E$ and $(x,y), (y,z) \notin E$.

\begin{restatable}[Adapted from Pach-Tomon~\cite{PachT20}]{theorem}{ThmPachTomonAdaptation}
\label{thm:Pach-Tomon-adapted}
An ordered graph $(G,<)$, with $G=(V,E)$, is a constrained monotone string graph if and only if there exists a second ordering $<_h$ such that
for every triplet of vertices $u < v < w$ that create the cocomparability pattern, it holds that $v <_h u$ and $v <_h w$.
\end{restatable}

\begin{figure}[!h]
    \centering
    \input{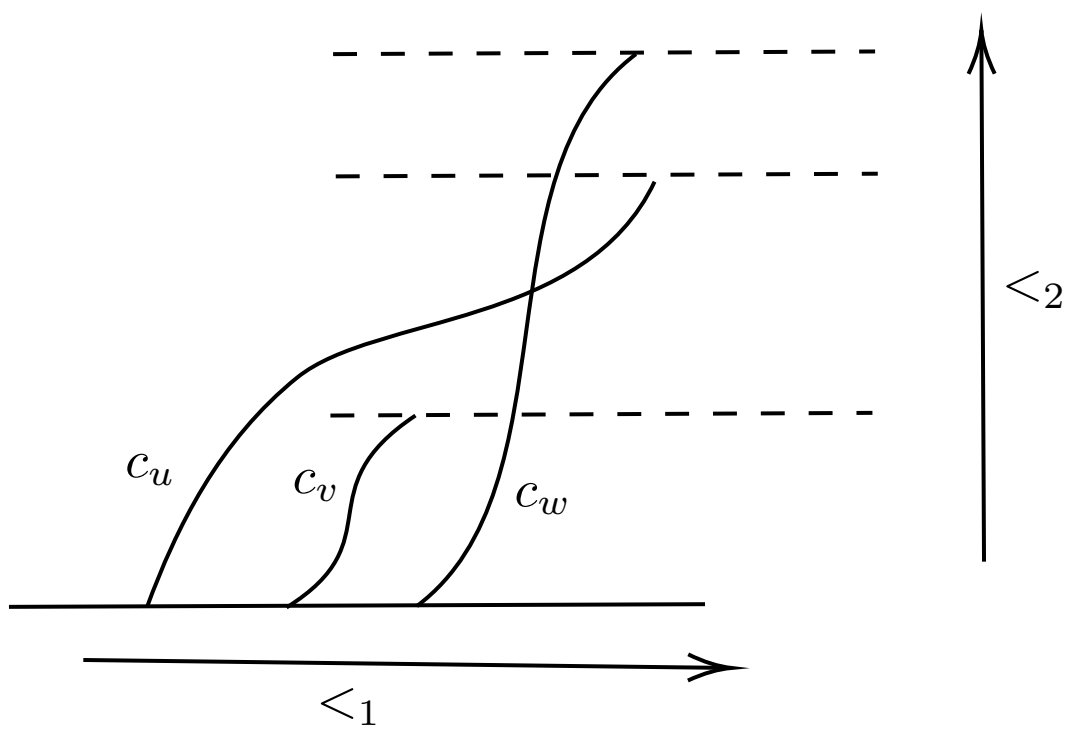}
    \caption{Illustration of Theorem~\ref{thm:Pach-Tomon-adapted}. Consider three grounded monotone strings, $c_u,c_v,c_w$ grounded in this order, that represent the vertices $u<v<w$. 
    If $(u,v,w)$ is a cocomparability pattern, it means that the only intersection between these strings is between $c_u$ and $c_w$. 
    Because the grounding point of $c_v$ is between these two curves, it must be that the non-grounding endpoint of $c_v$ is below the intersection point of $c_u$ and $c_w$, and by monotonicity it is also below the non-grounding endpoints of these curves. 
    When we take the second ordering $<_h$ as the height of the non-grounding endpoint, we directly get that $v <_h u$ and $v <_h w$. 
    Theorem~\ref{thm:Pach-Tomon-adapted} establishes that this is an equivalence.}
    \label{fig:two-orderings}
\end{figure}

See Figure~\ref{fig:two-orderings}, for an intuitive explanation of this theorem.

\begin{figure}[!h]
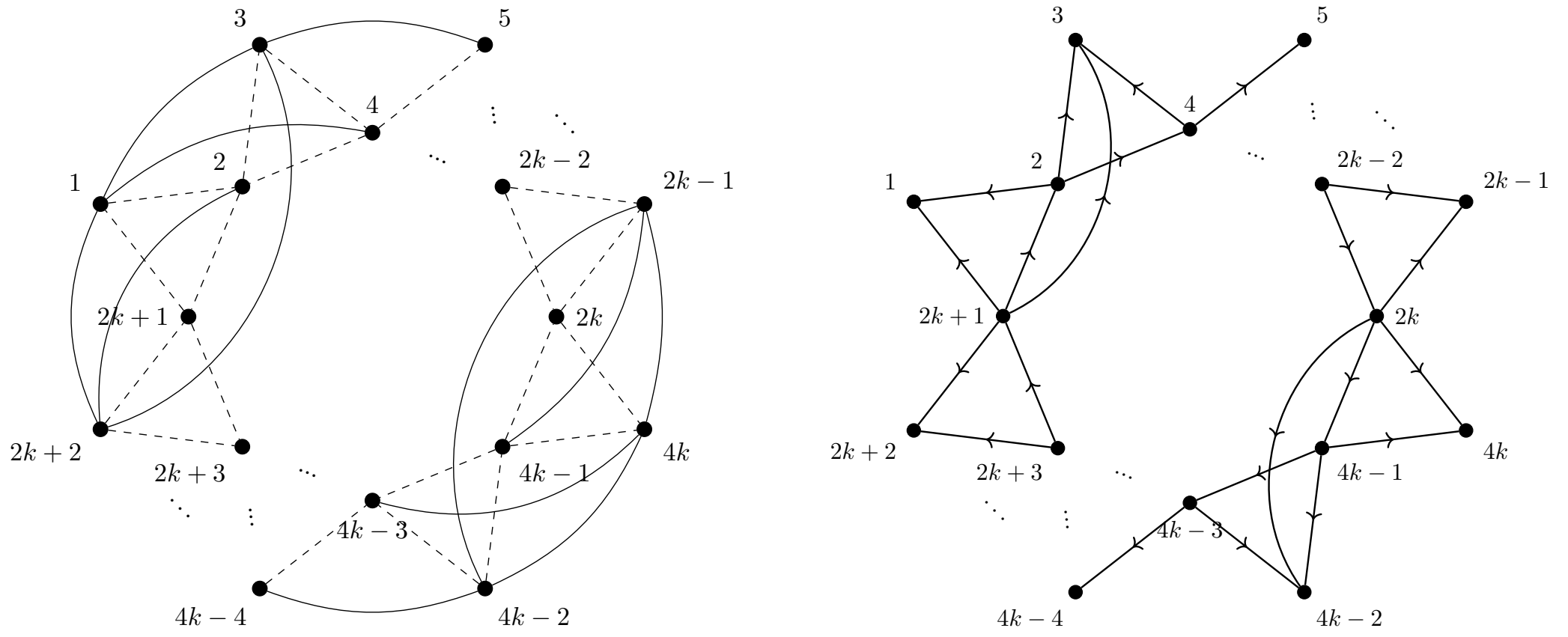

    \centering
    \begin{subfigure}{0.48\textwidth}
    \scalebox{0.65}{
        \input{pictures-and-tikz/cocomp-sun}}
    \end{subfigure} \hspace{2ex}
    \begin{subfigure}{0.48\textwidth}
    \scalebox{0.6}{
        \input{pictures-and-tikz/Dh}}
    \end{subfigure}
    \caption{The ordered graph $(G,<)$ (left) and its directed height graph $D_h(G,<)$ (right). All edges of the join between vertices in $\{4,5,6,...,2k \}$ and vertices in $\{2k+1, 2k+2, ..., 4k-3 \}$ are also edges of $(G,<)$, though omitted from the figure. The arrows in $D_h(G,<)$ correspond to cocomparability triplets in $(G,<)$, take as an example the cocomparability triplet $(1, 2, 3)$ in $(G,<)$ inducing arcs $(2 \rightarrow 1)$ and $(2 \rightarrow 3)$ in $D_h(G,<)$. We prove that the arcs of $D_h(G,<)$ are the ones depicted (refer to Proposition \ref{propos:cocomparability-sun}). Then, it is direct to see that the smallest, and only, directed cycle is the one formed by vertices $2, 4, ..., 2k, 4k-1, 4k-3, 2k+1$.}
    \label{fig:Cocomparability-sun-and-Dh-intro}
\end{figure}

We change perspective on Theorem~\ref{thm:Pach-Tomon-adapted}, and phrase it in the following way. To check whether a graph is a constrained monotone string graph, one can consider all the occurrences of the cocomparability pattern, deduce all the necessary ordering constraints on $<_h$, and check whether there is a directed cycle in these constraints. 
If there is no directed cycle, then we can build a total ordering $<_h$, and the graph is in the class, whereas if there is a directed cycle, then the graph cannot be in the class. 
We make this concrete by defining a directed graph $D_h(G,<)$, whose vertex set is the same as $G$, and where for every cocomparability triple $x < y < z$ in $(G,<)$, we add the arcs $(y \rightarrow x)$ and $(y \rightarrow z)$ to $D_h(G,<)$. Intuitively, the arc $(u \rightarrow v)$ in $D_h(G,<)$ represents the fact that $u <_h v$, and this is why we call $D_h(G,<)$ the \emph{directed height graph} of $(G,<)$. This is illustrated by Figure \ref{fig:Cocomparability-sun-and-Dh-intro}.

Next, we prove that there is an infinite family of ordered graphs, that we call the \emph{cocomparability suns}, such that the corresponding directed graphs $D_h$ have arbitrarily large directed girth (that is, their smallest directed cycle can be arbitrarily large). 
Cocomparability suns are described in Figure~\ref{fig:Cocomparability-sun-and-Dh-intro}, and the (long) formal definition is Definition~\ref{def:cocomparability-sun} in Section~\ref{sec:no-characterization-by-patterns}.

We conclude with the following argument: 
    
\begin{enumerate}
    \item Assume for the sake of contradiction that there is a finite family of patterns $\mathcal{F}$ that characterizes the class.
    \item There exists a cocomparability sun $(G,<)$ such that $D_h(G,<)$ has directed girth larger than the maximum number of vertices in these patterns. 
    \item This cocomparability sun is not in the class, because $D_h(G,<)$ contains a directed cycle. 
    \item Then, by the hypothesis, it must contain one of the patterns in $\mathcal{F}$. 
    \item Let $(K,<_K)$ be an ordered subgraph of $(G,<)$, induced by an occurrence of a pattern in $\mathcal{F}$.
    \item As $(K,<_K)$ contains a forbidden pattern, it is not in the class. 
    \item By construction, $D_h(K,<_K)$ is a directed subgraph of $D_h(G,<)$. But then, as the directed girth of $D_h(G,<)$ was strictly greater than the size of $D_h(K,<_K)$, the directed height graph $D_h(K,<_K)$ is necessarily acyclic. Hence, $(K,<_K)$ must be in the class.
    \item There is a contradiction between the last two items. 
    \end{enumerate}
    

\paragraph*{Pattern-defined class with no geometry characterization}

We have just established that some natural geometry-defined classes cannot be captured by forbidden patterns. 
We prove the opposite (but just as negative) type of statement: there are natural pattern-defined classes that cannot have a geometric characterization.  
More precisely, since ``geometric characterization'' is not a well-defined term, we will define a few constraints, or axioms, that a \emph{standard constrained intersection characterization} should satisfy, and prove that there are pattern-defined classes, for patterns of our hierarchy, for which there is no standard constrained intersection characterization. The constraints for such a standard characterization are that shapes should be compact and connected, and that the set of admissible geometric configurations should be stable under the following shape operations: removal, translation along the grounding line, uniform scaling (i.e. scaling while keeping the same shape proportions), and duplication (at a fresh grounding point). Note that this is satisfied for all the geometric classes considered in the paper.

Based on these axioms, we prove that $\Cao$ does not admit a standard constrained intersection characterization.

\begin{restatable}{theorem}{ThmNoGeometricCharac}
There is no standard constrained intersection characterization of the graph class~$\Cao$.
\end{restatable}

\begin{figure}[h!]
    \centering
    \input{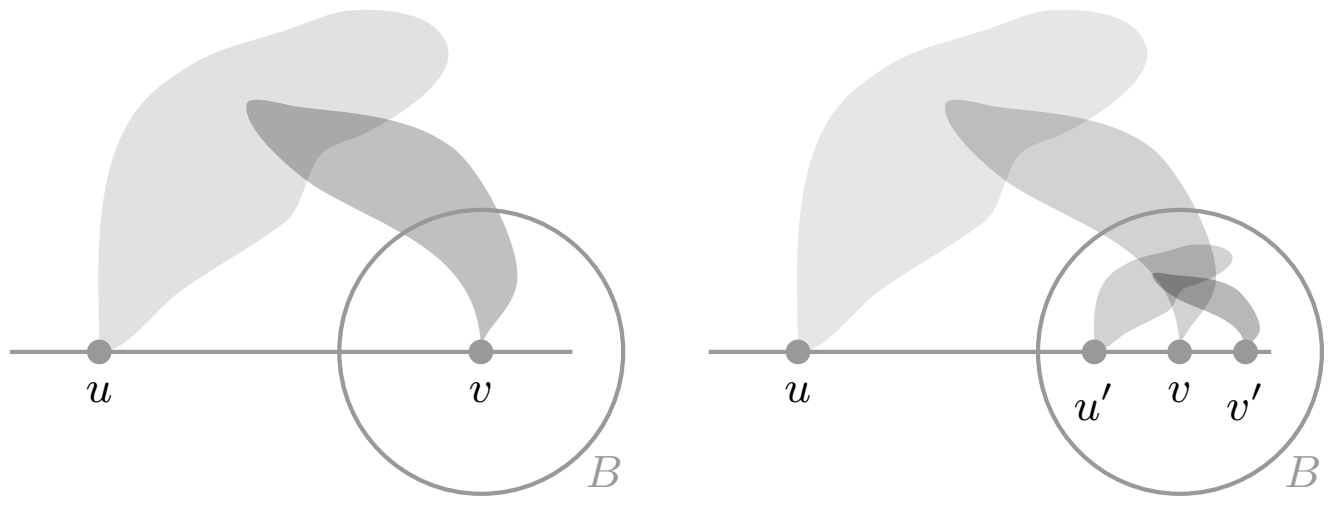}
    \caption{We start from a valid constrained representation of an ordered graph in $\Cao$, where the interval pattern is reproduced. Then, we claim the existence of two shapes $u$ and $v$ and of a ball $B$ around the grounding point of the shape $v$ that cannot be accessed by the shape $u$. We create a copy of shapes $u$ and $v$, denoted by $u'$ and $v'$, fully included in $B$, obtaining a valid constrained representation of an ordered graph in $\Cao$ whose grounding ordering reproduces pattern $P_a$ by vertices $(u,u',v,v')$, a contradiction.}
    \label{fig:no-model-c_a-intro}
\end{figure}

The proof consists in assuming such a standard characterization exists, and then taking an arbitrary geometric configuration that would avoid the pattern $P_a$, but not the interval pattern (i.e. the three-vertex pattern formed by vertices $\{1,2,3\}$, edge $(1,3)$, non-edge $(1,2)$ and undecided edge $(2,3)$). We know such a configuration must exist as the class of ordered graphs avoiding the interval pattern is strictly included in $\Cao$. Then, modifying this configuration via the allowed operations (removal, translation, uniform scaling, and duplication), we obtain a configuration of a graph in $\Cao$ that contains $P_a$, a contradiction. See Figure~\ref{fig:no-model-c_a-intro}.

We could not find another example of such an axiomatization approach in the intersection graph literature. 
What does exist is a set of tools to prove that a graph cannot be represented by (grounded) string graphs, arguably the most general class of (grounded) intersection graphs, for a natural definition of (grounded) intersection graphs. 
For example, outerstring graphs, hence also constrained string graphs, are $\chi$-bounded~\cite{RokW19} (i.e. their chromatic number is bounded by a function of the maximum clique size), implying that any class that is not $\chi$-bounded cannot be contained in outerstring graphs.    
There is some related work on intersection graphs with scaling, that we will discuss in the Section~\ref{sec:related-work}, but it is not comparable to our approach. 

\paragraph*{Beyond standard intersection graphs}

\begin{figure}[!h]
    \centering
    \input{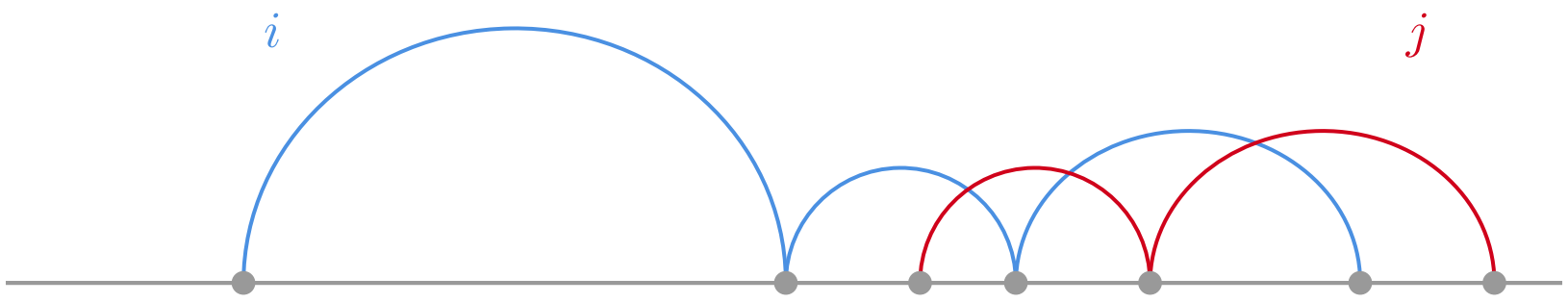}
    \caption{Illustration of two threads $i$ and $j$. The first-interval intersection constraints impose that thread $i$ must intersect the first half-circle of thread $j$. Note how threads $i$ and $j$ can have possibly multiple intersections.}
    \label{fig:threads-intro}
\end{figure}

The impossibility of having a standard geometric characterization for $\Cao$ is unfortunate, because, as explained earlier, geometry is a valuable tool for understanding pattern-defined classes. 
It is relevant to try to find a geometric representation that is not standard, yet still simple enough to be insightful. 
We introduce a new shape called \emph{thread}, defined as follows and illustrated by Figure~\ref{fig:threads-intro}.

\begin{restatable}[Grounded thread]{definition}{DefGroundedThread}
A grounded thread is a shape, defined on and above the grounding line, formed by a sequence of half-circles of increasing abscissas and connected by their extremities, such that the two extremities of each half-circle are anchored to the grounding line.
\end{restatable}

The grounding ordering of a grounded thread representation is the ordering defined by the leftmost grounding point of each thread.
Now we need to add specific requirements that make the intersection representation non-standard, and this is what we call the \emph{first-interval intersection} constraints. 
The constraints capture the fact that the first half-circle of a thread behaves differently than the rest. First, the first half-circles of such representations should not intersect nor be nested. Second, for every intersecting pair of threads $i$ and $j$, such that $i$ is smaller than $j$ in the grounding ordering, thread $i$ should intersect the first half-circle of $j$. We refer to Figure~\ref{fig:threads-intro} for a better explanation and illustration.

This first-interval intersection constraint is what makes this representation non-standard: by translating a thread, we could violate the constraint. We prove the following theorem.

\begin{restatable}{theorem}{ThmCaFilamentLinked}
The class $\Cao$ is exactly the class of ordered graphs admitting a constrained first-interval intersection representation of threads.
\end{restatable}

The proof shows both inclusions. First, for a graph ordered by $<$ avoiding the pattern $P_a$, we build a constrained thread intersection representation in which each vertex $i$ gets a first half-circle, plus an extra half-circle per neighbor $j$ to its right in $<$ (i.e. with $i<j$), intersecting the first half-circle of thread $j$. The representation is constructed such that any unwanted intersection between non-adjacent vertices would create pattern $P_a$, a contradiction. Conversely, in any first-interval intersection thread representation, an occurrence of $P_a$ on vertices $a<b<c<d$ would force the thread of $a$ to fully contain the first half-circle of $b$, which prevents any intersection between $b$ and $d$ without also intersecting $a$, also a contradiction.

We are aware of another non-standard geometry that characterizes pattern-defined classes. 
Outerplanar graphs, which correspond to the pattern with only the two interleaved compulsory edges $P_{\emptyset}$~\cite{BernhartK79}, can be characterized by some contact graphs~\cite{CatanzaroCFHHHS15}, that is, configurations where a shape can touch another but not cross it. 

\paragraph*{Transferring results to non-ordered classes} 

All the results above are for ordered graphs and constrained geometric configurations: the ordering is part of the object studied. This is not a new point of view (see related work). However, the most standard objects are unordered intersection graphs. 
For a pattern $P$, one usually studies the class of the graphs for which \emph{there exists} an ordering of the vertices that avoids the pattern. 
Likewise, a graph $G$ will belong to a grounded intersection graph class, if \emph{there exists} a grounding ordering that puts the graph in the class.  
In this paper, it feels more natural to study ordered graphs, because a priori, the only relevant cases are when the pattern-avoiding ordering is the same as the grounding ordering. 
But in order to get theorems that would be directly useful in the context of the existing literature, we would like to adapt our results to the unordered case.

The challenge in going from the ordered to the unordered case is that in the latter there can be many different vertex orderings that work for the geometry, and many different vertex orderings that avoid a given set of patterns, these two sets could be equal, overlapping, or even disjoint. Therefore, it is difficult to reason about all these orderings at once. 

\begin{figure}[h!]
    \centering
\begin{subfigure}{0.3 \textwidth}
	\centering
	\scalebox{0.75}{
    \input{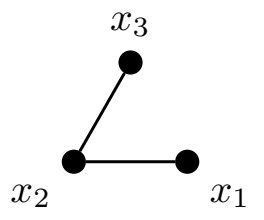}}
    \caption{\label{fig:original-graph-ordering-gadget} Ordered graph $(G,<)$, with vertices $x_1 < x_2 < x_3$.}
\end{subfigure}\hspace{5ex}
\begin{subfigure}{0.4 \textwidth}
	\centering
    \scalebox{0.75}{
    \input{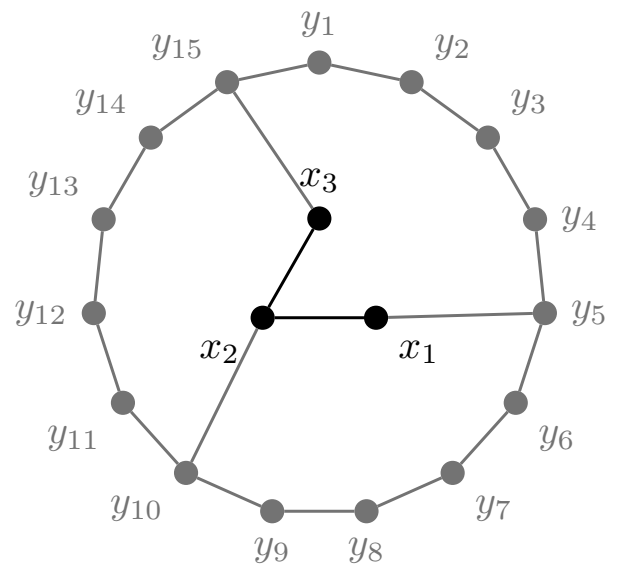}}
    \caption{\label{fig:cycle-extension-ordering-gadget} Unordered graph $H$, denoted as the \emph{cycle extension} of $(G,<)$.}
\end{subfigure}

\vspace{2ex}

\begin{subfigure}{0.7 \textwidth}
	\centering
    \scalebox{0.7}{%
    \input{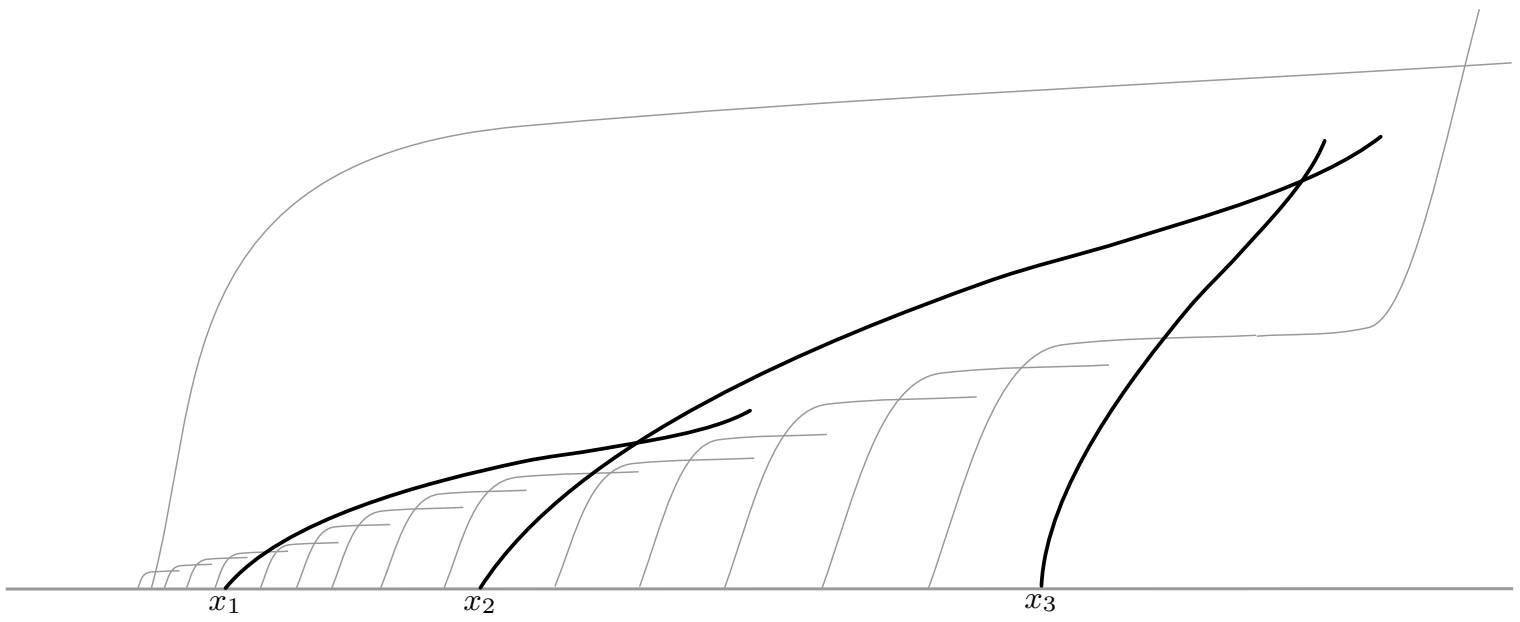}}
    \caption{\label{fig:rep-ordering-gadget} Grounded string representation of $H$. The vertices $x_1,x_2$ and $x_3$ are forced to maintain the original ordering $<$ up to cyclic shifts and inversions.}
\end{subfigure}
\caption{From an ordered graph $(G,<)$, we obtain its cycle extension $H$ by a well-defined construction. As the additional vertices in $H$ induce a cycle and the adjacencies between the cycle vertices and the original vertices of $G$ were carefully chosen, one can prove that the extension works as an ordering gadget for the vertices of $G$. That is, in every grounded string representation of $H$, the grounding ordering of vertices $x_1,x_2,x_3$ will be \emph{equivalent} (i.e. obtained by a cyclic shift and/or reversal) to ordering $<$.}
\label{fig:ordering-gadget-intro}
\end{figure}

A strategy here would be to argue about some notion of \emph{canonical ordering}: there would be one special ordering for geometric representation and a special ordering avoiding the pattern, and these would necessarily be equal.
The closest we can get to this is what is called an \emph{ordering gadget}~\cite{JelinekT19}.
It consists in the following procedure: one chooses a vertex ordering for a graph, and then adds a linear number of vertices and edges (to get what is called the cycle extension of the original ordered graph), such that in any grounded string representation of the new extended graph, the original graph can only appear in the original ordering (up to cyclic shifts and inversions).
See Figure~\ref{fig:ordering-gadget-intro} for an illustration of the ordering gadget.

Ordering gadgets are clearly not enough to transfer results from the ordered to the unordered setting: they force the order by modifying the graph, and they work only for the geometry side, since they constrain the grounding ordering of geometric representations.
Nevertheless, by adapting such a gadget to our setting, and reworking our arguments, we can get the following strengthening of Theorem~\ref{thm:no-characterization-monotone}.

\begin{restatable}{theorem}{thmNoCharacterizationGroundedMonotone}
There is no finite set of patterns $\mathcal{F}$, such that:
\begin{enumerate}
    \item For every grounded monotone string representation, the grounding ordering of the intersection graph avoids~$\mathcal{F}$.
    \item Any graph $G$ that admits an ordering avoiding $\mathcal{F}$ also admits a grounded monotone string representation. 
\end{enumerate}
\label{theo:no-finite-patterns-characterize-grounded-monotone}
\end{restatable}

Note that Item 1 is the same as in Theorem~\ref{thm:no-characterization-monotone} (any correct grounding ordering avoids the patterns), while Item 2 makes the theorem stronger: we only require that avoiding the patterns with some ordering implies that there exists a grounded monotone string representation, possibly with another ordering. 
The theorem rules out a scenario where the ordering of any representation should avoid the patterns, but only some specific pattern-avoiding orderings (for example the ones that are minimal for some criterion) would have a representation. 

We conjecture that the impossibility results is true in full generality. 

\begin{conjecture}
\label{conj:no-pattern}
There is no finite set of patterns that characterizes grounded monotone string graphs.
\end{conjecture}

Although the literature about forbidden patterns is rapidly developing, theorems stating such impossibilities are very rare. 
The closest to our perspective states that the class of graphs with no induced cycle of prime length has no finite pattern characterization~\cite{Guzman-ProH22}, which is an arguably less natural class than grounded monotone string graphs. 
Another approach to rule out a finite pattern characterization (for a hereditary class) is to argue about the complexity of the recognition. 
A finite pattern characterization implies that the recognition is in NP, since one can give the pattern-avoiding ordering as a certificate. 
Hence classes whose recognition is (conjectured to be) strictly harder cannot have such a characterization.
This is for example the case of grounded segment graphs, which are $\exists \mathds{R}$-hard to recognize~\cite{CardinalFMTV18}, as already noted by~\cite{JelinekT19}.  We observe that this type of argument cannot work for grounded monotone string graphs, because of the two-orderings characterization of~\cite{PachT20}.

Now that we have discussed the big picture, let us say a few words about the proof of Theorem~\ref{theo:no-finite-patterns-characterize-grounded-monotone}. First, we cannot use an off-the-shelf ordering gadget, instead we extend to general grounded string graphs the ordering gadget of~\cite{JelinekT19}, originally designed for representations where strings intersect at most once. Recall the approach used for the ordered version of the theorem (i.e. Theorem \ref{thm:no-characterization-monotone}) it relies on cocomparability suns, which are ordered graphs that admit no constrained monotone string representation. The main challenge for extending this strategy to the unordered setting is that, as mentioned earlier, using the ordering gadgets allows us to force the ordering, but only up to cyclic shifts and inversions. This is very problematic when applied to cocomparability suns. Indeed, the ordering of these graphs is tightly related to the graph's structure. In particular, the first and last three vertices were left out of the join (refer to Figure \ref{fig:Cocomparability-sun-and-Dh-intro} or to Section \ref{sec:no-characterization-by-patterns} for a formal definition) to guarantee the desired properties, and these properties are broken by cyclically shifting the vertex ordering. Consequently, the crucial fact that the associated directed height graphs $D_h$ have arbitrarily large girth is completely lost. 
The way we fix this is by first proving that the $D_h$-graphs of the cyclic shifts and inversions of cocomparability suns all contain directed cycles, even if we cannot control their size. As the ordering gadget will force the vertices of the cocomparability sun to follow an ordering equivalent to the original one in every grounded string representation, this proves that the extended graph cannot be in the class of grounded monotone string graphs. Finally, using arguments similar to those of the ordered case, we prove that there exists an ordering of the extended graph avoiding all patterns in the hypothetical finite pattern set and reach a contradiction.

We also use the extended ordering gadget to prove strict inclusions between some of the unordered graph classes. This way, we prove that the grounded monotone string graph class is strictly included in the class $\mathcal{C}_{abc}$. This allows us to distinguish the classes $\mathcal{C}_{ab}$ and $\mathcal{C}_{abc}$. Also, by applying the ordering gadget to the ordered Pach-Tomon graph (see Figure \ref{fig:pach-tomon-graph-intro}), we can prove that the class of grounded monotone string graphs is strictly included in the grounded string graph class. As we argue that the strict inclusions work, in particular, for the ordered classes, these results are reflected in the diagram of Figure \ref{fig:diagram-hierarchy}.

\subsection{Perspectives}

We started this project with great hopes of unraveling a strong connection between pattern-defined and geometry-defined classes. We described promising hierarchies of patterns and geometric representation, between which we planned to prove equalities. Unfortunately, our hopes have been crushed by our various findings: geometric classes that do not have any finite pattern characterization, patterns that do not have standard geometric characterizations. 
But in some sense, we also discovered a richer structure: if we allow for more creative geometries, then we capture new patterns, and if we allow more orderings we can capture new geometries. And so, as we parted the curtain, we caught a glimpse of a whole world waiting to be explored. 

A first direction for future work is to enrich our inclusion diagram with more geometric and pattern classes. We left out of this paper our preliminary results about intersection of grounded convex shapes and grounded segments, in order to get a simple and not-too-long story, but they offer hope for gaining a better understanding via our approach. 

A second direction is to transfer our results to the (fully) non-ordered setting, in particular prove Conjecture~\ref{conj:no-pattern}. 
More generally (and ambitiously), the sets of the orderings that correspond to correct grounding ordering and to pattern-avoiding ordering must have some additional structure, and it would be very insightful to get some understanding of this structure.

A third direction is to explore the characterizations with more than one ordering. We have seen that, in the case of constrained monotone strings, there is a simple characterization with two orderings, while no pattern characterization exists with one. 
Can we capture more geometries with more orderings? Strings can naturally be made into piece-wise curves, and the points of these curves can be used to define many orderings. 
One can also replace the 1D/2D objects of this paper by objects in higher dimensions with some grounding. 
For example, the constrained rectangle graphs could be generalized by boxes of higher dimension grounded in a hyperplane, and the configuration of the grounding points could capture interesting features of the graphs. 

A fourth direction is to explore non-standard intersections, such as the one we use for $\Cao$. Contact graphs are definitely something to look into, if one is interested in smaller classes. But there could be many other directions. For example, non-standard intersections of intervals have been studied: interval digraphs~\cite{DasSRW89}, where the intervals are somehow oriented, $t$-interval graphs, where a shape is the union of $t$ distinct intervals~\cite{TrotterH79}, overlap graphs, where an intersection counts only if it is not an inclusion~\cite{Gavril73}, and so on. 

\section{Related work}
\label{sec:related-work}


\paragraph*{The geometry-defined classes we study}

The \emph{constrained string graphs} of this paper are the ordered version of outer-string graphs from the literature. 
Outer-string graphs are the intersection graphs of curves that lie in a disk (or equivalently a half-plane) and have one endpoint on the boundary. These graphs are known to be $\chi$-bounded and more precisely, it is known that the chromatic number is at most doubly exponential in the clique size~\cite{RokW19}. It is also known that a representation might require an exponential number of segments~\cite{BiedlBD18}. 
There is also a literature focusing on the grounding ordering for this class. 
For both chordal and $(C_3,C_5)$-free graphs, it is known that the graph has an outerstring representation, if and only if it has an ordering that avoids an alternating pair of independent edges (what we call $P_{abcd}$ in this paper)~\cite{biedl2026constrained,antic2026two}.
It was also proved recently that a graph has an outer-string representation for any grounding ordering, if and only if it is co-chordal~\cite{biedl2026constrained}. 
It is known that outer-string graphs are NP-hard to recognize~\cite{Kratochvil91}, but the constrained case is open~\cite{biedl2026constrained}.  

The \emph{constrained monotone string graphs} of this paper do not have a well-identified counterpart in unordered graphs, but they have been studied indirectly.
They are an ordered grounded version of the monotone string graphs from~\cite{BiedlBD18}, which are known to be representable by a polynomial number of segments. 
They are also the complements of the grounded $x$-monotone disjointness graphs studied in~\cite{PachT20}. (We will use this connection extensively to establish Theorem~\ref{thm:no-characterization-monotone}. In Section~\ref{sec:no-characterization-by-patterns}, we precisely connect grounded $x$-monotone disjointness graphs and monotone string graphs.)
Monotone strings can be replaced by stairs, which alternate between nearly vertical and nearly horizontal segments, both slightly tilted to guarantee monotonicity. In this sense, they are a generalization of the grounded-\textsf{L} graphs of~\cite{JelinekT19}, where the stairs are just a vertical segment followed by a horizontal one.  

The \emph{constrained bigrounded string graphs} of this paper are the ordered versions of \emph{interval filament graphs}. There are several definitions for these, depending on whether the filament has to be monotone, or just stay between its endpoints, or just be bigrounded, but these are equivalent~\cite{BestW22}. These graphs are NP-hard to recognize~\cite{Pergel07}, but many optimization problems can be solved efficiently on them, for example maximum independent set in quadratic time~\cite{BestW22}. 

The intersection graphs of grounded rectangles we cited in the introduction have been defined multiple times, for multiple reasons. 
They were called \emph{hook graphs} in \cite{Hixon13}, and were motivated as a variant of grounded segment graphs. 
They were called \emph{max point-tolerance graphs} in \cite{CatanzaroCFHHHS15} and were studied as a new type of tolerance graphs~\cite{GolumbicT04} with applications to bioinformatics. They were called \emph{diagonal-touched rectangle graphs} in~\cite{CorreaFPS15}, where they were a special case of axis-aligned rectangle graphs, in the context of maximum independent set computation. They were called \emph{p-BOX(1) graphs} in \cite{SotoC15}, where the focus was on a variant of the boxicity parameter, with inspiration from sensor networks.

\paragraph*{Strings grounded on two lines}

We have hinted that there are other types of grounding than grounding a single horizontal line. 
One of them is to ground on two parallel lines. This was studied in the book~\cite{GolumbicT04}. Several classes can be characterized (or defined) this way: permutation graphs (where the double grounded shapes are segments), parallelogram graphs,  trapezoid graphs, and cocomparability graphs (where the shapes are strings). 

\paragraph*{Intersection graphs with prescribed orderings}

Our focus in this paper has mostly been on ordered graphs, and intersection graphs where the grounding ordering is fixed. As mentioned this is not the most-studied case: one usually seeks to have theorems that apply to any ordering. But there is a literature that corresponds to our setting. The first paper on outer-string graphs~\cite{Sinden66} actually studied the order in detail (in particular the co-hole obstruction). And the topic has recently been in the focus again in~\cite{biedl2026constrained, antic2026two}, that get stronger results but in restricted cases. 

\paragraph*{Intersection graphs and rescaling}

To argue that there is no classic intersection representation for $\Cao$, we have introduced a notion of standard representation (closure under translation, rescaling etc.). As far as we know this specific point of view is new, but there has been research related to intersection graphs and rescaling.
The two closest works to this approach are: \cite{AamandAKR21},
that studies conditions on pairs of shapes to decide whether they define the same intersection/contact graphs by copies and translations, and \cite{AbrahamsenW23} which studies a similar question permitting also rescaling and rotations. 

\paragraph*{Pattern-defined classes}

We have already reviewed part of the literature on (ordered) graphs defined by forbidden ordered subgraphs/patterns. 
The seminal papers observing that forbidden ordered subgraphs are interesting objects are from the 1980' and early 1990'~\cite{Skrien82, Damaschke90}. 
In \cite{Ginn99}, the relation between ordered and non-ordered forbidden subgraph was studied. 
The topic of recognizing and classifying classes defined by forbidden patterns was studied in~\cite{HellMR14,FeuilloleyH21} for small patterns, and in~\cite{DuffusGR95,KunN26} for arbitrary ones.
The complexity of detecting a pattern in an ordered graph was studied in~\cite{DucoffeFHP23}. 
Many classes have been characterized by forbidden patterns, beyond the ones of this paper: 2-thin and proper 2-thin graphs~\cite{BonomoBrabermanB23}, interval bigraphs~\cite{HellH04}, 12-representable graphs~\cite{Takaoka24} and so on. 
There are also generalizations and variants of the classic setting, in particular circular orderings and orientations, in the PhD thesis of Guzman-Pro~\cite{Guzman-Pro23} and tree orderings~\cite{PaulP24}.

\newpage{}

\section{Formal definitions}
\label{sec:definitions}

\subsection{Graphs and ordered graphs}
\label{subsec:graphs}

The graphs we consider are finite, loopless and simple (no multiple edges). A \emph{graph} is a pair $G=(V,E)$ where $V$ is the vertex set and $E \subseteq V^2$ is the set of edges. If we consider $G=(V,A)$ with $A$ a set of ordered pairs of vertices, called arcs, we say $G$ is a \emph{directed graph} or a \emph{digraph}. Note that a directed graph might contain 2-cycles, i.e., bidirected edges. If $u$ and $v$ are vertices of a graph (resp. digraph), we denote by $(u,v)$ the edge between $u$ and $v$ (resp. $(u \rightarrow v)$ the arc from $u$ to $v$). In the context of directed graphs, we will denote as a \emph{directed cycle of size $k$} the set of vertices $\{v_1, \dots, v_k\}$ such that $(v_i  \rightarrow v_{i+1})$,  for all $1\leq i < k$, and $(v_k  \rightarrow v_1)$ are arcs. An \emph{ordered graph} $(G,<)$ is a graph given with a total ordering of its vertices.

\subsection{Ordered subgraphs and patterns}
\label{subsec:preliminary-patterns}

\begin{definition}
We will say that an ordered graph $(G,<_G)$ is \emph{isomorphic} to another ordered graph $(H,<_H)$ if there exists an isomorphism from the vertices of the first graph to the vertices of the second graph preserving the graph's structure and relative ordering between vertices. More formally, there exists a bijection $f: V_G \rightarrow V_H$ such that $(u,v) \in E_G$ if and only if $(f(u),f(v)) \in E_H$, and $u <_G v$ if and only if $f(u) <_H f(v)$.
\label{def:ordered-isomorphic}
\end{definition}

\begin{definition}
The ordered graph $(H,<_H)$ is an induced ordered subgraph of the ordered graph $(G,<_G)$ if $H$ is an induced subgraph of $G$ and ordering $<_H$ preserves the ordering $<_G$ for the vertices in $H$.
\end{definition}

\begin{definition}
A \emph{trigraph} $T$ is a $4$-tuple $(V(T),E(T),N(T),U(T))$ where $V(T)$ is the vertex set and every unordered pair of vertices belongs to one of the three disjoint sets $E(T)$, $N(T)$, and $U(T)$ called respectively \emph{edges}, \emph{non-edges} and \emph{undecided edges}. A graph $G=(V(G),E(G))$ is a \emph{realization} of a trigraph $T$ if $V(G)=V(T)$ and $E(G)=E(T)\cup U'$, where $U'\subset U(T)$.
For us a \emph{pattern} is simply a trigraph together with a total ordering of its vertices.
We say that an ordered graph is a \emph{realization} of a pattern, if the ordered graph and the pattern share the same set of vertices, the same linear order, and the graph is a realization of the trigraph.
\end{definition}

When drawing a pattern, we will use plain edges for edges, dashed edges for non-edges, and nothing for undecided edges (just like we did in the introduction). Also, in our drawings, the patterns are ordered from left to right.

\begin{definition}
We say that an ordered graph $(G,<)$ avoids a pattern $P$ iff $(G,<)$ does not contain any ordered subgraph $(H,<)$ realization of $P$.
\end{definition}
\begin{definition}
Given a family of patterns $\mathcal{F}$, we define the class $\CC_{\FF}$ as the set of graphs that have an ordering avoiding the patterns in $\mathcal{F}$. 
\end{definition}

If $\mathcal{F}$ consists of only one pattern $P$, we abuse notation an simply write $\CC_{P}$, instead of $\CC_{\{P\}}$.
Note that one can equivalently consider forbidden ordered subgraphs instead of forbidden ordered trigraphs (which are the patterns). 
Like in \cite{FeuilloleyH21}, we prefer patterns, because they give more compact characterizations of the obstructions. 

As already noticed in \cite{Damaschke90, FeuilloleyH21, Wood06}, inclusions of patterns imply inclusions of classes.

\begin{observation}
\label{obs:inclusion-patterns}
Consider two patterns $P$ and $P'$. 
We say that $P$ is included in $P'$ if $V(P)\subseteq V(P')$, $E(P)\subseteq E(P')$, and $N(P)\subseteq N(P')$. 
In this case, $\CC_P \subseteq \CC_{P'}$.  
\end{observation}

There is one type of pattern that will be important in this paper, and that we define now.

\begin{figure}[!h]
\centering
\scalebox{1}{
\input{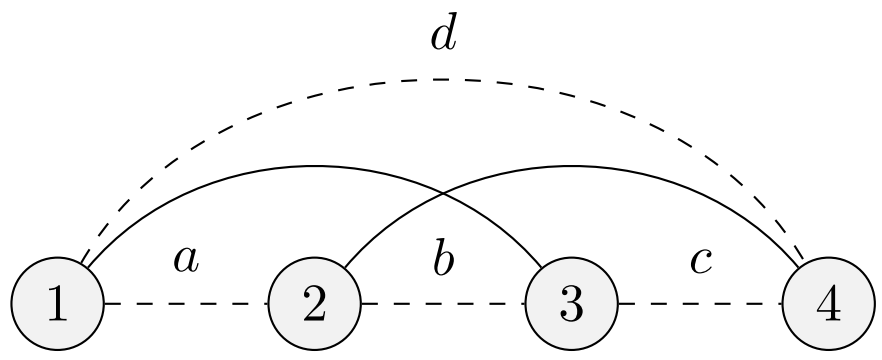}}
\caption{\label{fig:abcd-patterns}
Illustration of $P_{abcd}$ (Definition~\ref{def:Pabcd}).}
\end{figure}

\begin{definition}
\label{def:Pabcd}
Consider patterns on four nodes $\{1,2,3,4\}$. We use the following names for some pairs of nodes: $a=(1,2)$, $b=(2,3)$, $c=(3,4)$, $d=(1,4)$ (see Figure~\ref{fig:abcd-patterns}).
Let $S \subseteq \{a,b,c,d\}$, the pattern $P_{S}$ is the pattern on four vertices, with edges $(1,3)$ and $(2,4)$, and non-edge set $S$. The class $\CC_{S}$ is the class where $P_{S}$ is forbidden.
\end{definition}

\begin{remark}
\label{Sym}
It is easy to see that a pattern and its mirror (the same pattern but with reversed order) define the same class.
Consequently $\CC_{a}=\CC_c$, $\CC_{a, d}=\CC_{c,d}$, $\CC_{a, b}=\CC_{b, c}$, $\CC_{b,c,d}=\CC_{a, b, d}$. 
Therefore, we can first restrict our attention to the following classes: $\CC_{\emptyset}$, $\CC_a$, $\CC_b$, $\CC_d$, $\CC_{a, b}$, $\CC_{a,c}$, $\CC_{a, d}$, $\CC_{b, d}$, $\CC_{a, b, c}$, $\CC_{a, b, d}$, $\CC_{a, c, d}$ and $\CC_{a, b, c, d}$.
\end{remark}

\subsection{Constrained/ordered/grounded geometric graphs}
\label{subsec:constrained-representations}

We say a configuration of shapes is \emph{grounded} if either there exists a horizontal straight line (grounded circle) such that all the shapes are above the line (inside the circle),
and are touching it in at least one point.

We say that an unordered graph $G$ admits a \emph{grounded representation} with a given shape if $G$ is the intersection graph of a family of grounded shapes of that type, one for each vertex. An ordered graph $(G,<)$ admits a \emph{constrained representation} for a given shape if $G$ admits a grounded representation for this shape whose grounding-line ordering is $<$. In other words we use \emph{constrained} as a short cut for \emph{both grounded and ordered}.

\begin{definition}
\emph{Constrained bigrounded string graphs} are the ordered graphs $(G,<)$ that admit a bigrounded string intersection representation whose grounding-line ordering is $<$. Such representations are defined by taking a collection of intervals $I$ on a line $L$. In the plane above $L$, construct for each interval $i\in I$ a continuous $x$-monotone curve $c_i$ connecting its two endpoints, such that $c_i$ remains within the limits of $i$. A curve is $x$-monotone if every vertical line in the plane intersects it in at most one point.
\end{definition}

The unordered counterpart, that is, the class of graphs admitting a bigrounded string intersection representation, is the class of interval filament graphs.

\begin{definition}
\emph{Constrained monotone string graphs} are the ordered graphs $(G,<)$ that admit a grounded intersection representation of monotone curves (i.e. increasing functions), intersected at most one point by every vertical or horizontal line in the plane, whose grounding-line ordering is $<$.
\end{definition}

Constrained monotone string graphs are not known in the literature, even unordered, as far as we know. 

\begin{definition}
\emph{Constrained string graphs} are the ordered graphs $(G,<)$ that admit a grounded intersection representation of curves having arbitrary trajectories whose grounding-line ordering is $<$.
\end{definition}

Outer-string graphs, the unordered counterpart of this class, were formally introduced in \cite{Kratochvil82}, although the underlying concept already appears in \cite{Sinden66}.


\section{Interleaving of pattern-defined and geometry-defined classes}

In this section, we show that the hierarchies of pattern-defined and geometry-defined classes we study are interleaved.

\subsection{Geometric classes included in pattern-defined classes}
\label{subsec:geometry-in-pattern}

For a start, we establish inclusions of the form: if an ordered graph has some constrained geometric representation, then it should avoid a given pattern. 

\thmGeometryInPatterns*

\begin{proof}
\begin{enumerate}
    \item Consider a constrained bigrounded string graph $(G,<)$. We claim that $(G,<)$ avoids the pattern~$P_{ab}$.
For the sake of contradiction, suppose that the pattern appears in this ordering, on some vertices $a< b< c< d$. 
Let $\alpha_a,\alpha_b,\alpha_c,\alpha_d$ and $\beta_{a},\beta_{b},\beta_{c},\beta_{d}$ be respectively the left endpoints and  the right endpoints of the filaments corresponding to $a$, $b$, $c$ and $d$. 
Since $b$ and $d$ are adjacent, it must be that $\alpha_d< \beta_b$. 
Similarly, because $a$ and $c$ are adjacent, $\alpha_c< \beta_a$. 
Since $a$ and $b$ are not adjacent, we must have $\alpha_a< \alpha_b< \beta_b< \beta_a$.
Therefore, given the inclusions above, we must have $\alpha_a < \alpha_b < \alpha_c <  \alpha_d < \beta_b< \beta_a$. 
Consequently, $\alpha_c$ is in the region of the plane delimited by the grounding line and the filament of $b$. 
As $c$ is not adjacent to $b$, it cannot reach the filament of $a$ (whose extremities are outside the interval of $b$), a contradiction.

\item As already argued, this statement can be deduced from \cite{PachT20}. For completeness, we provide an alternative intuitive proof. Consider a constrained monotone string representation of an ordered graph $(G,<)$. 
We claim that $(G,<)$ avoids the pattern~$P_{abc}$. 
Suppose that it is not the case, and the pattern appears by some vertices $a<b<c<d$. We will study the configuration of strings corresponding to vertices $a,b,c$ and $d$.
Consider the region of the plane between the grounding line, the string $a$ and the string $c$, below their first intersection, whose coordinates will be denoted by $(x,y)$.
This region contains the grounding point of $b$.
Because intersections between  string $b$ and strings $a$ and $c$ are forbidden, the string $b$ must be fully contained in this region.
Now, the string $d$ must appear strictly on the right of the string $c$ until ordinate $y$, because these two do not intersect either. 
Finally, the rest of the string $d$ is in the quarter of the plane that is on the top right of $(x,y)$. For both parts of the string $d$, the intersection with the string of $b$ is impossible. This is a contradiction, and proves the inclusion.

\item 
This statement was already proved in \cite{PachT19}. We reprove it here formally for completeness.

We again reason by contradiction, and consider the grounding ordering in a representation by grounded strings. Suppose we have four vertices $a<b<c<d$, such that the only edges between these vertices are $(a,c)$ and $(b,d)$. 
Then, consider the closed region of the plane that is inside the Jordan curve defined by the string of $a$, the string of  $c$, and the grounding line. 
The shape $b$ must be inside this region because its root is, and it cannot intersect any of the boundaries, by Jordan curve theorem. 
On the other end $d$ must be outside this region, because likewise its root is outside and it cannot cross the curve. Then $b$ and $d$ cannot intersect, which is a contradiction. 

Note that another way to see this last statement is that the complement of grounded string graphs avoids the complement of $P_{abcd}$, which is an induced cycle in the natural order. 
This is actually not very surprising, as it is known that the class of complement of triangle-free intersection graphs of grounded curves is exactly the class of Hasse diagrams~\cite{MiddendorfP93}, 
and that even without the triangle-free assumption, these complement graphs avoid an infinite family of ordered cycles~\cite{Sinden66}. 

\end{enumerate}
\end{proof}

\subsection{\texorpdfstring{$\Cao$}{Cao} in constrained bigrounded string graphs}
\label{subsec:Ca-Interval-Filament}

\thmCaIntervalFilament*

\begin{proof}

Consider a graph $G$ in $C_a$. We show how to compute an interval filament representation of $G$.
To do so, we first create a set of filaments that is not a representation of $G$ for the general case nor satisfy the $x$-monotonicity constraint for interval filaments. We then modify this representation a number of times, until it becomes a valid interval filament representation of $G$.
The outcome of our construction is illustrated on an example in Figure~\ref{fig:filaments}. 

\begin{figure}[!h]
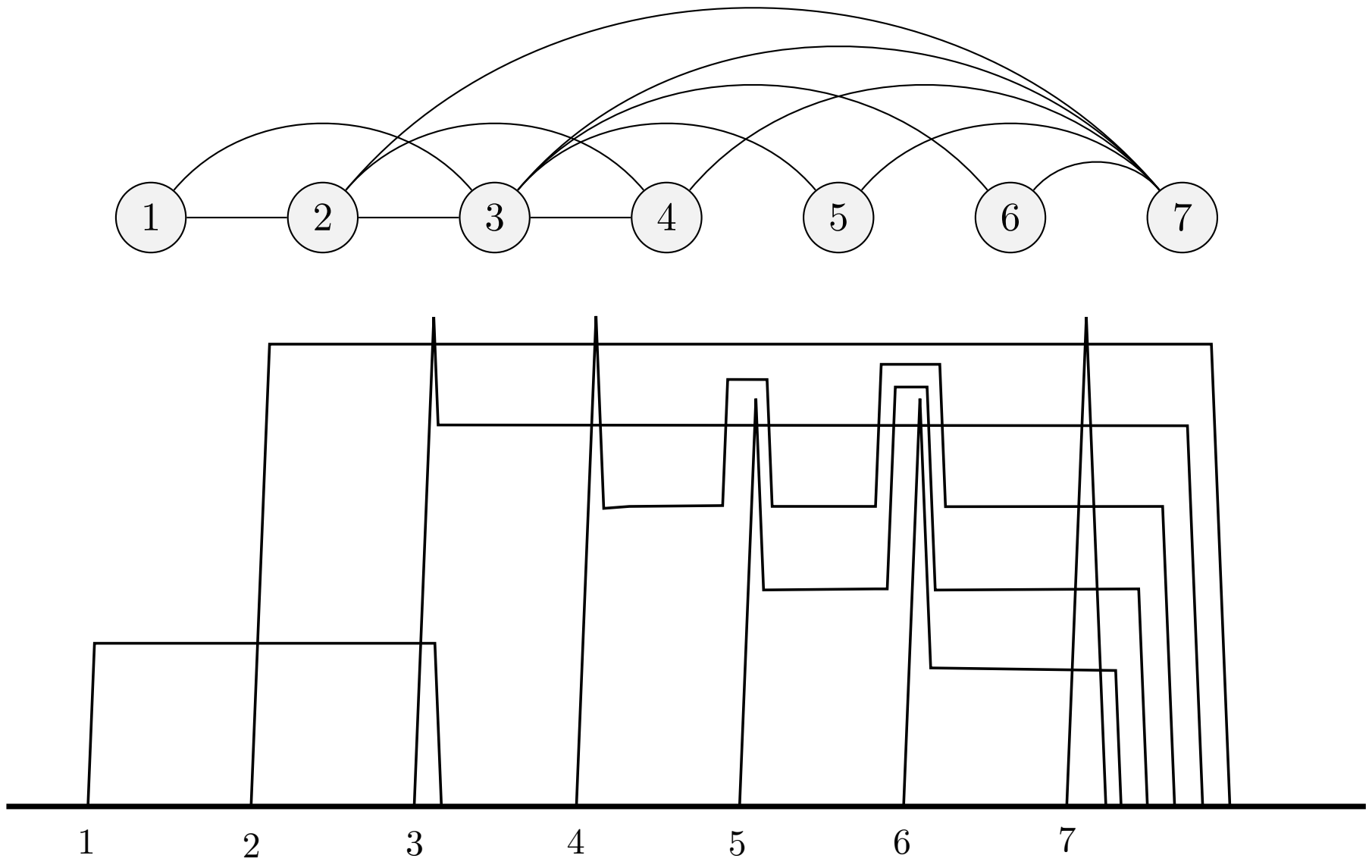

\begin{center}
\input{
			pictures-and-tikz/filament-graph.tex}
\vspace{0.5cm}

\scalebox{0.9}{			
\input{pictures-and-tikz/fig-filaments}}
\end{center}
\caption{
\label{fig:filaments}
On the top picture, a graph whose vertices are ordered to avoid $P_a$. 
On the bottom picture, the interval filament representation built by our construction. 
(A few modifications of the construction have been used for readability, including using a smaller scale on the vertical axis than on the horizontal axis. Note how the vertical segments were slightly tilted in order to satisfy the $x$-monotonicity constraint.)}
\end{figure}

Consider an ordering of the vertices of $G$ avoiding~$P_a$.
We will abuse notation and write $u$ instead of ``the rank of $u$ in the ordering''. 
For vertex $u$, we denote its rightmost neighbor by $r_u$ (with $r_u=u$ when $u$ has no neighbor on the right).
Let $\varepsilon_u=\frac{n-u+1}{n+1}$. 
We note that for two vertices $u<v$, we get that $0<\varepsilon_v<\varepsilon_u<1$. In $R_1$, the filament of $u$ starts at position $(u,0)$, goes up to point $(u,r_u-u+\varepsilon_u )$, then goes right to the position $(r_u+\varepsilon_u,r_u-u+\varepsilon_u)$, and then goes down to the grounding line at $(r_u+\varepsilon_u,0)$. Note that at this point each filament forms a square with the grounding line. For further reference, we denote the altitude of the horizontal segment as $h_u=r_u-u+\varepsilon_u$. 
Let us denote this representation by $R_1$.

\medskip

\begin{claim}
In $R_1$, all the intersections between filaments correspond to edges in the graph, and at any intersection point, there are exactly two filaments intersecting, one that is locally vertical and one that is locally horizontal. 
\end{claim}

\medskip

\begin{claimproof}
Let us start by proving the first part of the claim.
Consider the filaments of two vertices $u<v$. We make a case analysis depending on the placement of $r_u$ and~$r_v$.
\begin{itemize}
\item If $u\leq r_u<v\leq r_v$, then the two filaments do not intersect.
\item If $u<v \leq r_v\leq r_u$, then again the filaments do not intersect. Indeed, the square corresponding to $v$, starts at position $v$ ($>u$) and stops at $r_v+\varepsilon_v$ ($<r_u+\varepsilon_u$), thus is fully included into the square of $u$.
\item If $u<v<r_u<r_v$, then the filaments intersect in $R_1$, but in this case, since by definition $(u,r_u)$ and $(v,r_v)$ are edges in $G$, either $(u,v)$ is an edge in $G$, or we get that $u<v<r_u< r_v$ form a $P_a$ in the ordering, which is a contradiction.
\end{itemize} 

Now, for the rest of the claim, note that any horizontal segment is at an altitude of the form $k+\varepsilon_u$, where $k$ is an integer, and $\varepsilon_u\in (0,1)$ is specific to the node $u$ where the filament starts. Thus, no two horizontal parts of two different filaments can intersect. 
The analogue reasoning holds for the vertical parts. 
The claim follows. 
\end{claimproof}

In general, $R_1$ is not a representation of $G$, because there are edges in the graph that do not correspond to filament intersections.
We now modify this first representation $R_1$ into a second representation $R_2$.

For a vertex $v$, we select (when it exists) the vertex $w$ such that:
\begin{itemize}
\item $w$ is on the left to $v$,
\item $w$ is adjacent to $v$ in $G$,
\item the filaments of $w$ and $v$ do not intersect in $R_1$
\item among the vertices satisfying the three points above, $w$ is the one that has the highest horizontal segment (that is, the largest $h_w$).  
\end{itemize} 

We modify the shape of $v$ with a spike at the beginning: the first vertical line goes up to $h_w$, and then goes down to $h_v$, and continues the shape as before. 
(If we want the filament to not self-intersect, we can go down with some steep enough slope, like in Figure~\ref{fig:filaments}.)

\medskip

\begin{claim}
For every edge of the graph, there is an intersection of filaments in $R_2$. 
Moreover, if the filaments corresponding to some vertices $y<v$ intersect in $R_2$, and $y$ and $v$ are not adjacent in $G$, then there must exist a vertex $x$, with $x<v$, such that $(x,v)\in G$ and $h_y<h_x$. 
\end{claim}

\medskip

\begin{claimproof}
First, note that all filaments intersections from $R_1$ also appear in $R_2$. Also, note that the spikes appear on different zones of $x$-axis, thus they do not interfere.
Now, consider a pair of nodes $v<u$, such that $(v,u)\in G$, but the two corresponding filaments were not intersecting in $R_1$. 
By definition of $r_v$, $u\leq r_v$, thus $v<u <r_v+\varepsilon_v$. 
Therefore, in $R_1$ the horizontal part of the filament of $v$ goes above the left grounding point of $u$.
The vertex $u$ satisfies the three first conditions of the vertex $w$ defined above, and because $h_w$
is chosen to be maximal, we get that the spike of $u$ necessarily touches the filament of $v$.
Thus the intersection between the filaments of $v$ and $u$ appears in $R_2$, which proves the first part of the claim.

Now, if there is an unwanted intersection, that is an intersection in $R_2$ that does not appear in $G$, it must be because of a spike of some vertex $v$, that intersects the horizontal part of the filament of a vertex $y$ with $(v,y)\notin E$, while going up to touch~$u$. 
This can only happen if the conditions of the claim are satisfied. 
\end{claimproof}

We now define $R_3$, which will be our final filament representation.
Let us first define an $\varepsilon$-\emph{chimney}. 
Consider a horizontal segment $[(x_1,y),(x_2,y)]$ and a vertical segment $[(x,y_1),(x,y_2)]$ that intersect at position $(x,y)$. 
Then an $\varepsilon$-chimney is a modification of the horizontal segment, to avoid the vertical segment. More precisely, it is the piece-wise segment going through the following points: $(x_1,y), (x-\varepsilon,y), (x-\varepsilon, y_2+\varepsilon), (x+\varepsilon, y_2+\varepsilon), (x+\varepsilon, y), (x_2,y)$.

Let us call \emph{a conflict} an intersection between two filaments that does not correspond to an edge in $G$.
As mentioned earlier, a conflict is necessarily between the spike of a filament and the horizontal plateau of another filament. 
For any given filament $u$, we show how to solve all the conflicts that involve the spike of $u$, without losing the intersection that are needed in $G$. 
We do so by only modifying the filaments in a small vertical column around the abscissa of $u$. 
We will maintain the invariant that when we consider a filament $v$, every filament~$w$ above it has been treated: either $(u,w)\in E$ and the intersection has been preserved, or $(u,w)\notin E$ and the intersection has been removed.  
We consider the filaments that intersect the spike of $u$ ordered by the altitude of their horizontal plateaus, from the highest to the lowest.
For each such filament $v$, we do the following:
\begin{itemize}
\item If $(u,v)\in E$, do nothing. 
\item If $(u,v)\notin E$:
\begin{itemize} 
\item If there is a filament $w$ above $v$, such that $(v,w)\notin E$, and $w$ intersects the spike of $u$, then we claim that there is a contradiction. 
Indeed, if $(v,w)$ is not in~$G$, and  $h_v<h_w$, necessarily, $w<v<u<r_v<r_w$. Then $w<v<u<r_v$ is an occurrence of $P_a$, because $(w,v)\notin E$ (by hypothesis), $(w,u)\in E$ (by the invariant) and $(v,r_v)\in E$ (by definition).   
\item If there is no filament $w$ above $v$, such that $(v,w)\notin E$, and $w$ intersects the spike of $u$, then we transform the filament of $v$, by building a chimney around the spike of $u$. 
The $\varepsilon$ of this chimney is taken to be smaller than all the $\varepsilon$ used for chimneys of filaments modified so far around the spike of $u$. 
If this is the first chimney, then we take $\varepsilon$ to be the minimum between 1/4, and the space between the top of the spike of $u$ and the next plateau above (if it exists).  
\end{itemize}
\end{itemize}

We claim that this construction gives a filament intersection representation of~$G$. First note that, the conflict resolution is independent at all vertices, because the chimney we build are taken with a sufficiently small $\varepsilon$ around the spikes. 
Hence, it is sufficient to show for each spike that: we have not lost correct intersections, we have removed all the incorrect intersections, and have not created new ones.

The correct intersections that were present in $R_2$ are of two types: the ones coming from~$R_1$ and the ones between a spike and horizontal plateaus. 
We have only modified the graphs around the spikes, thus the first type is safe. 
Moreover, when we add an intersection corresponding to an edge $(u,v)\in E$, we have not modified the filament, hence the second type is also safe.

All the incorrect intersections at some given spike in $R_2$ have been removed thanks to our chimney constructions.

Finally, we have not created any new intersection. Indeed, while building a chimney, the only new incorrect intersection that we might have created is with another chimney, but this cannot happen as we have taken care of reducing the $\varepsilon$ parameter at every chimney. We showed that $R_3$ is an filament intersection representation for $G$. However, it is easy to see that $R_3$ is not an interval filament representation as the filaments contain vertical segments and therefore fail to satisfy the $x$-monotonicity constraint imposed to such representations. Nevertheless, this can be fixed by slightly tilting all vertical lines as shown in Figure \ref{fig:filaments}, in order to finally obtain an interval filament representation of $G$. \end{proof}


\subsection{\texorpdfstring{$\Cabo$}{Cab} in constrained monotone string graphs}
\label{subsec:Cab-in-Gstairs}

\ThmCabInGstairs*

\begin{proof}
Let us consider a graph $G=(V,E)$ in $C_{ab}$. 
We build a constrained monotone string representation incrementally, by adding the strings one by one, from left to right, in the ordering of $G$ avoiding the pattern~$P_{a,b}$. 
Actually, for every new vertex $i$, allow ourselves to modify the strings 1 to $i-1$, such that at the end of the process, the adjacency of the vertices 1 to $i$ in the graph is the same as the one in the constrained monotone string configuration. 

For any new vertex $i$, we create an infinite vertical line, grounded at position $i$. We will cut the line into a finite segment at the end of this step of the construction, and maintain the invariant that all the strings are finite.
We will maintain the invariant that the representation of vertices 1 to $k$ only uses abscissas in $[1,k]$, thus, at that point, the line has no intersection. 
Consider a string $j$, with $j<i$ in the ordering, and such that $(i,j)\in E$. 
We claim that it is possible to modify the string of $j$, so that they intersect the string of $i$, without creating an unwanted intersection with the string of any vertex $k$, with $(j,k)\notin E$.
Note that proving that we can make the strings of $j$ and $i$ intersect simply means that we can extend the string of $j$ beyond the abscissa $i$.
 
For the sake of contradiction, suppose there are some string $j$, such that the extension to $i$ is no possible.
It means that there is a position $x$ on the horizontal axis, such that the string $j$ cannot be extended beyond abscissa $x$. 
We extend the string $j$ to a position $p=(x-\varepsilon,t)$ for a small $\varepsilon$. 
Now, as by invariant, all the strings of index smaller than $i$ are finite, there must be an ordinate $y$, such that if we make string $j$ go up from $p$, it cannot go further than $y$. (If such an ordinate $y$ would not exist, then we could just go high enough to be above all the strings except $i$, and then go straight to $i$, which would contradict that fact that we cannot go beyond abscissa $x$.)

In this situation, the position $(x,y)$ must be the intersection of two strings $\ell$ and $r$, $\ell$ having a horizontal segment at height $y$ , and $r$ having a vertical segment at abscissa $x$. And necessarily $\ell$ and $r$ are not adjacent to $j$ in $G$.
We claim that in the ordering, $\ell < j < r <i $. 
Indeed, $i$ has the rightmost string by definition, and given the configuration described above, any other ordering would imply an intersection between the string $\ell$ or $r$, and the string of $j$. 
Then, the vertices $\ell < j < r <i $ correspond the pattern $P_{a,b}$ because we have edges $(\ell,r)$ and $(j,i)$, and non-edges $(\ell,j)$ and $(j,r)$. This is a contradiction. Therefore, we can always extend the string of $j$ to the string of $i$, and get the same adjacency in the representation and in the graph for vertices from 1 to $i$. Finally, we cut the infinite vertical line of $i$ just above its last intersection to make it a finite segment. We can slightly \enquote{tilt} all segments, in order obtain monotone strings.

\end{proof}

\section{Grounded intersection graphs with no pattern characterization}
\label{sec:no-characterization-by-patterns}

Since our goal is to draw parallels between the grounding ordering of an intersection representation (here of grounded $x$-monotone strings) and the orderings that avoid a set of patterns, in this section we discuss \emph{ordered} grounded intersection graphs that admit no characterization by a finite set of patterns. We prove the following:

\thmNoCharacterizationStairs*

In order to prove this, we adapt a result of~\cite{PachT20}.
To allow easy verification of the adaptation, we will proceed in the following way. First we introduce the terminology of~\cite{PachT20}, copy the result of the paper in this terminology, and sketch the proof. Second, we introduce our vocabulary, state our adaptation, and prove the equivalence.

\paragraph*{Description of the Pach-Tomon result}

\begin{figure}[!h]
    \centering
   \input{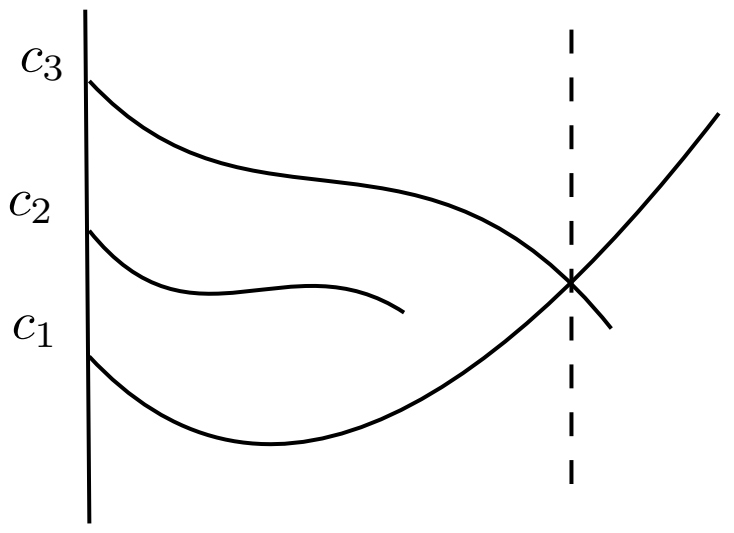}
    \caption{Illustration of the proof of Lemma~\ref{lem:PachTomon1}. Here we have three curves $c_1$, $c_2$ and $c_3$. They appear in this order on the grounding line, so the corresponding vertices are in that order for $<_1$: $v_1 <_1 v_2 <_1 v_3$. Now, if $v_1v_2$ and $v_2v_3$ are edges in the disjointness graph and $v_1v_3$ is not, it means that only $c_1$ and $c_3$ intersect. Then $c_2$ is trapped between these two curves and the grounding line. Because of the monotonicity, this means that the $x$-coordinate of the right-most endpoint $c_2$ must be strictly smaller than the ones of both $c_1$ and $c_3$. That is $v_2 <_2 v_1$ and $v_2 <_2 v_3$. This is exactly what magical means. }
    \label{fig:pach-tomon-proof}
\end{figure}

Paper~\cite{PachT20} is about the \emph{disjointness graph of grounded $x$-monotone curves}. 
In the terminology of that paper, a grounded $x$-monotone curve is a curve grounded on a \emph{vertical line}, that is monotone in the $x$-coordinate (see \emph{e.g.} Figure~\ref{fig:pach-tomon-proof}). The disjointness graph of a set of curves is the unordered graph where we take a vertex for each curve, and put an edge between two vertices if their curves \emph{do not} intersect.
We see that there are several differences with our setting, but as we will see the adaptation is fairly natural. 

Let us start with the definitions and lemmas from~\cite{PachT20}.

\begin{definition}[Definition 11 in \cite{PachT20}]
A double-ordered graph $G_{<_1,<_2}$ is called \emph{magical} if for any three distinct vertices $a, b, c \in V (G)$ with $a <_1 b <_1 c$, if $ab, bc \in E(G)$ and $ac \notin E(G)$, then $b <_2 a$ and $b <_2 c$.
A graph G is said to be magical if there exist two total orders $<_1, <_2$ on $V(G)$ such that $G_{<_1,<_2}$ is magical. 
In this case, we say that the pair $(<_1, <_2)$ witnesses~$G$.
\end{definition}

\begin{lemma}[Lemma 12 in \cite{PachT20}] 
\label{lem:PachTomon1}
If $\mathcal{C}$ is a collection of grounded $x$-monotone curves, then the disjointness graph of $\mathcal{C}$ is magical.
\end{lemma} 

In the proof of this lemma, the key element is the definition of the two orders, given a set of curves. The first ordering is the grounding ordering on the vertical line: exactly the ordering we consider. The second ordering corresponds to the ordering in which the curves \emph{end}, that is, the ordering of the $x$-coordinate of the endpoint that is not on the grounding line. See Figure~\ref{fig:pach-tomon-proof} for illustration and sketch of the proof.

\begin{lemma}[Lemma 13 in \cite{PachT20}]
\label{lem:PachTomon2}
Let $G$ be a magical graph. Then there exists a family $\mathcal{C}$ of grounded $x$-monotone curves such that the disjointness graph of $\mathcal{C}$ is isomorphic to $G$.
\end{lemma}

The proof of this statement is constructive and cannot be summarized in a few lines, but the important aspect for us is that the interpretation of the orderings is the same as for the previous lemma: the first ordering is the grounding ordering of the curves, and the second ordering is for the $x$-coordinate of the right-most endpoint.

\paragraph*{Adaptation to our setting}

We will now adapt the results above to our setting and terminology. 

We restate our adaptation of the theorem (remember that given an ordered graph $(G,<)$, with $G=(V,E)$, a triplet of ordered vertices $x<y<z$ will reproduce the cocomparability pattern if $(x,z) \in E$ and $(x,y), (y,z) \notin E$).

\ThmPachTomonAdaptation*

We will fix all the misatches between our setting and the one of \cite{PachT20}, one by one. We start with the orientation of the grounding line.

\begin{lemma}
\label{lem:adaptation1}
Configurations of $x$-monotone curves grounded on a vertical line or an horizontal line define the same class of intersection graphs and disjointness graphs. The top-down grounding ordering in the first setting corresponds to the left-right grounding ordering in the second.
\end{lemma}

\begin{figure}[!h]
    \centering
    \input{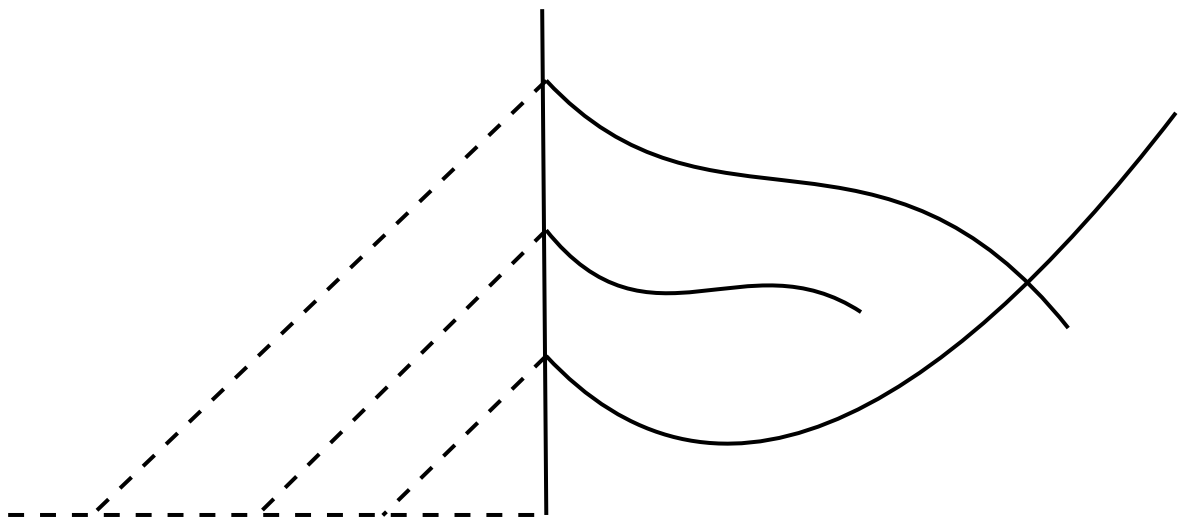}
    \caption{An illustration that a set of $x$-monotone curves grounded on a vertical line can be ``regrounded'' on an horizontal line. The plain curves are the original curves, the plain straight line is the original vertical axis. The dashed lines represent the modification: we simply add diagonal lines on the left to reach down to the horizontal line. Note that the left-right grounding ordering on the horizontal axis corresponds to the top-down grounding ordering in the vertical axis. A similar construction can be made to go from horizontal grounding to vertical grounding.}
    \label{fig:adaptation1}
\end{figure}

For the proof, see Figure~\ref{fig:adaptation1}.

Now, we fix the monotonicity assumptions: in our case, the curves are monotone in both $x$ and $y$ (they are analogue of increasing functions), while for \cite{PachT20}, they are only assumed to be $x$-monotone (they are analogue of functions). 

\begin{lemma}
\label{lem:adaptation2}
The grounded $x$-monotone and grounded monotone curves define the same (ordered) graphs.
\end{lemma}

\begin{figure}[!h]
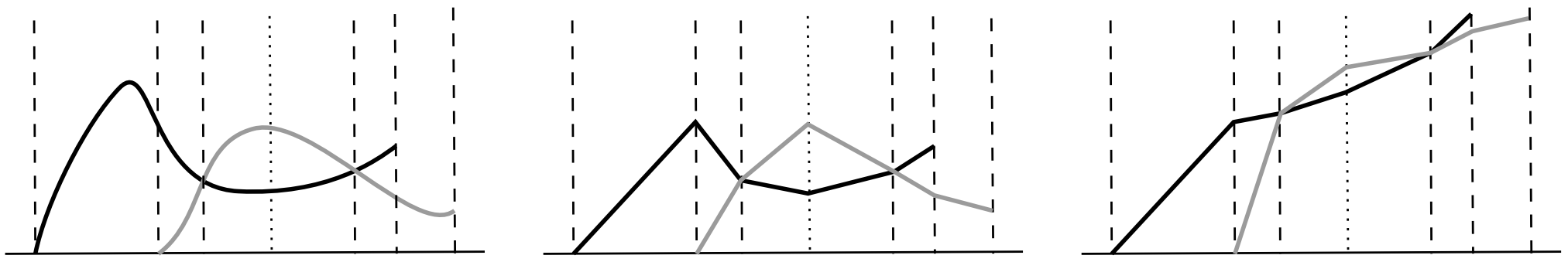

    \centering
    \begin{subfigure}{0.32\textwidth}
    \scalebox{0.7}{
    \input{pictures-and-tikz/adaptation2a}}
    \caption{Original configuration}
    \label{fig:adaptation2a}
    \end{subfigure}
    \begin{subfigure}{0.32\textwidth}
    \scalebox{0.7}{
    \input{pictures-and-tikz/adaptation2b}}
    \caption{Piece-wise version}
    \label{fig:adaptation2b}
    \end{subfigure}
    \begin{subfigure}{0.32\textwidth}
    \scalebox{0.7}{
    \input{pictures-and-tikz/adaptation2c}}
    \caption{Final configuration}
    \label{fig:adaptation2c}
    \end{subfigure}
    \caption{Illustration of the proof of Lemma~\ref{lem:adaptation2} with two curves. The dashed and dotted vertical lines are the stakes. The dotted line is a special stake added because otherwise the two curves would have intersected at two consecutive stakes.} 
    \label{fig:adaptation2}
\end{figure}

\begin{proof}
We modify a configuration of $x$-monotone curves into a set of monotone curves, keeping the same grounding points and intersections. See Figure~\ref{fig:adaptation2}.

First we replace the curves by piece-wise linear curves. To do so we identify the points on the $x$ axis where either (1) at least two curves intersect, or (2) a curve starts or terminates. 
(It is standard to assume that curves do not intersect in any other way than in points, for example they do not merge.) We note that because of the $x$-monotonicity, a curve can only have at most one point for any given $x$.
We call the vertical lines at these special axis the \emph{stakes}. 
For technical reasons, we also add stakes in a special case: if two curves meet at two consecutive stakes, we add one stake at an arbitrary place between these two stakes. 
The observation here is that the intersection between the curves only depend on which curves meet at which stake.  
Then, for each curve, we keep the position of the points of intersection with all the stakes. 
For each curve, for every pair of consecutive stakes, we replace the curve by a straight line. 
This process does not create new intersections, because an intersection of segments between stakes means that the ordering in which the two curves intersect the stakes would have been different before the operation, and this would have
implied an intersection between the stakes, which is impossible. Also, because of the special stakes we added, we keep the property that two curves do not meet for a segment, but only at isolated points. 

Now, we transform this configuration of piece-wise $x$-monotone curve into a configuration of piece-wise monotone curves. This is easy to do: (1) consider the first pair of consecutive stakes $S_1,S_2$ where a curve is decreasing, (2) increase the $y$ coordinate of all intersection points of the curves with $S_2$, in such a way that they are higher than their analogue on $S_1$, while keeping the same order on along the stake.  At that point, all curves are monotone, until $S_2$. We can repeat the process, until the curves are monotone up to the right-most stake.
\end{proof}

Before we finish the equivalence between the two settings, let us note that Lemma~\ref{lem:adaptation2} can also be used to prove the following statement, mentioned in the introduction, since constrained bigrounded string graphs are $x$-monotone.

\begin{corollary}
\label{cor:bigrounded-in-monotone}
Constrained bigrounded string graphs are included in constrained monotone string graphs.
\end{corollary}

Finally, to get from the characterization of~Lemma~\ref{lem:PachTomon1} and~\ref{lem:PachTomon2} to Theorem~\ref{thm:Pach-Tomon-adapted}, it suffices to note that: (1) the first ordering in the lemmas can be fixed to be the grounding order, (2) taking an ordering or the opposite does not impact the statements, (3) the local conditions of the magical graph definitions translate exactly to the ones of Theorem~\ref{thm:Pach-Tomon-adapted} when one takes the complement, which is consistent with the fact that we consider the intersection graph, while \cite{PachT20} considers the disjointness graph. 

\paragraph*{Introducing an intermediate directed graph}

In order to capture the relation between the two orders of Theorem~\ref{thm:Pach-Tomon-adapted}, we introduce an intermediate directed graph that we will call the \emph{directed height graph} of an ordered graph $(G,<)$, denoted by $D_h(G,<)$. Then, each arc $(u \rightarrow v)$ in $D_h(G,<)$ can be interpreted as a restriction for every constrained monotone string representation of $(G,<)$: the height of string $u$, i.e. its greater $y$-value, will be strictly smaller than the height of string $v$.

\begin{definition}[Directed height graph]
Given an ordered graph $(G,<)$, we define the ordered directed height graph $D_h(G,<)=(V,A)$ in the following way, $V=V(G)$ and for every ordered triplet $x<y<z$ reproducing the cocomparability pattern in $(G,<)$, we add the arcs $(y\rightarrow x)$ and $(y \rightarrow z)$.
\end{definition}

Then, we can rephrase Theorem~\ref{thm:Pach-Tomon-adapted} with the following proposition.

\begin{proposition}
\label{prop:PachTomonDh}
$(G,<)$ admits a constrained monotone strings representation if and only if $D_h(G,<)$ is a directed acyclic graph.
\end{proposition}

\paragraph*{Cocomparability suns and their directed girths}

Let us call \emph{directed girth} the smallest directed cycle in a directed graph. 
We will introduce a family of ordered graphs, that we will call \emph{cocomparability suns}, such that for any positive integer $k$, there exists an ordered graph $(G,<)$ in this family such that its directed height graph $D_h(G,<)$ has directed girth $2k$.
This will be then used to prove that no finite family of patterns can characterize constrained monotone string graphs. This family was created with the idea of extending the interesting properties of the Pach-Tomon graph to graphs of arbitrary size.

\begin{figure}
    \centering
    \input{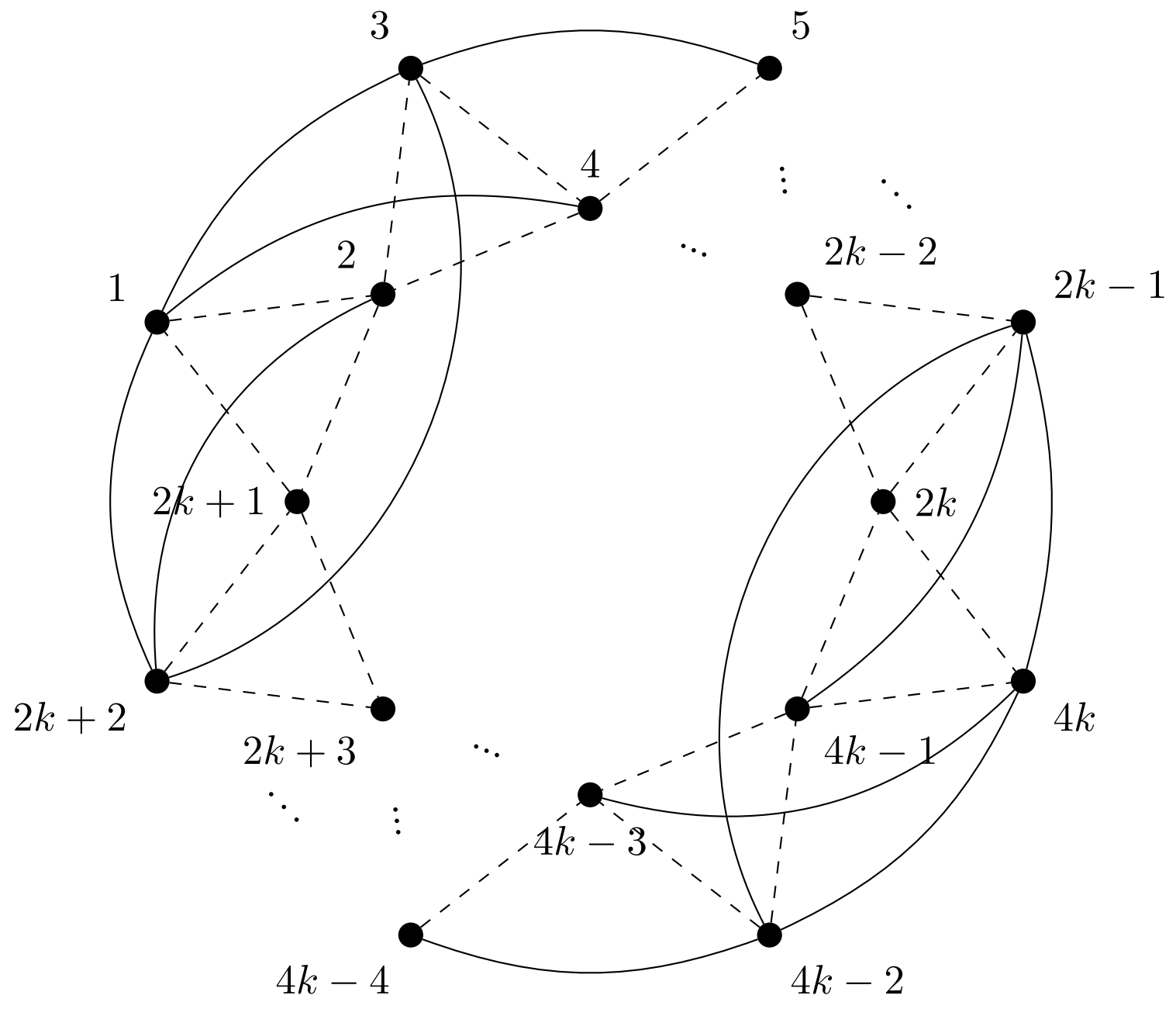}
    \caption{cocomparability sun of $4k$ vertices. All edges of the join between vertices in $\{4,5,6,...,2k \}$ and vertices in $\{2k+1, 2k+2, ..., 4k-3 \}$ are also edges in the graph though omitted from the image.}
    \label{fig:cocomparability-sun}
\end{figure}

\begin{definition}[Cocomparability sun]
\label{def:cocomparability-sun}
An ordered graph $(G,<)$ is a cocomparability sun if $n=4k$, for a given $k>2$, and $(G,<)$ has the following adjacencies, and no additional one (see Figure~\ref{fig:cocomparability-sun}):

\begin{itemize}
    \item $(i,i+2)$ and $(i,i+3)$ for every odd $i$, with $i < 2k-1$.
    \item $(i,i-2)$ and $(i,i-3)$ for every even $i$, with $i > 2k+1$.
    \item all edges of the join between vertices in $\{4,5,6,...,2k \}$ and vertices in $\{2k+1, 2k+2, ..., 4k-3 \}$.
    \item $(2k-1, 4k)$
    \item $(2k-1, 4k-1)$
    \item $(2k+2,1)$
    \item $(2k+2,2)$
    \item $(3,2k+2)$
    \item $(2k-1,4k-2)$
\end{itemize}
\end{definition}

\begin{proposition}
If $(G,<)$ is a cocomparability sun of $4k$ vertices, then the directed girth of $D_h(G,<)$ is $2k$.
\label{propos:cocomparability-sun}
\end{proposition}

\begin{proof}
The proof consists in describing precisely the arcs in $D_h(G,<)$. As the arcs in the directed height graph of $(G,<)$ are associated to the cocomparability triplets in $(G,<)$, this is equivalent to enumerating all such triplets. To do so, we will start by partitioning the vertices of $(G,<)$ into four sets $A, B, C$ and $D$, defined as follows.

\begin{itemize}
    \item $A = \{ 1,2,3 \}$
    \item $B = \{ 4,5,..., 2k\}$
    \item $C = \{ 2k+1, 2k+2, ..., 4k-3\}$
    \item $D = \{ 4k-2,4k-1,4k\}$
\end{itemize}

Note that if we extend the ordering to the sets we have $A<B<C<D$.
The two following claim state that some triplets cannot be comparability triplets.

\begin{claim}\label{clm:cocomp-sun-1}
No cocomparability triplet in $(G,<)$ can have (at least) a vertex in $B$ and (at least) another one in $C$.
\end{claim}

\begin{claimproof}
Recall that all edges of the join between vertices in $B$ and vertices in $C$ belong to $(G,<)$.
Suppose $x,y,z$ is a cocomparability triplet in $(G,<)$, with $x < y < z$. Due to their order relation, there are three cases to consider:
\begin{itemize}
    \item If $x \in A, y \in B, z \in C$, the cocomparability pattern does not appear because $(y,z) \in E$.
    \item If $x \in B, y \in C, z \in D$ , the cocomparability pattern does not appear because $(x,y) \in E$. 
    \item If $\{x,y,z\} \subseteq B \cup C$, there are at least two edges between vertices $x,y$ and $z$ this also implies a contradiction.
\end{itemize}
\end{claimproof}

\begin{claim}\label{clm:cocomp-sun-2}
No cocomparability triplet in $(G,<)$ can have (at least) a vertex in $A$ and (at least) another one in $D$.
\end{claim}

\begin{claimproof}
Suppose $x,y,z$ is a cocomparability triplet in $(G,<)$, with $x < y < z$. Then, necessarily $x \in A$ and $z \in D$. But this implies a contradiction, since there is no edge between $A$ and $D$, i.e. $(x,z) \notin E$.
\end{claimproof}

Below we list all the cases that are not covered by the claims, and for each list the corresponding cocomparability triplets.

\begin{enumerate}
    \item $ \{ x,y,z \} = A$. This is a cocomparability triplet.
    \item $ \{ x,y,z \} \subseteq B$. The cocomparability triplets of this form are $\{ x,y,z \} = \{ i-1,i,i+1 \}$ or $\{ x,y,z \} = \{ i-1,i,i+2 \} $ for some even $i$ such that $4<i<2k$.
    \item $ \{ x,y,z \} \subseteq C$. The cocomparability triplets of this form are $\{ x,y,z \} = \{ i-2,i,i+1 \}$ or $\{ x,y,z \} = \{ i-1,i,i+1 \} $ for some odd $i$ such that $2k+1<i<4k-3$.
    \item $ \{ x,y,z \} = D$. This is a cocomparability triplet.
    \item $ \{ x,y,z \} \subseteq A \cup B$ (with $\{ x,y,z \} \nsubseteq A$ and $\{ x,y,z \} \nsubseteq B$). The cocomparability triplets of this form are $\{3,4,5\}$, $\{3,4,6\}$ and $\{ 1,2,4\}$. 
    This is due to $(3,5)$, $(3,6)$ and $(1,4)$ being the only edges between $A$ and $B$ and therefore, due to the ordering constraint, the vertices 2 and 4 are the only candidates to form a cocomparability triplet with the edges' endpoints, respectively.
    \item $ \{ x,y,z \} \subseteq A \cup C$ (with $\{ x,y,z \} \nsubseteq A$ and $\{ x,y,z \} \nsubseteq C$). The cocomparability triplets of this form are $\{1,2k+1,2k+2\}$, $\{2,2k+1,2k+2\}$ and $\{ 3,2k+1,2k+2\}$. 
    This is due to $(1,2k+2)$, $(2,2k+2)$ and $(3,2k+2)$ being the only edges between $A$ and $C$ and therefore, due to the ordering constraint, the vertex $2k+1$ is the only candidate to form a cocomparability triplet with the edges' endpoints. 
    \item $ \{ x,y,z \} \subseteq B \cup D$ (with $\{ x,y,z \} \nsubseteq B$ and $\{ x,y,z \} \nsubseteq D$). The cocomparability triplets of this form are $\{2k-1,2k,4k-2\}$, $\{2k-1,2k,4k-1\}$ and $\{ 2k-1,2k,4k\}$. 
    This is due to $(2k-1,4k-2)$, $(2k-1,4k-1)$ and $(2k-1,4k)$ being the only edges between $B$ and $D$ and therefore, due to the ordering constraint, the vertex $2k$ is the only candidate to form a cocomparability triplet with the edges' endpoints.
    \item $ \{ x,y,z \} \subseteq C \cup D$ (with $\{ x,y,z \} \nsubseteq C$ and $\{ x,y,z \} \nsubseteq D$). The cocomparability triplets of this form are $\{4k-3,4k-1,4k\}$, $\{4k-4,4k-3,4k-2\}$ or $\{ 4k-5,4k-3,4k-2\}$. 
    This is due to $(4k-3,4k)$, $(4k-4,4k-2)$ and $(4k-5,4k-2)$ being the only edges between $C$ and $D$ and therefore, due to the ordering constraint, the vertices $4k-1$ and $4k-3$ are the only candidates to form a cocomparability triplet with the edges' endpoints, respectively.
\end{enumerate}

Now recall that if $x,y,z$ is a cocomparability triplet in $(G,<)$, then $(y \rightarrow x)$ and $(y \rightarrow z)$ are arcs of $D_h(G,<)$. The arcs in $D_h(G,<)$ can be deduced from the cocomparability triplets listed above and are illustrated by Figure \ref{fig:comparability-sun-ug}.

\begin{figure}[H]
    \centering
\input{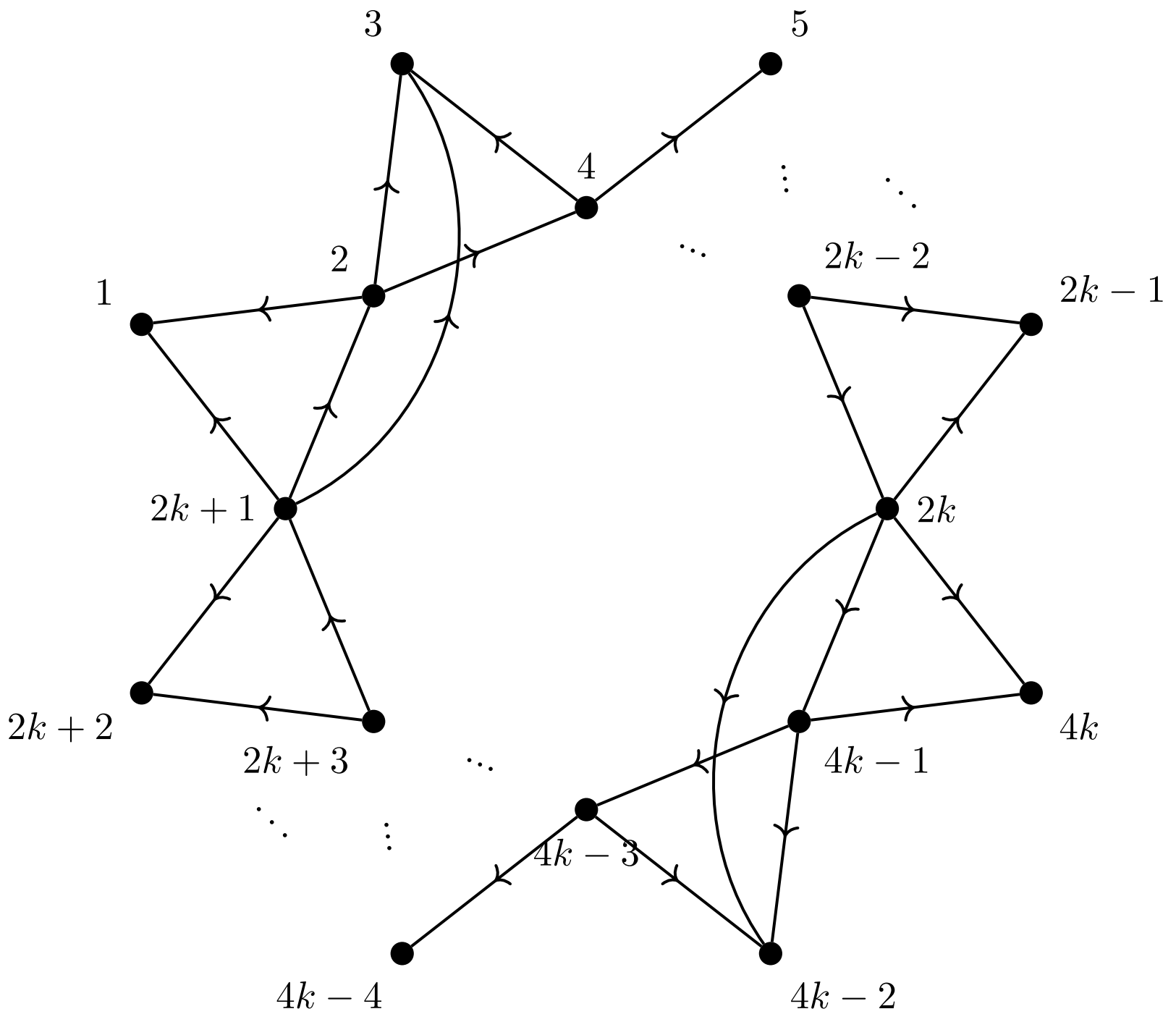}
    \caption{$D_h(G,<)$ when $G$ is the cocomparability sun of $4k$ vertices.}
    \label{fig:comparability-sun-ug}
\end{figure}

Then, we claim the following.

\begin{claim}
Vertices $2,4,6,8,...,2k, 4k-1,4k-3,4k-5,...,2k+3,2k+1$ form a directed cycle in $D_h(G,<)$ of length $2k$.
\end{claim}

\begin{claimproof}
As already stated, by construction we get the following:

\begin{itemize}
    \item For all even $i$ such that $2\leq i<2k$, $\{i-1,i,i+2\}$ is a cocomparability triplet in $(G,<)$ and therefore $(i  \rightarrow i+2)$ is an arc in $D_h(G,<)$.
    \item For all odd $i$ such that $2k+1 < i \leq 4k-1$, $\{i-2,i,i+1\}$ is a cocomparability triplet in $(G,<)$ and therefore $(i \rightarrow i-2)$ is an arc in $D_h(G,<)$.
    \item $\{2,2k+1,2k+2\}$ is a cocomparability triplet in $(G,<)$ and therefore $(2k+1 \rightarrow 2)$ is an arc in $D_h(G,<)$.
    \item $\{2k-1,2k,4k-1\}$ is a cocomparability triplet in $(G,<)$ and therefore $(2k \rightarrow 4k-1)$ is an arc in $D_h(G,<)$.
\end{itemize}
\end{claimproof}

And finally:

\begin{claim}
The length of the smallest directed cycle in $D_h(G,<)$ is of $2k$.
\end{claim}

\begin{claimproof}
Let's show that the directed cycle $2,4,6,8,...,2k, 4k-1,4k-3,4k-5,...,2k+3,2k+1$, that we will denote by $C$, is the smallest.
By arc enumeration in $D_h(G,<)$, we get that all the vertices in $D_h(G,<)$ that are not contained in $C$ are sinks, i.e. do not have out-coming arcs. Therefore, the only vertices that can be contained in a directed cycle are the ones in $C$. In addition, also by arc enumeration, we state that there is no \enquote{shortcut} in $C$. Then, it is not possible to obtain a smaller directed cycle induced by a strict subset of the vertices in $C$. We conclude that $C$ is the smallest directed cycle in $D_h(G,<)$.
\end{claimproof}
\end{proof}

\paragraph*{Putting all the pieces together to prove Theorem~\ref{thm:no-characterization-monotone}}

Using Proposition \ref{propos:cocomparability-sun}, we can now prove that no finite family of patterns can characterize the class of constrained monotone string graphs.

\begin{proof}[Proof of Theorem~\ref{thm:no-characterization-monotone}]

For the sake of contradiction, suppose that there exists a finite family $\mathcal{F}$ of patterns characterizing constrained monotone string graphs. Let $M$ be the number of vertices of the largest pattern in $\mathcal{F}$.

By Proposition~\ref{propos:cocomparability-sun}, there exists an ordered graph whose directed height graph has directed girth strictly larger than $M$: it suffices to take a cocomparability sun $(C,<)$ of size $4k$, with $k > 2$, such that $2k > M$. By Proposition~\ref{prop:PachTomonDh}, this ordered graph is \emph{not} a constrained monotone string graph. By our hypothesis, this implies $(C,<)$ must contain a pattern $P$ from $\mathcal{F}$. Let $V_{P}$ be a set of vertices of $(C,<)$ reproducing pattern $P$, and $(K,<)$ be the ordered subgraph of $(C,<)$ induced by $V_P$. 
Since $P$ has size at most $M$, so does $(K,<)$. By our hypothesis, $(K,<)$ is not a constrained monotone string graph, as it induces pattern $P$. We now consider its directed height graph $D_h(K,<)$. By definition, $D_h(K,<)$ must be a directed subgraph of $D_h(C,<)$, and by construction, $D_h(K,<)$ must be of size at most $M$. Then, we argue $D_h(K,<)$ has no directed cycle, because $D_h(C,<)$ has directed girth strictly larger than $M$. But if $D_h(K,<)$ has no directed cycle, by Proposition~\ref{prop:PachTomonDh}, $(K,<)$ admits a constrained monotone string representation. This contradicts the opposite statement we derived above, showing that the hypothesis was false. We conclude there is no finite set of patterns that characterizes constrained monotone string graphs.
\end{proof}

\section{Pattern-defined classes with no grounded intersection characterization}
\label{sec:no-characterization-by-geometry}

We say that a graph class has a \emph{standard grounded intersection characterization} if every graph in the class admits an intersection representation made of \emph{finite, compact and connected} shapes, and the set of valid representations is closed under removal, translation along the grounding line, uniform scaling and duplication. That is, applying any of these operations to a grounded intersection representation of a graph in the class must always yield a valid representation for the class. There can be some additional grounding rules specifying how the shapes interact with the grounding line, and these rules must not be violated when the operations are applied. For example, when duplicating a shape, the grounding point of the new shape must not coincide with the one of an existing shape. Note that all the geometric classes considered in this paper admit a standard grounded intersection characterization.

We use this definition to argue that the class $\Cao$ does not admit a standard constrained intersection characterization.

\ThmNoGeometricCharac*

\begin{proof}

For contradiction, suppose that there exists a standard constrained intersection characterization for $\Cao$. Then, we propose the following claim.

\begin{claim} There exists an ordered graph $(G,<)$ in $\Cao$ and a valid constrained geometric representation of it, having two intersecting shapes $u<v$, such that there is a ball around the grounding point of shape $v$, that the shape of $u$ does not intersect.
\label{claim:ball-cao-interval-pattern}
\end{claim}

In other words, it is not true that for any two intersecting shapes, the leftmost shape intersects the rightmost one at its grounding point. In the context of the paper, this means that we must enforce the first-interval intersection explicitly in our definition: it cannot just follow from the right choice of shapes.

\begin{proof}[Proof of Claim]

We start by stating that $\Cao$ is strictly larger than the class of ordered graphs avoiding the interval pattern in Figure \ref{fig:interval-pattern}, as we can take the graph in Figure \ref{fig:extended-claw} as an example. Then, let $(G,<)$ be an ordered graph in $\Cao$, such that $(G,<)$ reproduces the interval pattern in Figure \ref{fig:interval-pattern} by a triplet of vertices $u<v<w$. Let $R$ be the constrained geometric representation of $(G,<)$. Shapes $u$ and $w$ must intersect in $R$. If shape $u$ does not touch the grounding point of shape $w$, then $u$ and $w$ is our pair of shapes satisfying the claim, see Figure \ref{fig:beyond-standard-model-ca-case1}. Suppose now that shape $u$ touches the grounding point of shape $w$. As shapes $u$ and $v$ cannot intersect, we claim that there exists a ball around the grounding point of shape $v$ that the shape of $u$ does not intersect, the compactness of the shapes can be used as argument. Then, the shape $u$ must go above the ball surrounding the grounding point of shape $v$, in order to reach the grounding point of shape $w$. Now, remove all shapes in $R$ except shape $u$, and create a copy of $u$, denoted by $u'$, having as grounding point the former grounding point of shape $v$. From the hypothesis on shape $u$, this shape touches the grounding line at its grounding point and the former grounding point of $w$ (even if we removed shape $w$ from $R$, we will keep denoting by $w$ this point as reference). Hence, as $u'$ is a copy of $u$, we get that $u'$ will also touch the grounding line at a point $w'$, such that $u < u' < w < w'$. These conditions force the intersection of shapes $u$ and $u'$, as we can see in Figure \ref{fig:beyond-standard-model-ca-case2}. As shape $u$ was not modified, it is still true that there exists a ball around the grounding point of shape $u$, which is the former grounding point of shape $v$, that the shape of $u$ does not intersect. The claim follows as we obtained a new representation whose intersection graph must be in $\Cao$, where shapes $u$ and $u'$ satisfy the desired requirements.

\begin{figure}[!h]
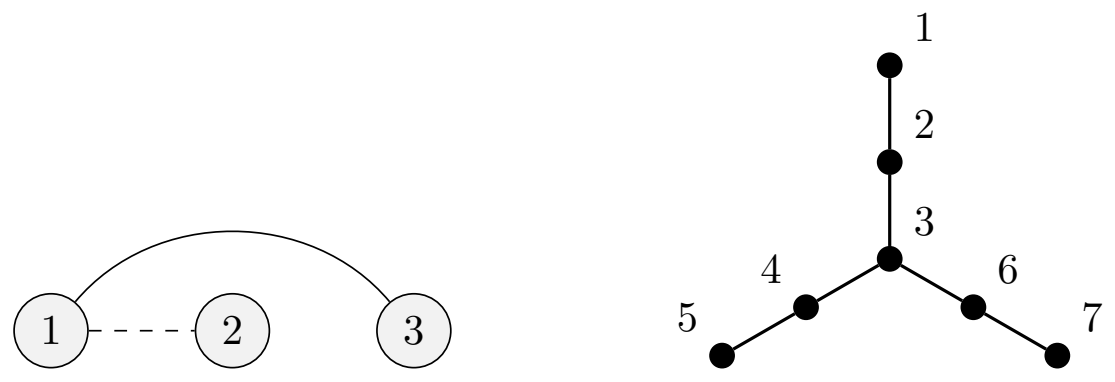

\centering
\captionsetup[subfigure]{skip=15pt}
\captionsetup{skip=12pt}
\begin{subfigure}{0.2\textwidth}
	\centering
	\scalebox{0.8}{
    \input{pictures-and-tikz/interval-pattern}}
	\caption{\label{fig:interval-pattern}Interval pattern.}
\end{subfigure}
\hspace{5ex}
\begin{subfigure}{0.3\textwidth}
	\centering
    \scalebox{0.8}{
    \input{pictures-and-tikz/extended-claw-graph}}
	\caption{\label{fig:extended-claw} Ordered graph in $\Cao$.}
\end{subfigure}
\caption{The ordering of the graph reproduces the interval pattern by vertices (3,5,6).}
\end{figure}
\end{proof}

\begin{figure}[!h]
\centering
\captionsetup[subfigure]{skip=15pt}
\captionsetup{skip=12pt}
\begin{subfigure}{\textwidth}
	\centering
	\scalebox{0.8}{%
	\input{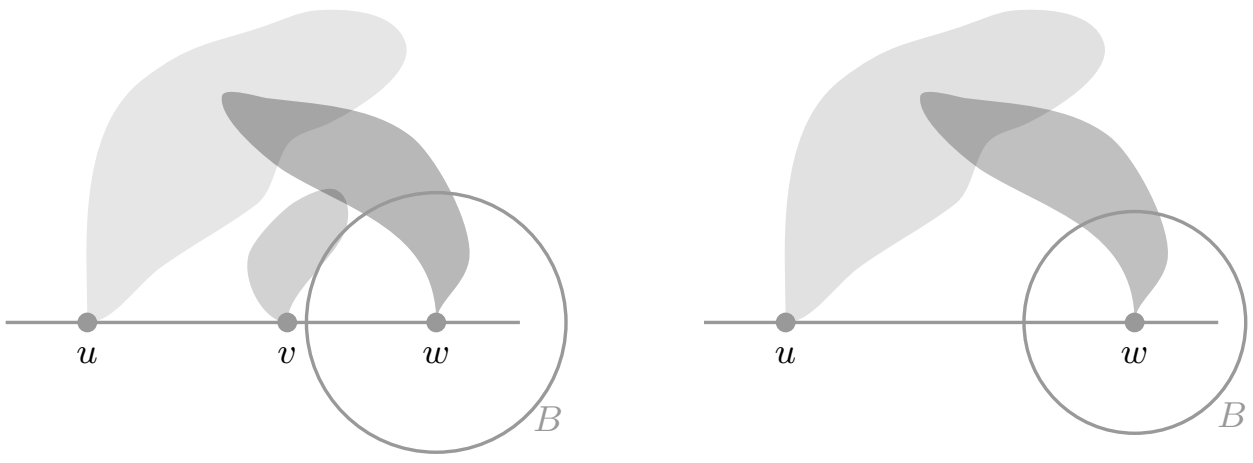}}
	\caption{\label{fig:beyond-standard-model-ca-case1} Case 1: shape $u$ does not touch the grounding point of shape $w$.}
\end{subfigure}

\bigskip

\begin{subfigure}{\textwidth}
	\centering
	\scalebox{0.8}{%
	\input{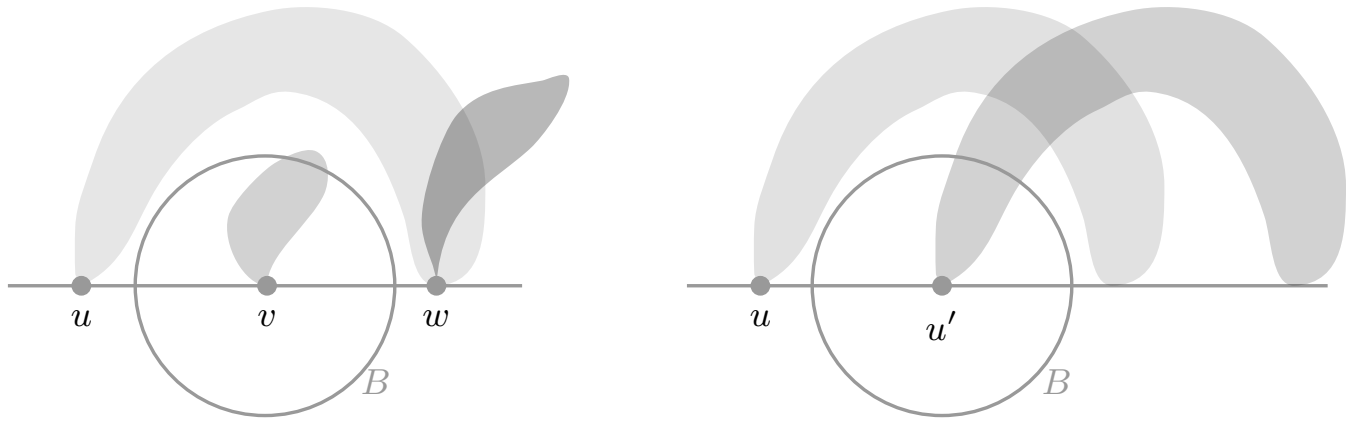}}
	\caption{\label{fig:beyond-standard-model-ca-case2} Case 2: shape $u$ touches the grounding point of shape $w$.}
\end{subfigure}
\caption{Obtaining a representation satisfying Claim \ref{claim:ball-cao-interval-pattern} of a representation of an ordered graph in $\Cao$ reproducing the interval pattern by a triplet of vertices $u,v,w$.}
\end{figure}

Now consider a geometry that satisfies the claim for two intersecting shapes $u$ and $v$ and remove all the other shapes. Let $B$ be the ball around $v$ that $u$ does not access. Copy both shapes and their relative position, and denote the new shapes by $u'$ and $v'$. Let the grounding points of the newly created shapes be such that $u<u'<v<v'$. Uniformly scale both shapes $u'$ and $v'$ such that both shapes are fully included in $B$, note that this is possible because all shapes must be finite. Then, by construction, we get that $u$ and $u'$ do not intersect. Also, $u$ and $v$ must intersect and so do $u'$ and $v'$. The new configuration must be the representation of a graph in the class $\Cao$, as only valid operations were applied to the initial geometry. However, we reach a contradiction since the ordered vertices $(u,u',v,v')$ form the pattern $P_a$. The strategy is shown by Figure \ref{fig:no-model-c_a}.

\begin{figure}[h!]
    \centering
    \input{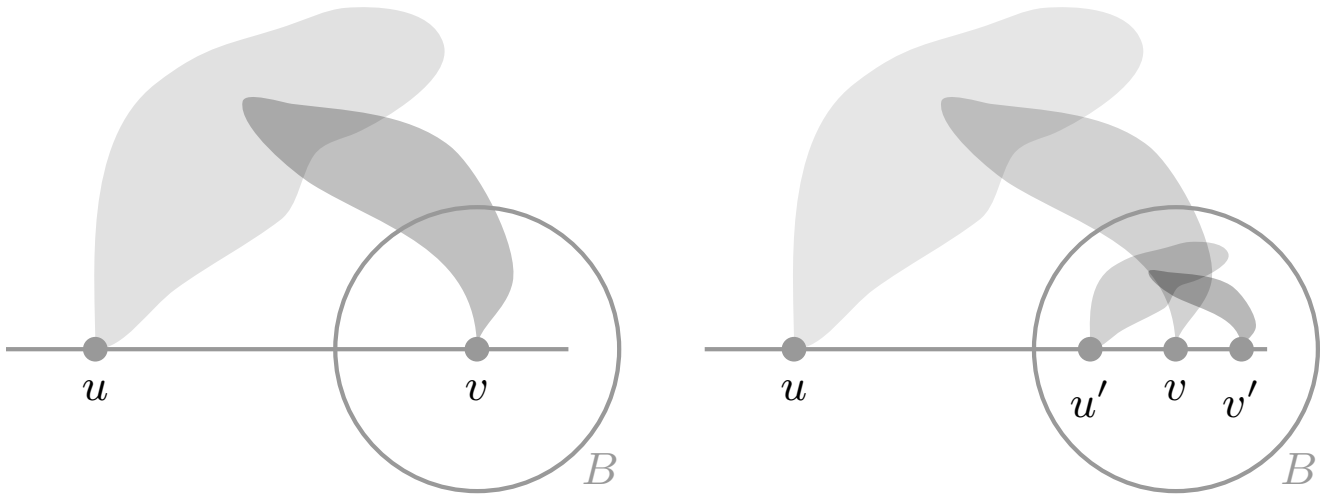}
    \caption{We obtain a valid constrained representation of an ordered graph in $\Cao$ whose grounding ordering reproduces pattern $P_a$, a contradiction.}
    \label{fig:no-model-c_a}
\end{figure}
\end{proof}

As an intuitive result to close this section, we state the following proposition.

\begin{proposition}
Any ordered graph admitting a standard constrained intersection representation avoids pattern $P_{abcd}$.
\end{proposition}

\begin{proof}
For contradiction, suppose there exists a valid standard geometric representation of an ordered graph such that the pattern $P_{abcd}$ is induced by four ordered vertices $u,v,w,x$. As $v$ is adjacent to neither $u$ nor $w$, its shape must be completely contained in the region between shape $u$, shape $w$ and the grounding line. Shape $x$ cannot enter this region without intersecting either shape $u$ or shape $w$. Thus, an intersection between shapes $v$ and $x$ is impossible in a standard representation.
\end{proof}
\section{Beyond standard intersection graphs: a constrained intersection model for~\texorpdfstring{$\Cao$}{Ca}}

\label{sec:first-intersection-filament-linked-interval-ca}

\begin{figure}[H]
    \centering
    \input{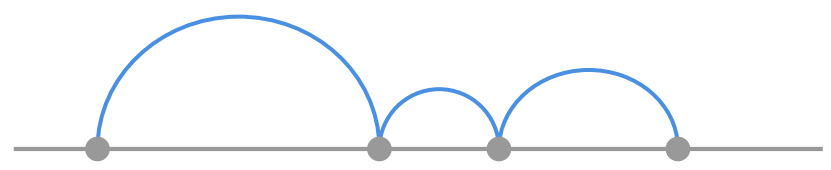}
    \caption{A grounded thread shape.}
    \label{fig:grounded-thread}
\end{figure}

As we saw in the previous section, it is not possible to define a standard constrained geometric characterization for the class $\Cao$. As it was previously argued, having a geometric characterization provides powerful insight that can be useful to better understanding pattern-defined classes and to produce efficient specialized algorithms. Therefore, this section is dedicated to the definition of a non-standard constrained intersection characterization for the class $\Cao$, hoping that it will be simple enough to still be insightful. In order to do this, we introduce a new shape called \emph{thread}. As for all the shapes considered in this paper, the threads will be anchored to the grounding line, see Figure \ref{fig:grounded-thread} for an illustration of a grounded thread. Recall that when working with ordered graphs, we will refer to their grounded intersection representations as \emph{constrained}.

\DefGroundedThread*

What makes the constrained intersection representation to be non-standard is the restriction we will impose to two intersecting thread shapes. This restriction will be denoted by \emph{first-interval intersection}, and we formally define it as follows.

\begin{definition}[First-interval intersection grounded thread representation]
A first-interval intersection grounded thread representation is an intersection representation of grounded threads such that, for any two threads $i$ and $j$, with $i < j$ in the grounding ordering, we get that:
\begin{itemize}
    \item The first half-circles of $i$ and $j$ do not intersect, nor one is contained inside the other. 
    \item If threads $i$ and $j$ intersect, then necessarily thread $i$ intersects the first half-circle of thread $j$.
\end{itemize}
\end{definition}

\begin{figure}[H]
    \centering
    \input{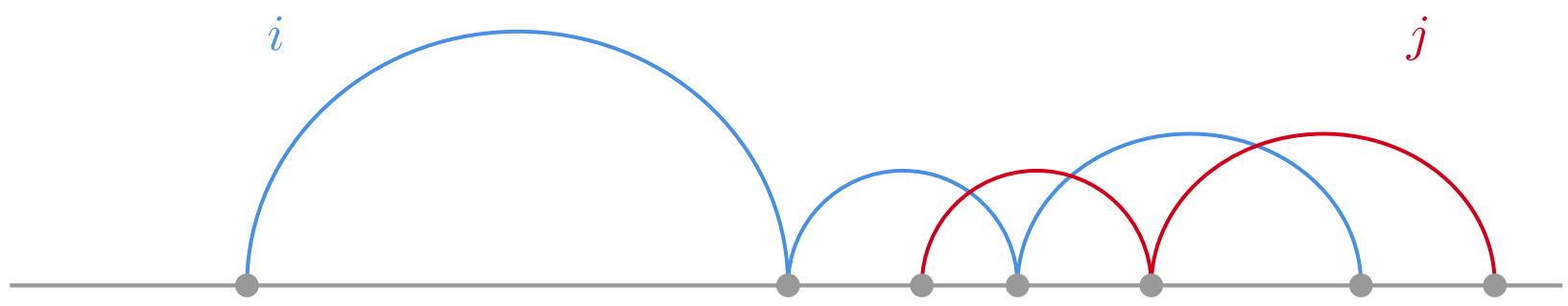}
    \caption{First-interval intersection grounded thread representation having two intersecting threads.}
    \label{fig:first-intersection-grounded-thread-model}
\end{figure}

Note that the grounding ordering between a set of grounded threads is defined by the threads' leftmost abscissa.

\ThmCaFilamentLinked*

\begin{proof}
For the first inclusion, let $(G,<)$, with $G=(V,E)$, be an ordered graph such that the vertices of $G$ ordered by $<$ avoid pattern $P_{a}$. We will construct a constrained first-interval intersection thread representation $M$ of $(G,<)$. For convenience let us denote $<$ as the natural numbering of the vertices from 1 to $n$.

For each vertex $i$, with $1 \leq i \leq n$, we define its first half-circle as the one going from coordinate $(i,0)$ to coordinate $(i+\frac{1}{2},0)$. We will define an additional half-circle of thread $i$ for each vertex $j$ adjacent to $i$ such that $i < j$. For this, let us define first $\varepsilon_i = \frac{1}{2} - \frac{i}{2n+2}$. Note that whenever $i < j$, we get $\varepsilon_i > \varepsilon_j$. Then, we follow the grounding ordering to take each $j$ with $(i,j) \in E$ and $i < j$, and to add a half-circle going from the right extremity of the last added half-circle of thread $i$ to coordinate $(j+ \varepsilon_i,0)$. This is depicted in Figure \ref{fig:ca-model-construction-1}.

\begin{figure}[H]
    \centering
    \input{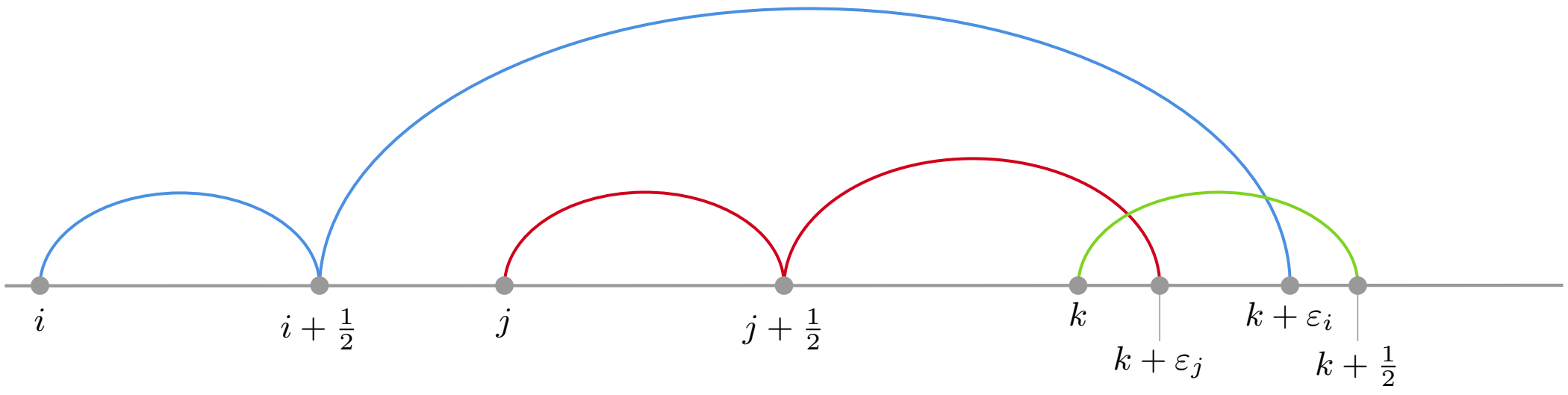}
    \caption{Constrained first-interval intersection thread representation of vertices $i < j < k$ such that $(i,j) \notin E$, $(i,k) \in E$ and $(j,k) \in E$.}
    \label{fig:ca-model-construction-1}
\end{figure}

From the construction of $M$, we get that if two vertices $i$ and $j$ are adjacent in $G$, then thread $i$ intersects thread $j$ in $M$ in its first half-circle. Note that threads $i$ and $j$ might have other intersections as well. 

Additionally, we claim the following:

\begin{claim}
If two vertices $i$ and $j$ are not adjacent in $G$, then their threads do not intersect in $M$.
\end{claim}

\begin{claimproof}
For contradiction, suppose threads $i$ and $j$ intersect, with $(i,j) \notin E$. Let $c_i$ and $c_j$ be the half-circles of $i$ and $j$ respectively, corresponding to the first intersection of the threads, i.e. the leftmost intersection of $i$ and $j$. From construction of $M$, $c_i$ cannot be the first half-circle of thread $i$, and $c_j$ cannot be the first half-circle of $j$. Without loss of generality, suppose the left extremity of $c_i$ has a strictly smaller abscissa value than the one of $c_j$. Then, necessarily the first half-circle of $j$ is contained in the region formed by $c_i$ and the grounding line, as otherwise either the intersection of $c_i$ and $c_j$ was not the leftmost intersection of threads $i$ and $j$ or $c_j$ was the first half-circle of $j$, both situations imply a contradiction. From the latter, we deduce $i < j$. The configuration is the illustrated by Figure \ref{fig:ca-model-construction-2}.

\begin{figure}[H]
    \centering
    \input{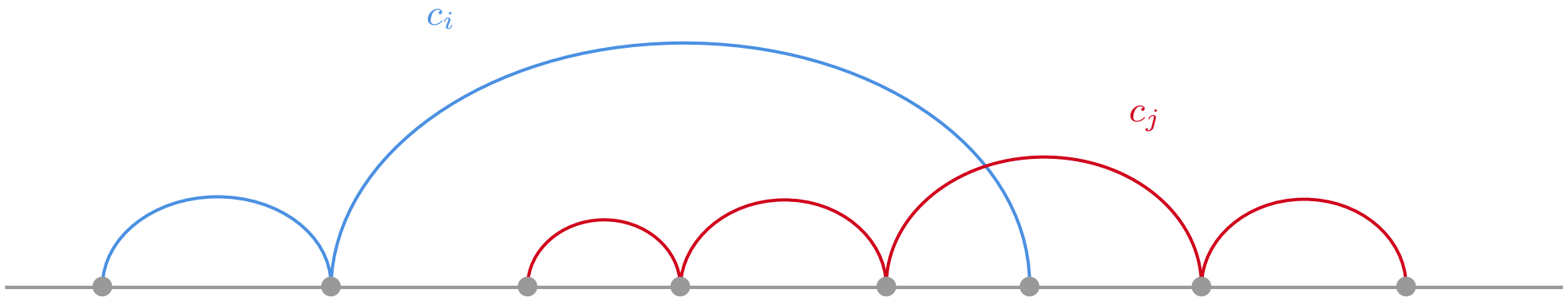}
    \caption{First half-circle of $j$ is contained in region formed by $c_i$ and the grounding line.}
    \label{fig:ca-model-construction-2}
\end{figure}

Now, from construction of $M$, if half-circle $c_i$ was added to thread $i$, there must exist a vertex $k$ such that $(i,k) \in E$ and $i < j < k$. Similarly, the presence of half-circle $c_j$ implies the existence of a vertex $l$ such that $(j,l) \in E$ with $i < j < l$. Note that it is not possible for $k$ to be equal to $l$, as $i < j$ implies $k + \varepsilon_i > k + \varepsilon_j$ and therefore the intersection would not be produced. Hence, we get $i < j < k < l$. But then, pattern $P_a$ is produced by vertices $(i,j,k,l)$, which implies a contradiction. 
\end{claimproof}

We conclude that $M$ is a constrained first-interval intersection thread representation of $(G,<)$.

For the inverse inclusion, we take a first-interval intersection grounded thread representation and prove that the grounding ordering, denoted by $<$, avoids pattern $P_a$. For contradiction, suppose there exist a set of ordered vertices $a,b,c,d$ reproducing pattern $P_a$. We know the extremities of the first half-circles of threads $a$, $b$, $c$ and $d$ define ordered and pairwise disjoint intervals of the grounding line. Also, as vertices $a$ and $c$ are adjacent, thread $a$ needs to intersect thread $c$ at its first half-circle. As thread $a$ cannot intersect thread $b$, this implies the existence of a half-circle of thread $a$ delimiting a closed region with the grounding line which contains the first half circle of thread $b$. The latter is illustrated in Figure \ref{fig:ca-model-proof}. 

\begin{figure}[H]
    \centering
    \input{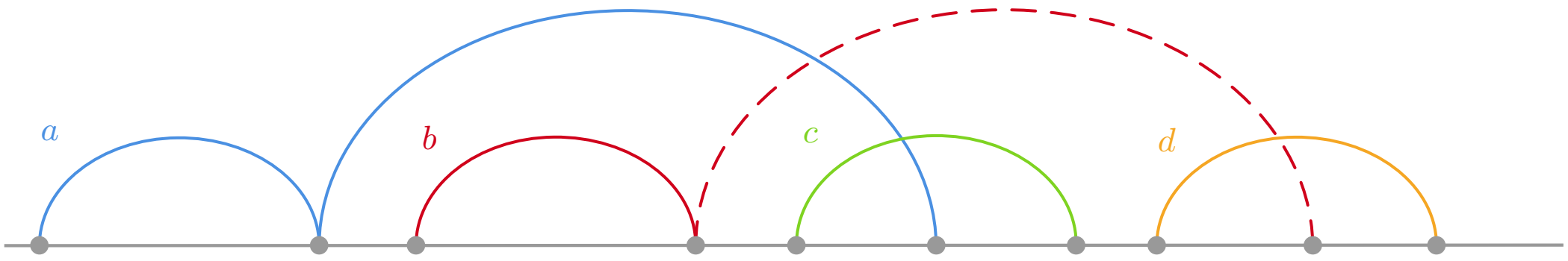}
    \caption{Pattern $P_a$ implies a contradiction in the representation.}
    \label{fig:ca-model-proof}
\end{figure}

We see that it is impossible then for thread $b$ to intersect thread $d$ without creating an unwanted intersection with thread~$a$. \end{proof}


\section{Toolbox for extending some results to the unordered setting}
\label{sec:toolbox-cycle-lemma}

Until now, we focused on establishing a class hierarchy for ordered graphs, and therefore constrained configurations. Now, we wish to extend our results to the unordered setting. To do so, we will generalize a powerful lemma presented in \cite{CardinalFMTV18}, which in turn was an adaptation to the grounded setting of the ordering gadget used in \cite{schaefer2009complexity}.
In its original form \cite{JelinekT19}, this lemma is called \emph{The Cycle Lemma} as it considers the \emph{cycle extension} of an ordered graph $(G,<)$ to an unordered graph $H$, containing the underlying unordered graph $G$ as induced subgraph. The particularity of the extension, is that it guarantees that the ordering of vertices in $V_G$, i.e. the vertices of $G$, will be equivalent to $<$ in any grounded 1-string representation of $H$. Note that a grounded 1-string representation is a representation where each pair of strings have at most one intersection and that two linear orderings are \emph{equivalent} if one can be obtained from the other by applying a cyclic shift and / or a reversal. 
These types of tools are referenced in the literature as \emph{ordering gadgets}, as they fix the ordering (up to reversal and cyclic shift) of a subset of vertices in any such representation. As we wish to consider monotone representations where strings intersect potentially multiple times, we will generalize this lemma such that the ordering of vertices in $V_G$ will be equivalent to $<$ in any grounded string representation of $H$.

\begin{figure}[h!]
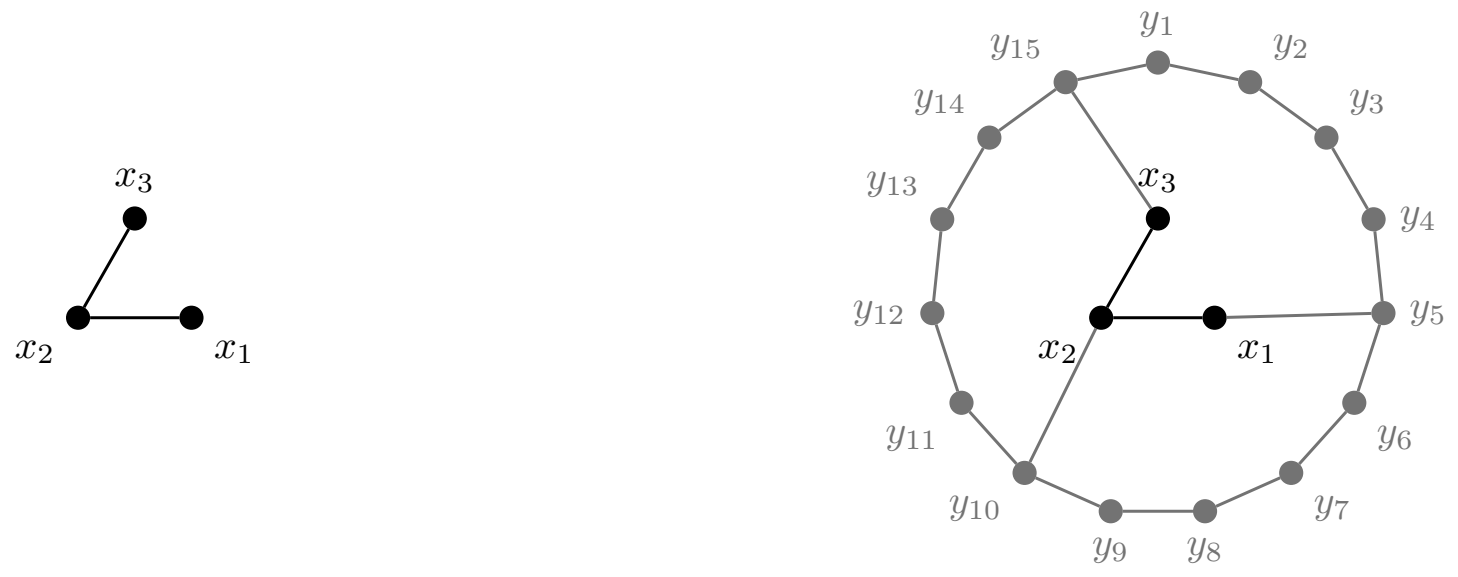

    \centering
\begin{subfigure}{0.45 \textwidth}
	\centering
	\scalebox{0.75}{
    \input{pictures-and-tikz/cycle-extension}}
	\end{subfigure}
\begin{subfigure}{0.5 \textwidth}
	\centering
    \scalebox{0.75}{
    \input{pictures-and-tikz/cycle-extension2}}
\end{subfigure}
    \caption{An ordered graph $(G,<)$ (left) and its unordered cycle extension $H$ (right).}
    \label{fig:cycle-extension-figure}
\end{figure}

First, we propose a slight variation of the cycle extension of an ordered graph $(G,<)$ defined by \cite{JelinekT19}. This extension is illustrated by Figure \ref{fig:cycle-extension-figure}. 

\begin{definition}[Cycle extension]
A cycle extension of the ordered graph $(G,<)$, with $G=(V_G,E_G)$, is an unordered graph $H=(V_H,E_H)$, such that:
\begin{itemize}
    \item $V_H = V_G \cup V_C$ with $V_G = \{ x_1, ..., x_n \}$ and $V_C = \{ y_1, ..., y_{5n} \}$.
    \item $E_H = E_G \cup \{ (x_i, y_{5i}) \hspace{0.5ex} | \hspace{0.5ex} i = 1,...,n \} \cup \{ (y_j,y_{j+1}) \hspace{0.5ex} | \hspace{0.5ex} j = 1,...,5n \}$ 
\end{itemize}
\label{def:cycle-extension}
\end{definition}

For simplicity, throughout this subsection we consider $5n + 1 = 1$.

In order to prove the following lemma, we follow an analogous strategy to the ones in \cite{JelinekT19,CardinalFMTV18}. As we aim to consider grounded string representations, and not grounded 1-string representations as in the previous works, we will handle carefully the potential multiple intersections between every pair of intersecting strings, and this will be the distinctive aspect of our proof.

\begin{lemma}
Let $(G,<)$ be an ordered graph and $H$ its cycle extension. Then, in every grounded string representation of $H$, the ordering of the vertices of $G$ induced by the representation will be equivalent to the ordering $<$.
\label{lem:cycle-extension-grounded-string-representation}
\end{lemma}

\begin{proof}
Let $(G,<)$ be an ordered graph with vertices denoted by $V_G= \{ x_1, ..., x_n \}$ and edges denoted by $E_G$. Let $H$ be its cycle extension, having vertices $V_H = V_G \cup V_C$ with $V_C = \{ y_1, ..., y_{5n} \}$ and edges $E_H$ as detailed in Definition \ref{def:cycle-extension}. Suppose $H$ has a grounded string representation. For simplicity, we will transform the representation into an outer-string representation, while preserving the induced vertex ordering up to equivalence. Then, suppose an outer-string representation $\mathcal{R}$ of $H$ is given, inside a disk of boundary $B$. We will denote by the \emph{anchor} of a string its intersection point with $B$. Let $c_j$ and $s_i$ be the strings in $\mathcal{R}$ respectively representing vertices $y_j$ and $x_i$, with $1 \leq i \leq n$ and $1 \leq j \leq 5n$. Note that string $c_j$ will only intersect, an arbitrary number of times, strings $c_{j-1}$, $c_{j+1}$ and when $j=5i$ also string $s_i$. The \emph{first intersection} of a string $c_j$ and a string $c_{j+1}$ will be defined with respect to the anchor of $c_j$. We define the \emph{central} part of a curve $c_j$, denoted by $center(j)$, as the subcurve of $c_j$ going from the first intersection of $c_j$ with $c_{j-1}$ to the first intersection of $c_j$ with $c_{j+1}$. This is illustrated in Figure \ref{fig:cycle-extension-substring-center}.

\begin{figure}[h!]
    \centering
    \input{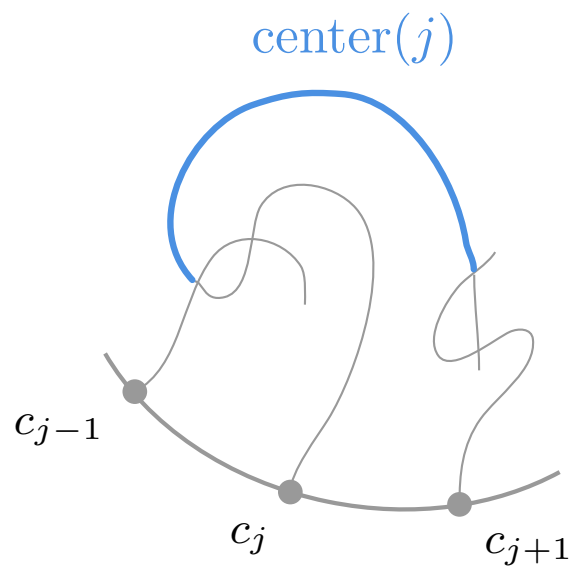}
    \caption{Defining the center of string $c_j$ as the subcurve going from the first intersection of $c_j$ with $c_{j-1}$ to the first intersection of $c_j$ with $c_{j+1}$.}
    \label{fig:cycle-extension-substring-center}
\end{figure}

We will now define a closed Jordan curve formed by subcurves of every $c_j$. This curve will allow us in turn to define $n$ interior disjoint regions containing the anchor points of strings $s_1, ..., s_n$ and therefore forcing a particular cyclic ordering on vertices $x_1,...,x_n$ in $\mathcal{R}$.

Consider the curve formed by $center(1),center(2),...,center(5n)$, that we will denote by $\mathscr{C}$. We claim that $\mathscr{C}$ is a closed curve, as $center(j)$ and $center(j+1)$ share a common endpoint: the one corresponding to the first intersection of strings $c_j$ and $c_{j+1}$. From construction, $\mathscr{C}$ contains subcurves of strings $c_1,...,c_{5n}$ in their natural ordering. We note that $\mathscr{C}$ is not necessarily a simple curve as, one the one hand, we allow strings to self-intersect and, on the other hand, $center(j)$ might intersect $center(j+1)$ multiple times, forming \emph{loops} in $\mathscr{C}$, see Figure \ref{fig:non-simple-closed-curve}. However, as every $c_j$ intersects exclusively strings $c_{j-1}$ and $c_{j+1}$ of the string set $\{c_1,...,c_{5n}\}$, we argue that loops are formed by at most two strings $c_{j}$ and $c_{j+1}$. Then, we can remove all the loops formed by one or two strings, by jumping from the first visit of an intersection point to the last visit of that same point in $\mathscr{C}$, to obtain a simple closed subcurve that we will denote by $J(\mathscr{C})$. As no more than two strings in $\{c_1,...,c_{5n}\}$ intersect at the same point, this loop removal will not produce the complete removal of a string $c_j$ in $J(\mathscr{C})$. Then, $J(\mathscr{C})$ will contain subcurves of every string $c_j$, with $1\leq j\leq 5n$, in the natural cycle ordering of $V_C$.

\begin{figure}[h!]
    \centering
    \input{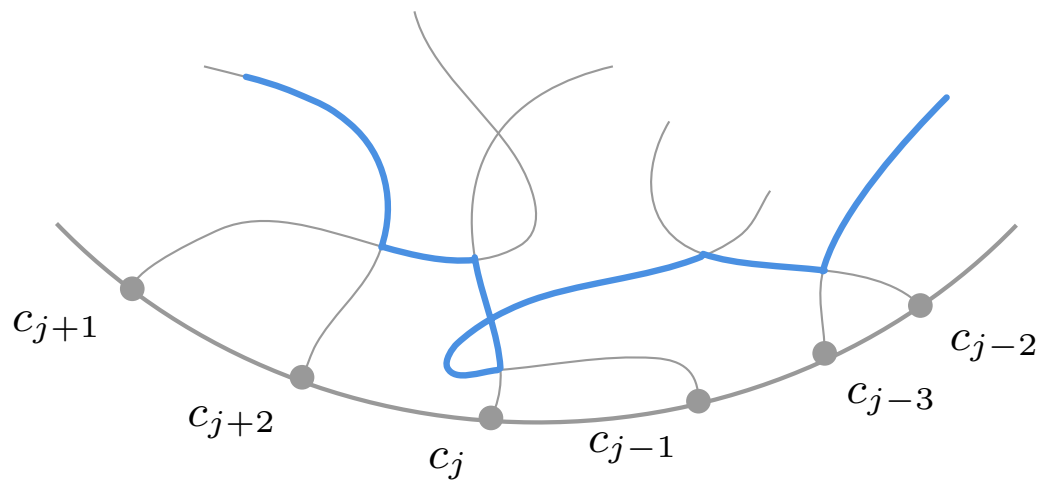}
    \caption{The closed curves formed by $center(1),center(2),...,center(5n)$ can contain loops, as shown by the subcurves of strings $c_j$ and $c_{j-1}$. }
    \label{fig:non-simple-closed-curve}
\end{figure}

We define as the \emph{start} of a string $c_j$, the subcurve of $c_j$ going from its anchor point to its first intersection  with $J(\mathscr{C})$ and denote it by $start(j)$. Following the notation in \cite{CardinalFMTV18}, given three distinct points $p,q,r$ on $J(\mathscr{C})$, we denote by \emph{path}($p,q,r$) the portion of $J(\mathscr{C})$ bounded by $p$ and $q$ containing point $r$. Similarly, if $p,q,r$ are three points in the boundary $B$, we denote by \emph{circle}($p,q,r$) the portion of $B$ bounded by $p$ and $q$ containing $r$. For simplicity, for a string $c_j$, we will denote by $j$ its anchor point and by $p_j$ its first intersection with $J(\mathscr{C})$. Then, we define the region $R_i$ (depicted in Figure \ref{fig:cycle-extension-region-ri}), for $1 \leq i \leq n$, as the planar region bounded by the union of the following subcurves:

\begin{itemize}
    \item $start(5i-3)$.
    \item $start(5i+2)$.
    \item \emph{circle}($5i-3,5i+2,5i$).
    \item \emph{path}($p_{5i-3},p_{5i+2},p_{5i}$).
\end{itemize}

\input{pictures-and-tikz/cycle-extension-ri-region-tikz}

By definition of the $R_i$ regions, we get that regions $R_i$ and $R_{i+1}$, need to be contiguous, as they both contain the subcurve $start(5i + 2)$ in their boundaries. This implies that all $R_i$ and $R_j$ with $j\neq i$ are interior disjoint and that $B$ is partitioned by these regions, i.e. $B=(B\cap R_1) \cup ... \cup (B\cap R_n)$. Also, as region $R_i$ has boundaries delimited by fragments of strings $c_{5i-3}, ..., c_{5i+2}$ and $B$, we deduce $s_i$ is the only string among $s_1,...,s_n$ that can cross, potentially many times, the boundary of $R_i$ and it can do so in points contained in $J(\mathscr{C}) \cap c_{5i}$. Then, as $s_i$ can only access region $R_i$ from regions $R_1,...,R_n$, we deduce its anchor point must be contained in $R_i$. Therefore, the anchor points of $s_1, ..., s_n$ appear on $B$ in the ordering which, up to equivalence, corresponds to the ordering $<$ on $V_G$, concluding the proof. 
\end{proof}

The following lemma is an adaptation of yet another lemma in~\cite{JelinekT19}. It will be useful when proving the strict inclusion between different classes, as we will see.

\begin{lemma}
Let $\mathcal{R}$ be a  grounded monotone string representation of a graph $G=(V_G,E_G)$ inducing a grounding ordering $<$ on $V_G$. Then, a grounded  monotone string  representation of the cycle extension $H=(V_H,E_H)$ of the ordered graph $(G,<)$ can be constructed by adding into $\mathcal{R}$ the  monotone strings  corresponding to vertices of $V_H \setminus V_G$.
\label{lem:cycle-extension-grounded-stairs-representation}
\end{lemma}

\begin{figure}[h!]
    \centering
    \input{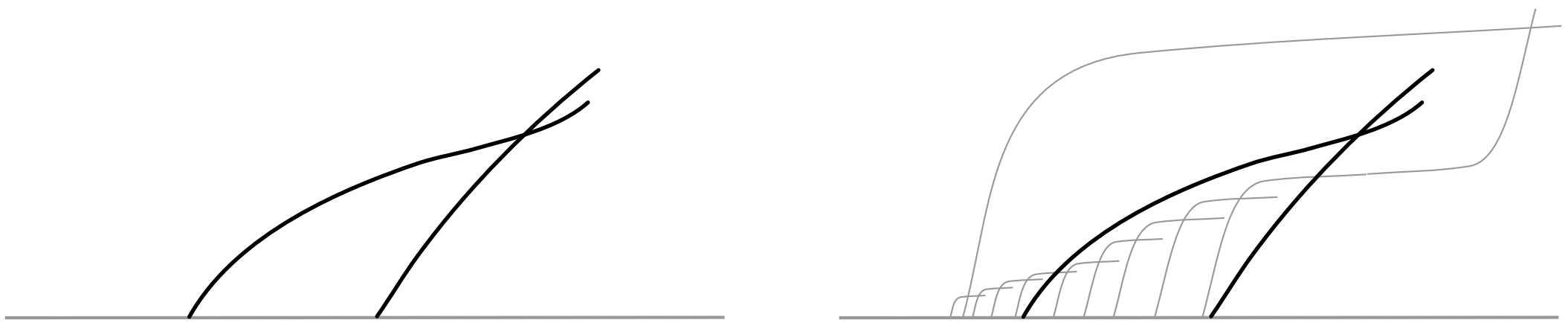}
    \caption{Transforming a grounded string representation of $G$ (left) inducing the grounding ordering $<$ to a grounded string representation of $H$ (right), with $H$ being the cycle extension of $(G,<)$.}
    \label{fig:cycle-extension-rep}
\end{figure}

The idea is that any grounded monotone string representation $\mathcal{R}$ of $G$ can be extended with the monotone strings of $V_H \setminus V_G$, by placing the corresponding strings sufficiently close to the grounding line, as shown in Figure \ref{fig:cycle-extension-rep}. For a formal proof, we refer the reader to~\cite{JelinekT19}.

\section{Strict inclusions for unordered pattern-defined classes}
\label{sec:strict-inclusions-unordered-pattern-classes}

\begin{figure}[h!]
    \centering
    \input{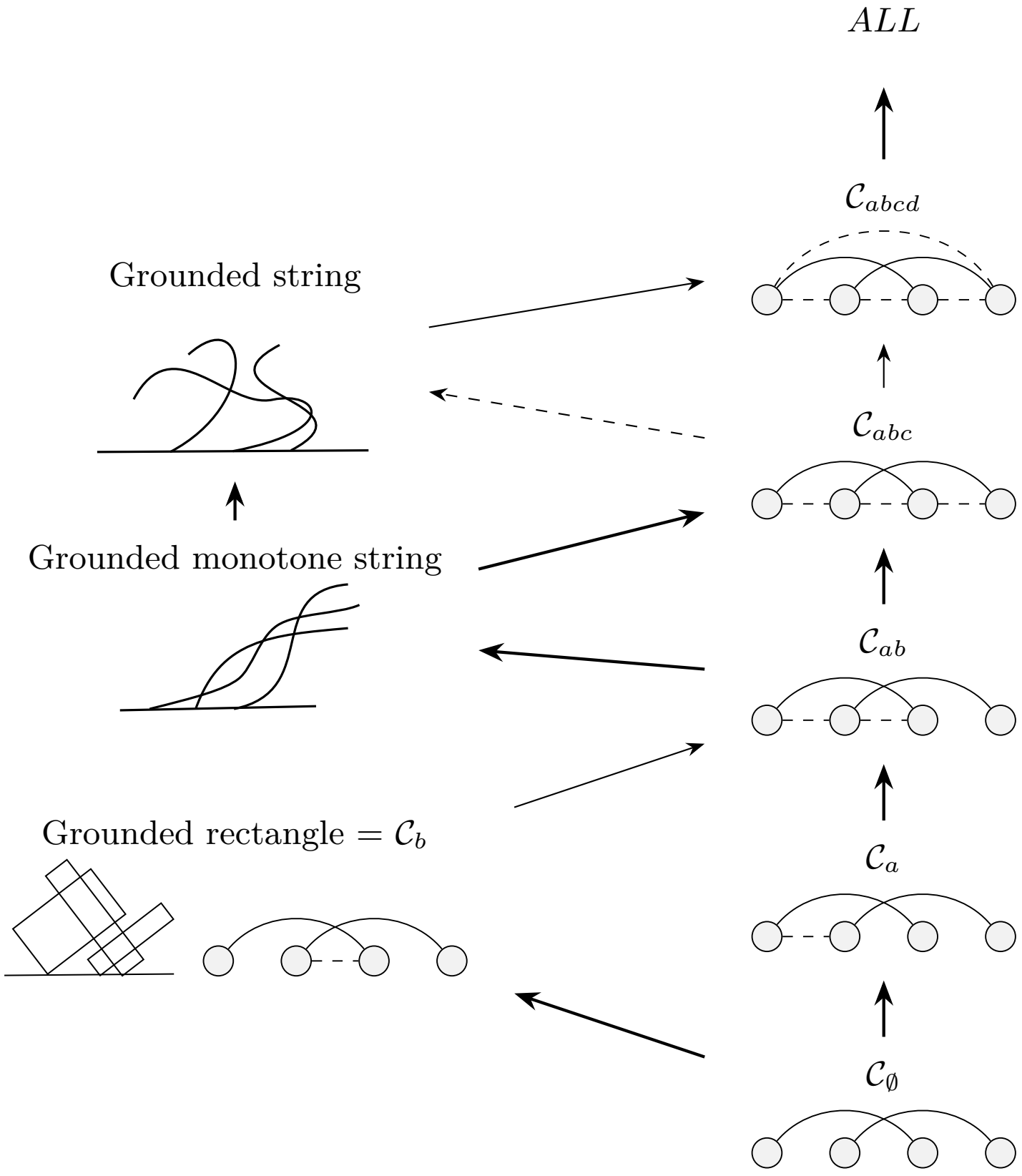}
    \caption{Strict inclusions for the unordered pattern-defined graph classes are represented by dark arrows in the diagram. We denote by $ALL$ the class of all graphs.} 
    \label{fig:strict-inclusions-diagram}
\end{figure}

In this section, we will discuss the separation of different unordered graph classes, as summarized in Figure \ref{fig:strict-inclusions-diagram}. We will do so by means of the lemmas presented in Section \ref{sec:toolbox-cycle-lemma} (Subsections \ref{subsec:strict-inclusions-for-grounded monotone-strings} and \ref{subsec:classes-c_abc-c_abcd}) and computer enumeration (Subsection \ref{subsec:more-strict-inclusions}).  These results are also valid for the ordered setting, as we argue next. Let $X$ and $Y$ be two unordered graph classes, admitting characterizations of the form \enquote{a graph belongs to $X$ (resp. $Y$) if and only if it admits an ordering satisfying a certain property $\varphi_X$ (resp. $\varphi_Y$)}. Suppose now $X \subsetneq Y$. Then, necessarily there exists a graph $G$ such that $G \in Y$ and $G \notin X$. Let $<_G$ be the ordering of the vertices in $G$ certifying that $G$ is in $Y$. Necessarily $<_G$ does not satisfy $\varphi_X$ as otherwise $G$ would belong to $X$ as well. Then, the ordered graph $(G,<_G)$ belongs to the class of ordered graphs $Y^{<}$ but not to $X^{<}$.

\subsection{Strict inclusions for grounded monotone strings}
\label{subsec:strict-inclusions-for-grounded monotone-strings}

In order to apply the extended Cycle Lemma (i.e. Lemma \ref{lem:cycle-extension-grounded-string-representation}) presented in Section \ref{sec:toolbox-cycle-lemma}, we start by defining an operation that we will call \emph{mirrored duplication}, also introduced in \cite{JelinekT19}. From an ordered graph $(G,<)$ with vertices $x_1 < ... < x_n$, we construct a second ordered graph $(G',<')$ by taking a mirrored disjoint copy of $(G,<)$ with vertices $y_1 < ... < y_n$ and by placing the mirrored copy to the right in the ordering $<'$, i.e. defining $<'$ as $x_1 <' ... <' x_n <' y_n <' ... <' y_1$. 
Finally, we construct a third ordered graph $(G'',<'')$ by placing two disjoint copies of $(G',<')$ consecutively. Note that if $(G,<)$ reproduces a pattern $P$, then any equivalent ordering to $<''$ will reproduce $P$ in $G''$. 

We will now extend the results presented in Theorem \ref{theo:geometry-in-patterns} (2).


\begin{theorem} The following holds:
 grounded monotone strings $\subsetneq \mathcal{C}_{abc}$.
\label{theo:c_ab-strictly-in-gstairsstrictly-in-c_abc}
\end{theorem}

\begin{proof}
Let $(G,<)$ be the Pach-Tomon ordered graph depicted in Figure \ref{fig:pach-tomon-graph-intro}. We will denote by $(G'',<'')$ its mirrored duplication. Note that any equivalent ordering of $<''$ will induce a Pach-Tomon ordered subgraph in $G''$. 

On the one hand, we claim that the cycle extension $H$ of $(G'',<'')$ will be an unordered graph admitting an ordering avoiding $P_{abc}$. Let us prove the latter claim. As ordering $<$ of $G$ avoids pattern $P_{abc}$, and $P_{abc}$ is a symmetrical pattern, necessarily the ordering $<'$ of $G'$ will avoid the pattern as well. Finally it is easy to see that taking two disjoint copies of $(G',<')$ to produce $(G'',<'')$ implies that $<''$ is also a pattern-free ordering of $G''$. Let $v_1,...,v_{32}$ be the ordered vertices of $(G'',<'')$ and $c_1,...,c_{5\times 32} = c_{160}$ be the vertices added by the cycle extension, i.e. the vertices of $V_{H} \setminus V_{G''}$, ordered in the natural cycle ordering. Then, $ c_1 < ... < c_5 < v_1 < c_6 < ... < c_{10} < v_2 < c_{11} < ... < c_{5i} < v_i < c_{5i+1} < ... < v_{32} $, with $1 \leq i \leq 32$ is a $P_{abc}$-free ordering of $H$ (we denote the ordering by $<$ and not by $<_{H}$ to lighten the notation).

On the other hand, we argue that $H$ does not admit a grounded monotone string representation. By Lemma \ref{lem:cycle-extension-grounded-string-representation}, any grounded string representation of $H$ will induce an ordering of vertices of $G''$ equivalent to the ordering $<''$. In particular, if $H$ admits a grounded monotone string representation, $H$ will contain the Pach-Tomon graph as induced subgraph ordered by the grounding ordering. But the latter implies a contradiction as this graph was proven to be a monotone strings obstruction. We conclude that $H$ belongs to $\mathcal{C}_{abc}$ but not to the grounded monotone string graphs.

\end{proof}

As the class $\mathcal{C}_{ab}$ is included in the class of grounded monotone strings, we state the following as a natural corollary:

\begin{corollary}
The following holds: $\mathcal{C}_{ab} \subsetneq \mathcal{C}_{abc}$.
\end{corollary}

Note that these results are also valid for the ordered setting, i.e. the class $\Cabo$ is strictly contained in the class of constrained monotone strings, which in turn is strictly in the class $\Cabco$. 

The following is easily deduced using the extended Cycle Lemma and can be seen as a complement to the inclusion hierarchy established \cite{JelinekT19}.

\begin{theorem}
The following holds: grounded monotone strings $\subsetneq $ grounded strings.
\end{theorem}

\begin{proof}
Once again, we take $(G,<)$ as the Pach-Tomon ordered graph (see Figure \ref{fig:pach-tomon-graph-intro}). We define $(G'',<'')$ as the mirrored duplication of $(G,<)$ and $H$ as the cycle extension of $(G'',<'')$. 
As $G$ does not admit a grounded monotone strings representation having $<$ as grounding ordering, neither does $G''$ ordered by $<''$. Then, by the Cycle Lemma (Lemma \ref{lem:cycle-extension-grounded-string-representation}), $H$ does not belong to the grounded monotone string graph class. However, $G$ does admit a grounded string representation with $<$ as grounding ordering, see Figure \ref{fig:grounded-string-patchtomon}, and therefore it is easy to see that $G''$ admits a grounded string representation ordered by $<''$ as well. Then, by Lemma \ref{lem:cycle-extension-grounded-stairs-representation}, we state that $H$ belongs to the grounded string graph class. We state that graph $H$ serves to distinguish both classes, which concludes the proof.

\begin{figure}[h!]
    \centering
    \input{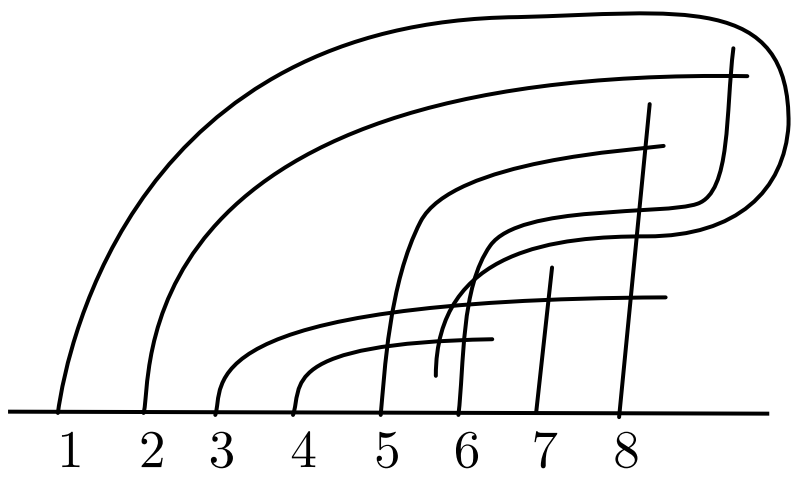}
    \caption{Constrained string representation of the Pach-Tomon ordered graph.}
    \label{fig:grounded-string-patchtomon}
\end{figure}
\end{proof}
\subsection{A note on classes  \texorpdfstring{$\mathcal{C}_{abc}$}{Pabc} and \texorpdfstring{$\mathcal{C}_{abcd}$}{Pabcd}}
\label{subsec:classes-c_abc-c_abcd}

We now show that if the class $\mathcal{C}_{abc}$ were to admit a characterization by a grounded intersection representation, then we could prove that classes $\mathcal{C}_{abc}$ and $\mathcal{C}_{abcd}$ are distinct as the inclusion of the first one into the second one would be necessarily strict.

\begin{proposition}
If $\mathcal{C}_{abc}$ admits a characterization by a grounded intersection representation, then $\mathcal{C}_{abc} \subsetneq \mathcal{C}_{abcd}$.
\end{proposition}

\begin{figure}[!h]
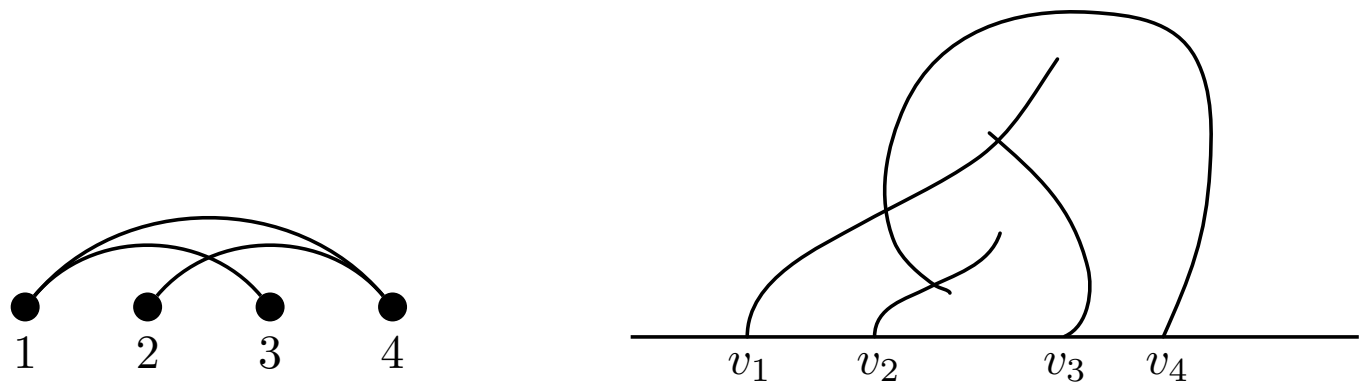

\centering
\begin{subfigure}{0.35 \textwidth}
	\centering
	\scalebox{0.9}{
    \input{pictures-and-tikz/c_abc-in-c_abcd-proof}}
	\end{subfigure}
\begin{subfigure}{0.35 \textwidth}
	\centering
    \scalebox{0.9}{
    \input{pictures-and-tikz/string-rep-for-non-convex-shapes}}
\end{subfigure}
\caption{Ordered graph $(G,<)$ (left) and its constrained string representation (right).}
\label{fig:prop-grounded-string-Pabc}
\end{figure}

\begin{proof}
We start by proving the existence an unordered graph in $\mathcal{C}_{abcd}$, such that any grounded string representation of $H$ induces a grounding ordering reproducing pattern $P_{abc}$.

Consider the ordered graph $(G,<)$ with vertices $v_1 < v_2 < v_3 < v_4$ and edge-set $\{(v_1,v_3), $ $(v_1,v_4), (v_2,v_4)\}$. Pattern $P_{abc}$ is clearly reproduced by vertices $v_1,v_2,v_3,v_4$. Then, if we define $(G'',<'')$ as the mirrored duplication of $(G,<)$, we get that any equivalent ordering of $<''$ will reproduce pattern $P_{abc}$ in $G''$. Also, as $(G,<)$ avoids pattern $P_{abcd}$, we state that any equivalent ordering of $<''$ will avoid the pattern in $G''$ as well.

Let $H$ be the cycle extension of $(G'',<'')$. As $(G,<)$ admits a grounded string representation, as shown in Figure \ref{fig:prop-grounded-string-Pabc}, it is easy to see that so will $(G'',<'')$. Then, by Lemma \ref{lem:cycle-extension-grounded-stairs-representation}, we argue that $H$ admits a grounded string representation as well. Then, as the class of grounded string graphs belong to $\mathcal{C}_{abcd}$, necessarily $H$ belongs to this class too. This can be certified by the following ordering. Let $v_1,...,v_{16}$ be the vertices of $(G'',<'')$ and $c_1,...,c_{5\times 16} = c_{80}$ be the vertices of $V_H \setminus V_{G''}$. Then, $ c_1 < ... < c_5 < v_1 < c_6 < ... < c_{10} < v_2 < c_{11} < ... < c_{5i} < v_i < c_{5i+1} < ... < v_{16} $, with $1\leq i\leq 16$, is a $P_{abcd}$-free ordering of $H$. However, by Lemma \ref{lem:cycle-extension-grounded-string-representation}, any grounded string representation of $H$ will induce a grounding ordering of the vertices of $G''$ equivalent to the ordering $<''$, and therefore reproducing pattern $P_{abc}$. 

Now, as strings are the most general shape, we conclude that if the class $\mathcal{C}_{abc}$ were to admit a grounded intersection representation, then graph $H$ could be used to distinguish both classes.
\end{proof}

Note that if $\mathcal{C}_{abc}$ were to admit a characterization by a grounded intersection representation, as strings are the most general shape, this would imply $\mathcal{C}_{abc}$ is contained in the class of grounded string graphs. This would resolve the missing inclusion in Figure \ref{fig:strict-inclusions-diagram}.

\subsection{More strict inclusions for classes \texorpdfstring{$\mathcal{C}_{a}$}{Ca}, \texorpdfstring{$\mathcal{C}_{b}$}{Cb} and \texorpdfstring{$\mathcal{C}_{abc}$}{Cabc}}
\label{subsec:more-strict-inclusions}

For this last subsection, we exhibit some graphs that further distinguish the unordered pattern-defined classes. These graphs were found using computer enumeration. 
Remember that inclusion of patterns imply inclusion of classes (Observation~\ref{obs:inclusion-patterns}), thus we have the following inclusions: $\mathcal{C}_{\emptyset} \subseteq \mathcal{C}_a  \subseteq \mathcal{C}_{ab} \subseteq \mathcal{C}_{abc} \subseteq \mathcal{C}_{abcd}\subseteq ALL$, where $ALL$ stands for the set of all graphs. Also $\mathcal{C}_{\emptyset} \subseteq \mathcal{C}_b  \subseteq \mathcal{C}_{ab}$.

Two inclusions are known to be strict:

\begin{observation}
\label{obs:Cempty-Ca-Cb}
The class $\mathcal{C}_{\emptyset}$, is strictly included into $\mathcal{C}_a$ and $\mathcal{C}_b$. 
For example $K_4$ has no ordering avoiding $P_{\emptyset}$, but all orderings avoid $P_a$ and $P_b$.
\end{observation}

We prove three other inequalities between included classes.

\begin{theorem}
\label{thm:pattern-strict}
$\mathcal{C}_a \subsetneq \mathcal{C}_{ab}$, $\mathcal{C}_b \subsetneq \mathcal{C}_{ab}$, and $\mathcal{C}_{abcd} \subsetneq ALL$
\end{theorem}

\begin{proof}
The proof of this theorem uses a computer program for enumerating vertex orderings. The source code  is available upon request.

For the first two inequalities, $\mathcal{C}_a \subsetneq \mathcal{C}_{ab}$ and $\mathcal{C}_b \subsetneq \mathcal{C}_{ab}$, we consider the graph of Figure~\ref{Cab-Ca}. 
This graph is in $\mathcal{C}_{ab}$, as the second pictures shows, and a computer enumeration shows that it is not in $\mathcal{C}_{a}$, nor in $\mathcal{C}_{b}$. 

\begin{figure}[h!]
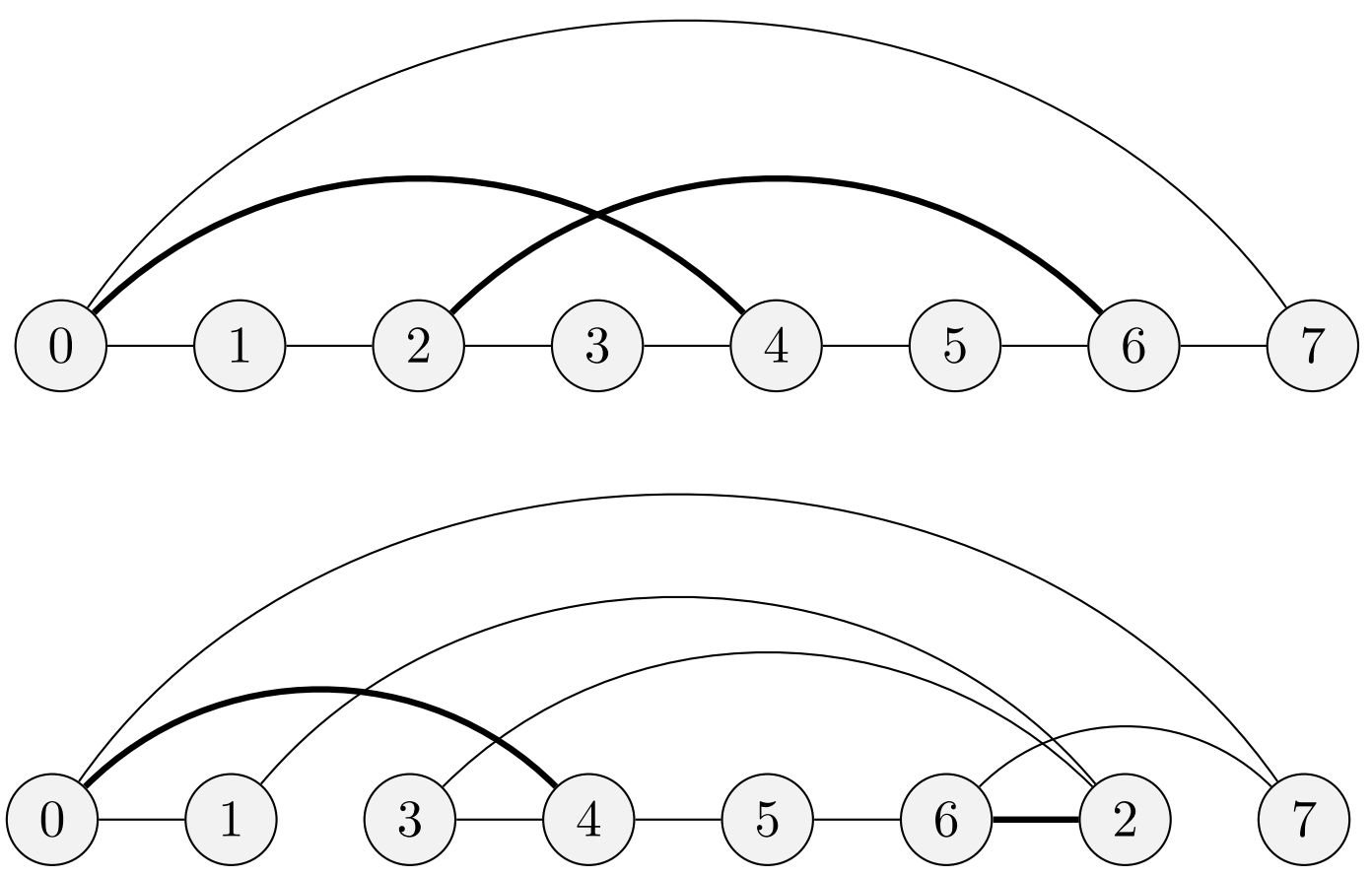

\begin{center}
\input{pictures-and-tikz/strict-a-ab.tex}
\input{pictures-and-tikz/strict-a-ab-2.tex}
\caption{\label{Cab-Ca}
Two vertex orderings of a graph that is in $\mathcal{C}_{ab}$, but not in $\mathcal{C}_a$, nor in $\mathcal{C}_b$. The first pictures illustrates how we discovered this graph: taking $P_{\emptyset}$ and then forcing some order conditions with intermediate nodes and a cycle. The second picture is an ordering avoiding $P_{ab}$. (Some edges are made thicker to increase readability.)}
\end{center}
\end{figure}

For the third inequality, $\mathcal{C}_{abcd} \subsetneq ALL$, we consider the graph of Figure~\ref{Cabcd}. 
Again a computer enumeration shows that no vertex ordering of this graph avoids the pattern $P_{abcd}$, thus the graph is not in $\mathcal{C}_{abcd}$.

\begin{figure}[h!]
\begin{center}
\input{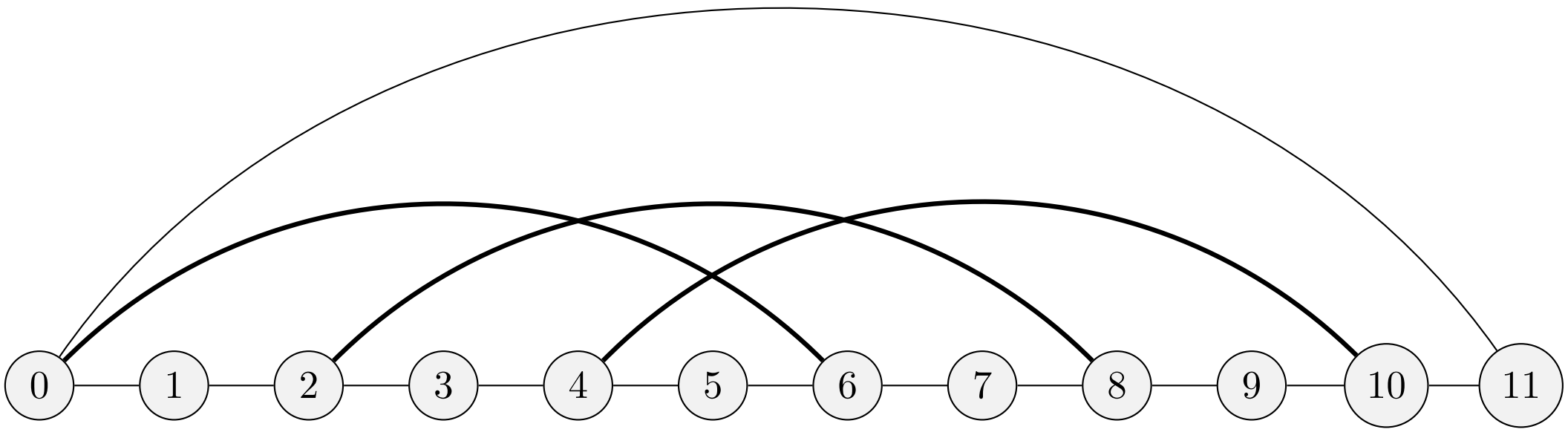}
\caption{\label{Cabcd}A graph that is not in $\mathcal{C}_{abcd}$, by computer enumeration. (We made some edges thicker to increase readability.)}
\end{center}
\end{figure}
\end{proof}

\section{Strengthening an ordered result: (unordered) monotone string graphs and pattern characterizations}
\label{sec:no-characterization-patterns-unordered-monotone-strings}

In this section, we use the toolbox introduced in Section \ref{sec:toolbox-cycle-lemma} to strengthen the result presented in Section \ref{sec:no-characterization-by-patterns}:

\thmNoCharacterizationGroundedMonotone*

We start by introducing an additional notation to the vertices of the underlying graph of a cocomparability sun $(G,<)$ of size $4k$, with $G=(V,E)$ and $k > 2$, as shown in Figure \ref{fig:cocomparability-sun}. We will denote the vertices of $G$ by $x_1,...,x_{4k}$ following the natural cyclic ordering of $G$. More formally, $x_i$ will correspond to vertex at index $i$ in ordering $<$, for $1 \leq i \leq 2k$. Then, we reach a breaking point in the notation, as $x_{2k+1}$ will correspond to the last vertex in $<$, i.e. vertex at index $4k$. This way, $x_{i}$ will denote vertex at index $6k-i+1$ in ordering $<$, for $2k+1 \leq i \leq 4k$. This is illustrated by Figure \ref{fig:cocomparability-sun-natural-cyclical-ordering}.

\input{pictures-and-tikz/cocomp-sun-natural-cyclical-ordering}

Now, let us propose two preliminary propositions that will be useful for proving Theorem \ref{theo:no-finite-patterns-characterize-grounded-monotone}. Let $(G,<)$ be a cocomparability sun. Our strategy will be to apply the Cycle Lemma to $(G,<)$, i.e. Lemma \ref{lem:cycle-extension-grounded-string-representation}, in order to obtain an unordered graph $H$ such that the ordering of vertices in $G$ will be equivalent to $<$ in any grounded monotone string representation of $H$. Recall that an ordering $<'$ is equivalent to an ordering $<$ if it can be obtained from $<$ by applying a cyclic shift and/or inversion.
We know, from Proposition \ref{propos:cocomparability-sun}, that $(G,<)$ does not admit a constrained monotone string representation. However, this is not sufficient, as we need to argue now that $(G,<')$, with $<'$ an equivalent ordering to $<$, does not admit a grounded monotone string representation neither. In order to do this, we introduce Proposition \ref{prop:cyclical-directed-height-graph-equivalent-ordering} and \ref{prop:inverse-ordering-cocomp-sun-isomorphic}, that will be useful when studying the directed height graph of $(G,<')$ when $<'$ is a cyclic shift or an inversion of $<$ respectively.

\begin{proposition}
Let $(G,<)$ be a cocomparability sun and let the ordering $<'$ be an arbitrary cyclic shift of the ordering $<$. Then, $D_h(G,<')$ has a directed cycle.
\label{prop:cyclical-directed-height-graph-equivalent-ordering}
\end{proposition}

\begin{proof}

Let $(G,<)$ be a cocomparability sun of size $4k$, with $G=(V,E)$ and $k > 2$, as shown in Figure \ref{fig:cocomparability-sun}. We start by distinguishing the vertices of $G$ into two classes. We will say a vertex is a $v$-vertex if its index in the natural cyclic ordering notation is odd (outer green vertices in Figure \ref{fig:cocomparability-sun-natural-cyclical-ordering}) and a $u$-vertex if its index is even (inner blue vertices in Figure \ref{fig:cocomparability-sun-natural-cyclical-ordering}).

Now, for every ordering $<'$, such that $<'$ is a cyclic shift of ordering $<$, we will also partition the vertices of $G$ into four sets $M_0, M_1, M_2$ and $M_3$. We define the sets to be such that every $M_i$ contains vertices that are consecutive both in the natural cyclic ordering and in $<'$, for $0 \leq i \leq 3$. Also, we ask that $M_0 <' M_1 <' M_2 <' M_3$ (for two sets $X$ and $Y$, $X <' Y$ implies that all the elements of $X$ are strictly smaller than all the elements in $Y$ in ordering $<'$). By observing the ordered graph $(G,<)$, we note that the ordering of the vertices $<$ does not follow the natural cyclic ordering of the underlying graph $G$. Hence, applying a cyclic shift to $<$ will produce a particular rotation of the indexes of the vertices in $G$. Figure \ref{fig:cocomp-sun-cyclical-shift} aims at representing this graphically.


\input{pictures-and-tikz/cocomp-sun-cyclic-index-flow}

Then, for $<'$ a cyclic shift of ordering $<$, we argue that there are only four possible configurations for the sets $M_0, M_1, M_2$ and $M_3$, depicted in Figure \ref{fig:cocomparability-sun-possible-configurations}. Note that sets $M_0, M_1, M_2$ and $M_3$ are not fixed for all cyclic shifts of $<$, their elements and size will depend on the shift applied to $<$. However, these sets will always serve as delimiters of consecutive vertices in ordering $<'$. Observe that by the means of cyclic shifts of the initial ordering $<$, we can go from the initial configuration to Configuration 2, then to Configuration 3, and Configuration 4, until reaching once again Configuration 1. Hence, if $<'$ is obtained from applying a cyclic shift to $<$, it necessarily is in one of the four possible configurations. Then, our strategy will consist of proving the existence of a directed cycle in the directed height graph $D_h(G,<')$, for $<'$ in each valid configuration.

For Configuration $i$, the sets $M_0, M_1, M_2$ and $M_3$ will be such that:
\begin{itemize}
    \item $M_{1-i}$ contains vertices $x_1, x_2, ..., x_{l}$ with $l$ such that $1 < l < 2k$.
    \item $M_{2-i}$ contains vertices $x_{l + 1}, ..., x_{2k}$.
    \item $M_{3-i}$ contains vertices $x_{2k+1}, ..., x_{m}$ with $m$ such that $2k+1 < m < 4k$.
    \item $M_{4-i}$ contains vertices $x_{m + 1}, ..., x_{4k}$.
\end{itemize}

The subindices are taken modulo 4. Note that, for Configuration $i$, if $x_l, x_j \in M_{1-i}$ or $x_l, x_j \in M_{2-i}$, with $l<j$, then $x_l <' x_j$. However, if $x_l, x_j \in M_{3-i}$ or $x_l, x_j \in M_{4-i}$, with $l<j$, then $x_j <' x_l$.

\usetikzlibrary{calc, backgrounds, arrows.meta}
\colorlet{sg}{black!40}   
\tikzset{sv/.style={draw=sg, circle, inner sep=2pt, fill=sg}}

\colorlet{colM0}{red!60!black}
\colorlet{colM1}{blue!60!black}
\colorlet{colM2}{green!45!black}
\colorlet{colM3}{orange!70!black}

\newcommand{\sunconfig}[5]{%
\begin{tikzpicture}[scale=0.25]

\def\r{5}
\def\R{8}

\node (v1) at ({\r*cos(0)}, {\r*sin(0)})     [sv, label=180:$x_{2k}$] {};
\node (v2) at ({\r*cos(45)}, {\r*sin(45)})   [sv] {};
\node (v3) at ({\r*cos(90)}, {\r*sin(90)})   [sv] {};
\node (v4) at ({\r*cos(135)}, {\r*sin(135)}) [sv] {};
\node (v5) at ({\r*cos(180)}, {\r*sin(180)}) [sv, label=0:$x_{4k}$] {};
\node (v6) at ({\r*cos(225)}, {\r*sin(225)}) [sv] {};
\node (v7) at ({\r*cos(270)}, {\r*sin(270)}) [sv] {};
\node (v8) at ({\r*cos(315)}, {\r*sin(315)}) [sv] {};

\node (v9)  at ({\R*cos(22.5)},  {\R*sin(22.5)})  [sv] {};
\node (v10) at ({\R*cos(67.5)},  {\R*sin(67.5)})  [sv] {};
\node (v11) at ({\R*cos(112.5)}, {\R*sin(112.5)}) [sv] {};
\node (v12) at ({\R*cos(157.5)}, {\R*sin(157.5)}) [sv, label=190:$x_1$] {};
\node (v13) at ({\R*cos(202.5)}, {\R*sin(202.5)}) [sv] {};
\node (v14) at ({\R*cos(247.5)}, {\R*sin(247.5)}) [sv] {};
\node (v15) at ({\R*cos(292.5)}, {\R*sin(292.5)}) [sv] {};
\node (v16) at ({\R*cos(337.5)}, {\R*sin(337.5)}) [sv, label=360:$x_{2k+1}$] {};

\node at (0,-14.5) {\underline{#1}};

\def\RA{10.5}
\def\RB{10.5}
\def\RC{10.5}
\def\RD{10.5}
\def\tick{0.22}
\def\signr{1.0}   
\def\labr{1.6}    

\draw[thick, colM#2]
    ({(\RA-\tick)*cos(157.5)}, {(\RA-\tick)*sin(157.5)}) -- ({(\RA+\tick)*cos(157.5)}, {(\RA+\tick)*sin(157.5)});
\node[colM#2, scale=0.8] at ({(\RA+\signr)*cos(157.5)}, {(\RA+\signr)*sin(157.5)}) {$-$};
\draw[-{Stealth[length=3mm,width=2mm]}, thick, colM#2]
    ({\RA*cos(157.5)}, {\RA*sin(157.5)}) arc (157.5:90:\RA);
\node[colM#2, scale=0.8] at ({(\RA+\signr)*cos(90)}, {(\RA+\signr)*sin(90)}) {$+$};
\node[colM#2] at ({(\RA+\labr)*cos(123.75)}, {(\RA+\labr)*sin(123.75)}) {$M_{#2}$};

\draw[thick, colM#3]
    ({(\RB-\tick)*cos(67.5)}, {(\RB-\tick)*sin(67.5)}) -- ({(\RB+\tick)*cos(67.5)}, {(\RB+\tick)*sin(67.5)});
\node[colM#3, scale=0.8] at ({(\RB+\signr)*cos(67.5)}, {(\RB+\signr)*sin(67.5)}) {$-$};
\draw[-{Stealth[length=3mm,width=2mm]}, thick, colM#3]
    ({\RB*cos(67.5)}, {\RB*sin(67.5)}) arc (67.5:0:\RB);
\node[colM#3, scale=0.8] at ({(\RB+\signr)*cos(0)}, {(\RB+\signr)*sin(0)}) {$+$};
\node[colM#3] at ({(\RB+\labr)*cos(33.75)}, {(\RB+\labr)*sin(33.75)}) {$M_{#3}$};

\draw[thick, colM#4]
    ({(\RC-\tick)*cos(270)}, {(\RC-\tick)*sin(270)}) -- ({(\RC+\tick)*cos(270)}, {(\RC+\tick)*sin(270)});
\node[colM#4, scale=0.8] at ({(\RC+\signr)*cos(270)}, {(\RC+\signr)*sin(270)}) {$-$};
\draw[{Stealth[length=3mm,width=2mm]}-, thick, colM#4]
    ({\RC*cos(337.5)}, {\RC*sin(337.5)}) arc (337.5:270:\RC);
\node[colM#4, scale=0.8] at ({(\RC+\signr)*cos(337.5)}, {(\RC+\signr)*sin(337.5)}) {$+$};
\node[colM#4] at ({(\RC+\labr)*cos(303.75)}, {(\RC+\labr)*sin(303.75)}) {$M_{#4}$};

\draw[thick, colM#5]
    ({(\RD-\tick)*cos(180)}, {(\RD-\tick)*sin(180)}) -- ({(\RD+\tick)*cos(180)}, {(\RD+\tick)*sin(180)});
\node[colM#5, scale=0.8] at ({(\RD+\signr)*cos(180)}, {(\RD+\signr)*sin(180)}) {$-$};
\draw[{Stealth[length=3mm,width=2mm]}-, thick, colM#5]
    ({\RD*cos(247.5)}, {\RD*sin(247.5)}) arc (247.5:180:\RD);
\node[colM#5, scale=0.8] at ({(\RD+\signr)*cos(247.5)}, {(\RD+\signr)*sin(247.5)}) {$+$};
\node[colM#5] at ({(\RD+\labr)*cos(213.75)}, {(\RD+\labr)*sin(213.75)}) {$M_{#5}$};

\foreach \t in {0.45,0.5,0.55} {
\node[sg] at ($(v7)!\t!(v14)$) {$\cdot$};
}
\foreach \t in {0.45,0.5,0.55} {
\node[sg] at ($(v10)!\t!(v3)$) {$\cdot$};
}
\foreach \t in {0.45,0.5,0.55} {
\node[sg] at ($(v2)!\t!(v3)$) {$\cdot$};
}
\foreach \t in {0.45,0.5,0.55} {   
\node[sg] at ($(v6)!\t!(v7)$) {$\cdot$};
}
\foreach \t in {0.45,0.5,0.55} {
\node[sg] at ($(v10)!\t!(v11)$) {$\cdot$};
}
\foreach \t in {0.45,0.5,0.55} {
\node[sg] at ($(v14)!\t!(v15)$) {$\cdot$};
}

\draw[sg] (v12) to[bend left=20] (v11);
\draw[sg] (v12) to[bend left=25] (v3);
\draw[sg] (v9)  to[bend left=15] (v16);
\draw[sg] (v9)  to[bend left=25] (v8);
\draw[sg] (v16) to[bend left=20] (v15);
\draw[sg] (v16) to[bend left=30] (v7);
\draw[sg] (v11) to[bend left=50] (v13);
\draw[sg] (v9)  to[bend right=50] (v15);
\draw[sg] (v13) to[bend left=25] (v12);
\draw[sg] (v13) to[bend left=35] (v4);
\draw[sg] (v10) to[bend left=35] (v1);
\draw[sg] (v14) to[bend left=35] (v5);
\draw[sg] (v10) to[bend left=20] (v9);
\draw[sg] (v14) to[bend left=20] (v13);

\draw[dashed, sg] (v12) -- (v4);
\draw[dashed, sg] (v4) -- (v11);
\draw[dashed, sg] (v11) -- (v3);
\draw[dashed, sg] (v3) -- (v4);
\draw[dashed, sg] (v2) -- (v1);
\draw[dashed, sg] (v2) -- (v9);
\draw[dashed, sg] (v8) -- (v1);
\draw[dashed, sg] (v8) -- (v16);
\draw[dashed, sg] (v1) -- (v16);
\draw[dashed, sg] (v1) -- (v9);
\draw[dashed, sg] (v5) -- (v12);
\draw[dashed, sg] (v13) -- (v5);
\draw[dashed, sg] (v13) -- (v6);
\draw[dashed, sg] (v5) -- (v6);
\draw[dashed, sg] (v8) -- (v7);
\draw[dashed, sg] (v4) -- (v5);
\draw[dashed, sg] (v15) -- (v7);
\draw[dashed, sg] (v15) -- (v8);
\draw[dashed, sg] (v10) -- (v2);
\draw[dashed, sg] (v6) -- (v14);
\end{tikzpicture}%
}

\begin{figure}[H]
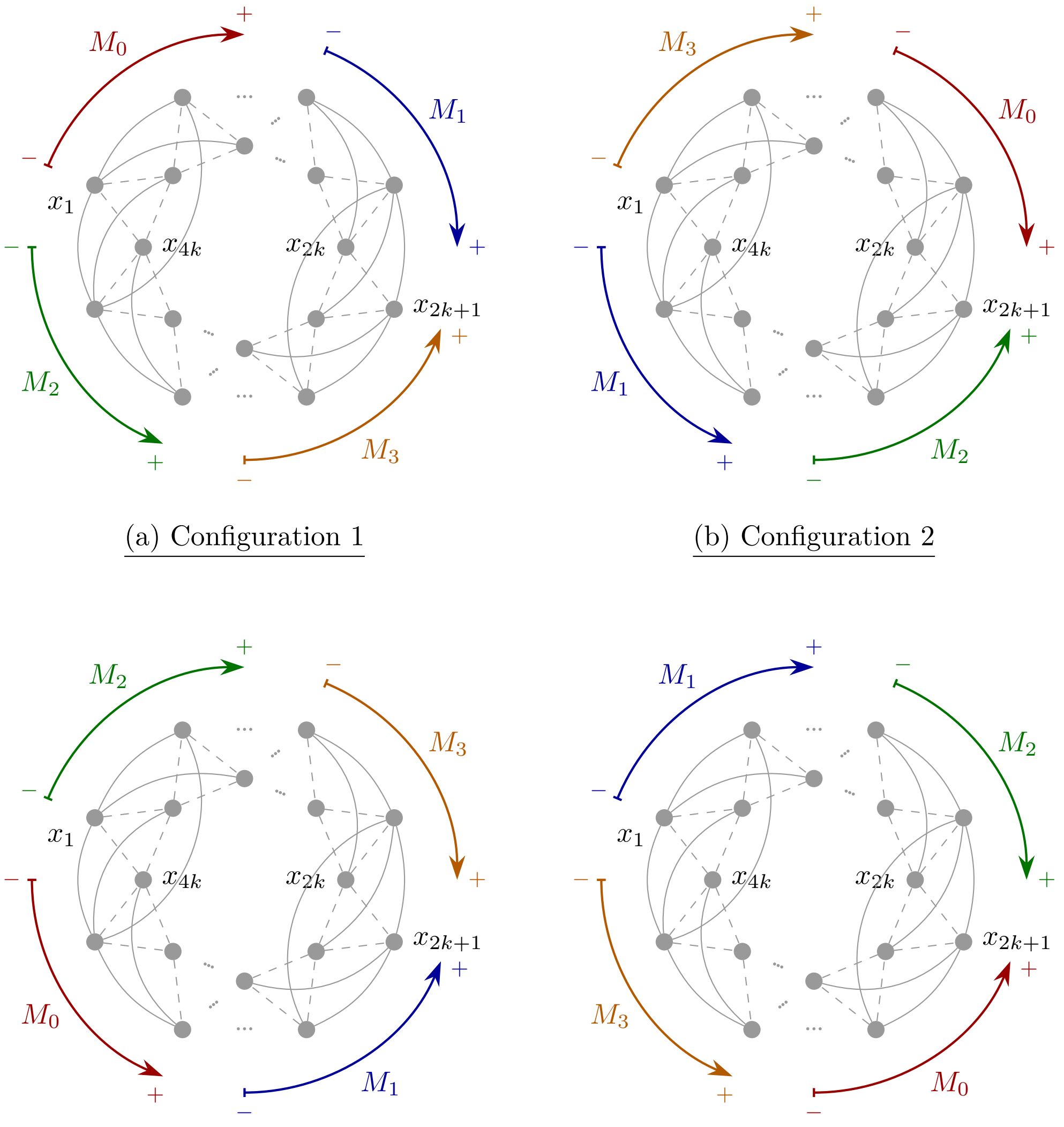

\centering
\sunconfig{(a) Configuration 1}{0}{1}{3}{2}\hspace{5ex}
\sunconfig{(b) Configuration 2}{3}{0}{2}{1}

\vspace{5ex}
\sunconfig{(c) Configuration 3}{2}{3}{1}{0}\hspace{5ex}
\sunconfig{(d) Configuration 4}{1}{2}{0}{3}
\caption{Possible configurations for the sets $M_0, M_1, M_2$ and $M_3$.}
\label{fig:cocomparability-sun-possible-configurations}
\end{figure}

Configuration 1, or the initial configuration, corresponds to the original ordering $<$, that is applying 0 (modulo $4k$) shifts, illustrated in Figure \ref{fig:cocomparability-sun-possible-configurations} $(a)$. Then, by Proposition \ref{propos:cocomparability-sun}, $D_h(G,<)$ has a directed cycle.
 
We now study Configuration 2, depicted in Figure \ref{fig:cocomparability-sun-possible-configurations} $(b)$. Let $v^3$ be the greatest $v$-vertex in $M_3$, and let $u_1^0$ be the $u$-vertex in $M_0$ adjacent to $v^3$. Note that $u_1^0$ is not necessarily the smallest $u$-vertex in $M_0$. Let $u^2$ be the greatest $u$-vertex in $M_2$. We claim there is a directed path in $D_h(G,<')$ from $u_1^0$ to $u^2$. Let $v_1^0$ be the $v$-vertex in $M_0$ immediately before $u_1^0$ in $<'$, note that $v_1^0$ exists from the choice of $v^3$. Starting from $v_1^0$, we denote in order the $v$-vertices of $M_0$ by $v_1^0, v_2^0, v_3^0, ..., v_r^0$ and, starting from $u_1^0$, the $u$-vertices of $M_0$ by $u_1^0, u_2^0, u_3^0, ..., u_r^0$, with $v_r^0$ and $u_r^0$ the greatest vertices of $M_0$. Then, from the structure of $G$, for $1 \leq i < r$, the triplet $v_i^0, u_i^0, u_{i+1}^0$, such that $v_i^0 <' u_i^0 <' u_{i+1}^0$, is a cocomparability triplet and therefore arc $(u_i^0 \rightarrow u_{i+1}^0)$ is in $D_h(G,<')$. Also, the vertices $v_r^0, u_r^0, u^2$, such that $v_r^0 <' u_r^0 <' u^2$ also form a cocomparability triplet, as $(v_r^0, u^2)\in E$, $(v_r^0, u_r^0)\notin E$ and $(u_r^0,u^2)\notin E$. Then, $(u_r^0 \rightarrow u^2) \in D_h(G,<')$. Now, consider vertices $u_1^0 <' u^2 <' v^3$. As $(u_1^0, v^3)\in E$, $(u_1^0, u^2)\notin E$ and $(u^2, v^3)\notin E$, this is a cocomparability triplet as well and therefore $(u^2 \rightarrow u_1^0)$ is an arc of the directed height graph. Then, $u_1^0, u_2^0, ..., u^2$ is a directed cycle in $D_h(G,<')$.
    
\input{pictures-and-tikz/cocomp-sun-configuration2}

Next, we will handle Configuration 3. Refer to Figure \ref{fig:cocomparability-sun-possible-configurations} $(c)$ for an illustration of this configuration. Let $v_1^0$ be the smallest $v$-vertex of $M_0$ and $v_1^2$ be the smallest $v$-vertex of $M_2$. Let $u^1$ be the greatest $u$-vertex of $M_1$ and $v^3$ be the greatest $v$-vertex of $M_3$. Consider the triplet $v_1^0 <' u^1 <' v_1^2$ with $(v_1^0,v_1^2)\in E$, $(v_1^0, u^1)\notin E$ and $(u^1, v_1^2)\notin E$. As these vertices form a cocomparability triplet, we state that $(u^1 \rightarrow v_1^2) \in D_h(G,<')$. Now, we see that triplet $u^1 <' v_1^2 <' v^3$ is also a cocomparability triplet as $(u^1,v^3)\in E$, $(u^1,v_1^2)\notin E$ and $(v_1^2,v^3)\notin E$. Then, $(v_1^2 \rightarrow u^1) \in D_h(G,<')$. The latter implies that vertices $v_1^2$ and $u^1$ form a directed cycle of length 2 in $D_h(G,<')$.
    
\input{pictures-and-tikz/cocomp-sun-configuration3}
    
Finally, we will consider Configuration 4, depicted in Figure \ref{fig:cocomparability-sun-possible-configurations} $(d)$. We will use a similar strategy as in configuration 2. Let $v_1^0$ be the smallest $v$-vertex of $M_0$ and let $u_r^3$ be the $u$-vertex in $M_3$ adjacent to $v_1^0$. Note that $u_r^3$ might not be the greatest $u$-vertex in $M_3$. We will denote by $v_r^3$ the $v$-vertex in $M_3$ immediately after in $<'$ to $u_r^3$. From $v_r^3$, we denote in decreasing $<'$ order the $v$-vertices of $M_3$ by $v_r^3, v_{r-1}^3, v_{r-2}^3, ..., v_1^3$ and, starting from $u_r^3$, the $u$-vertices of $M_3$ by $u_r^3, u_{r-1}^3, u_{r-2}^3, ..., u_1^3$, with $v_1^3$ and $u_1^3$ the smallest vertices of $M_3$. Then, from the structure of $G$, for $1 < i \leq r$, the triplet $v_i^3, u_i^3, u_{i-1}^3$, such that $u_{i-1}^3 <' u_i^3 <' v_i^3$, is a cocomparability triplet and therefore arc $(u_i^3 \rightarrow u_{i-1}^3)$ is in $D_h(G,<')$. Hence, there exists a directed path in $D_h(G,<')$ from $u_r^3$ to $u_1^3$. Now, if we let $u_1^1$ be the smallest $u$-vertex in $M_1$, we see that $u_1^1 <' u_1^3 <' v_1^3$ with $(u_1^1, v_1^3)\in E$, $(u_1^1, u_1^3)\notin E$ and $(u_1^3, v_1^3)\notin E$. Then, necessarily $(u_1^3 \rightarrow u_1^1) \in D_h(G,<')$. Finally, consider vertices $v_1^0 <' u_1^1 <' u_r^3$. As $(v_1^0, u_r^3) \in E$, $(v_1^0, u_1^1) \notin E$ and $(u_1^1, u_r^3) \notin E$, we get that $(u_1^1 \rightarrow u_r^3) \in D_h(G,<')$. Then, $u_r^3, u_{r-1}^3, ..., u_1^1$ form a directed cycle in $ D_h(G,<')$.
    
\input{pictures-and-tikz/cocomp-sun-configuration4}

For every possible configuration of $<'$, we found a directed cycle in $D_h(G,<')$, proving that the proposition holds.
\end{proof}

\begin{proposition}
Let $(G,<)$ be a cocomparability sun and let the ordering $<^{-1}$ be the inverse ordering of $<$. Then, $(G,<^{-1})$ is isomorphic to $(G,<)$.
\label{prop:inverse-ordering-cocomp-sun-isomorphic}
\end{proposition}

\begin{proof}
Let $(G,<)$ be a cocomparability sun of size $4k$, with $k > 2$. We will study $(G,<^{-1})$, in order to prove the existence of an isomorphism between this ordered graph and the original one, that is $(G,<)$. We start by observing that every vertex in the underlying graph $G$ having index $i$ in ordering $<$, with $1 \leq i \leq 4k$, will have index $4k-i+1$ in ordering $<^{-1}$. Then, by definition of the cocomparability sun $(G,<)$, we get that $(G,<^{-1})$ contains the following edges:

\begin{itemize}
    \item $(4k-i+1, 4k-i-1)$ and $(4k-i+1, 4k-i-2)$ for every odd $i$, with $i < 2k-1$.
    \item $(4k-i+1, 4k-i+3)$ and $(4k-i+1, 4k-i+4)$ for every even $i$, with $i > 2k+1$.
    \item all edges of the join between vertices in $\{4k-3, 4k-4, ..., 2k+1 \}$ and vertices in $\{2k, 2k-1, ..., 4 \}$
    \item $(2k+2,1)$
    \item $(2k+2,2)$ 
    \item $(2k-1,4k)$
    \item $(2k-1,4k-1)$
    \item $(4k-2,2k-1)$
    \item $(2k+2,3)$ 
\end{itemize}

Hence, after the conversion of the indices from ordering $<$ to ordering $<^{-1}$, we obtain the ordered graph $(G,<^{-1})$. We can see both ordered graphs in Figure \ref{fig:cocomp-sun-and-inversed-ordered-cocomp-sun}. Consider the natural cyclic notation of the underlying graph $G$, where vertices are denoted by $x_1,...,x_{4k}$. Then, we claim that the bijection defined by $f(x_i) = x_{i+2k}$, where $i+2k$ is taken modulo $4k$, is a valid ordered isomorphism, as it maintains both adjacencies and relative ordering between vertices. We conclude that $(G,<)$ and $(G,<^{-1})$ are isomorphic (see Definition \ref{def:ordered-isomorphic}).

\begin{figure}[h!]
    \centering
    \input{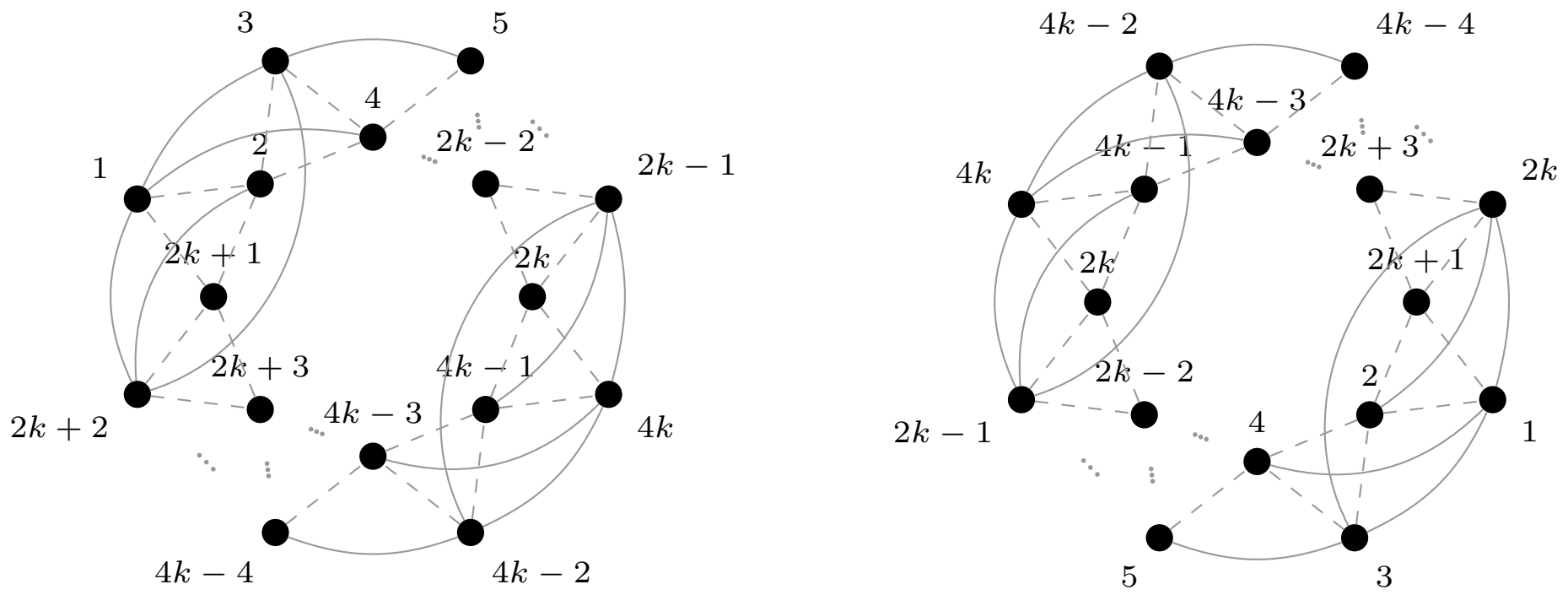}
    \caption{Ordered graphs $(G,<)$ (left) and $(G,<^{-1})$ (right).}
    \label{fig:cocomp-sun-and-inversed-ordered-cocomp-sun}
\end{figure}
\end{proof}

We are now ready to prove Theorem \ref{theo:no-finite-patterns-characterize-grounded-monotone}.

\begin{proof}[Proof of Theorem \ref{theo:no-finite-patterns-characterize-grounded-monotone}.]

For contradiction, suppose that $\mathcal{F}$ is a finite family of patterns
satisfying items~1 and~2 of
Theorem~\ref{theo:no-finite-patterns-characterize-grounded-monotone}, and let
$M$ be the number of vertices of the largest pattern in $\mathcal{F}$. 
Take a cocomparability sun $(G,<)$ of size $4k$, with $k>2$, such that $2k > M$. Let $H$ be the cycle extension of $(G,<)$. By Proposition \ref{prop:cyclical-directed-height-graph-equivalent-ordering}, for any ordering $<'$ obtained from applying a cyclic shift to $<$, the associated directed height graph $D_h(G,<')$ will have a directed cycle. 
Then, by Proposition \ref{prop:PachTomonDh}, $(G,<')$ will not have a constrained monotone string representation. By Proposition \ref{prop:inverse-ordering-cocomp-sun-isomorphic}, when inverting ordering $<$, we obtain $(G,<^{-1})$ ordered isomorphic to $(G,<)$. Then, we deduce that any ordering $<'$ equivalent to $<$, i.e. an ordering obtained from applying a cyclic shift and/or an inversion to $<$, will produce an ordered graph $(G,<')$ having no constrained monotone string representation. However, by Lemma \ref{lem:cycle-extension-grounded-string-representation} (Cycle Lemma), in any grounded monotone string representation of $H$, the vertices $V_G$ (i.e. induced by $G$) will be ordered following an equivalent ordering to $<$. Then, we deduce $H$ has no grounded monotone string representation.

Let $v_1, ..., v_{n}$ be the vertices of $G$, also denoted by $V_G$, ordered by $<$ and $y_1, ..., y_{5n}$ be the vertices added by the cycle extension operation, also denoted by $H \setminus V_G$, ordered in the cycle ordering. We start by defining the sub-ordering $i$, denoted by $s_i$, as the ordered set of vertices $y_{5i-4}, y_{5i-3}, y_{5i-2}, y_{5i-1}, y_{5i}, v_i$, for $1 \leq i \leq n$. Then, we define $<_H$ as the ordering of vertices in $H$ obtained by concatenating the $s_i$ in order, such that we finally get $y_1 <_H y_2 <_H y_3 <_H y_4 <_H y_5 <_H v_1 <_H y_6 <_H ... <_H y_{5n-1} <_H y_{5n} <_H v_n$. Observe that $<_H$ is such that:

\begin{itemize}
    \item The vertices in $V_G$ follow the relative ordering defined by $<$ in $<_H$.
    \item The vertices in $H \setminus V_G$ follow the cycle ordering in $<_H$.
    \item Every vertex in $V_G$ is consecutive in $<_H$ to its single adjacent vertex in $H \setminus V_G$.
\end{itemize}

Now, we claim the following about this ordering $<_H$:

\begin{claim}
No directed cycle in $D_h(H,<_H)$ contains vertices of $H \setminus V_G$.
\end{claim}

\begin{proof}[Proof of Claim]

For the sake of contradiction, suppose there is a directed cycle in $D_h(H,<_H)$ containing a vertex $y_j$ of $H \setminus V_G$, with $1 \leq j \leq 5n$. We argue that for vertex $y_j$ to belong to a directed cycle in $D_h(H,<_H)$, there needs to be an arc directed towards $y_j$. The latter implies $y_j$ is the \enquote{extremity} of a cocomparability triplet in $(H,<_H)$, i.e. there exist two vertices $u,v \in H$, with $(u,y_j)\in H$, $(u,v)\notin H$ and $(v,y_j)\notin H$, such that either $u <_H v <_H y_j$ or $y_j <_H v <_H u$. If $u \in V_G$, we reach a contradiction as $u$ and $y_j$ will be consecutive in $<_H$ by construction, implying that there is no $v$ such that $u,v$ and $y_j$ form a cocomparability triplet. Now, if $u \in H \setminus V_G$, then $u$ is either $y_{j-1}$ or $y_{j+1}$ (modulo $5n$). We reach a contradiction as well, as either $u$ and $y_j$ are consecutive in $<_H$, or there is only a vertex of $V_G$ between them, necessarily adjacent to one of them and therefore incapable of forming a cocomparability triplet with $u$ and $y_j$.
\end{proof}


We deduce that the only directed cycles in $D_h(H,<_H)$ are the ones formed by vertices in $V_G$. Note that, since the ordering $<$ of vertices in $V_G$ is maintained in $<_H$, we argue that $D_h(G,<)$ is an induced directed subgraph of $D_h(H,<_H)$. Then, by Proposition \ref{propos:cocomparability-sun} and the previous claim, $D_h(H,<_H)$ has directed girth of $2k$.

The fact that $H$ has no constrained monotone string representation implies that every ordering of $H$ should reproduce a pattern in $\mathcal{F}$ by item $(2.)$ of the theorem's statement. In particular, $<_H$ reproduces a pattern $P$ in $\mathcal{F}$. Let $V(H,<_H,P)$ be a set of nodes reproducing pattern $P$ in $H$ ordered by $<_H$, and let $(K,<_K)$ be the ordered subgraph induced by $V(H,<_H,P)$. Then, by item $(1.)$ of the theorem's statement, $<_K$ cannot be the grounding ordering of $K$ in a valid grounded monotone string representation, i.e. $(K, <_K)$ is not in the class of constrained monotone string graphs. By Proposition \ref{prop:PachTomonDh}, this implies $D_h(K,<_K)$ contains a directed cycle. However, as $D_h(K,<_K)$ must be an induced directed subgraph of $D_h(H,<_H)$ of size at most $M$ and $D_h(H,<_H)$ has directed girth of size $2k > M$, we reach a contradiction.
\end{proof}

\appendix
\bibliography{biblio.bib}{}
\bibliographystyle{plain}


\end{document}